\documentstyle[emulateapj,psfig]{article}

\begin{document}

\submitted{To appear in ApJ}
\title{
The Formation of the First Stars. I. The Primordial Star Forming
Cloud}

\author{Volker Bromm$^{1}$, 
Paolo S. Coppi and Richard B. Larson}
\affil{Department of Astronomy, Yale University, New Haven, CT 06520-8101;\\
volker@ast.cam.ac.uk, coppi@astro.yale.edu, larson@astro.yale.edu}

\begin{abstract}
To constrain the nature of the very first stars, we investigate the
collapse and fragmentation of primordial, metal-free gas clouds. We
explore the physics of primordial star formation by means of three-dimensional
simulations of the dark matter and gas components, using smoothed particle
hydrodynamics, under a 
wide range of initial conditions, including the initial spin, the total mass
of the halo, the redshift of virialization, the power spectrum of the DM
fluctuations, the presence of HD cooling, and the number of particles
employed in the simulation. We find characteristic values for the
temperature, $T\sim$ a few 100 K, and the density, $n\sim 10^{3}-10^{4}$ cm$^{-3}$,
characterising the gas at the end of the initial free-fall phase. These
values are rather 
insensitive to the initial conditions. The corresponding Jeans mass is $M_{J}\sim
10^{3}M_{\odot}$. The existence of these characteristic values has a
robust explanation in the microphysics of H$_{2}$ cooling, connected to the
minimum temperature that can be reached with the H$_{2}$ coolant, and to
the critical density at which the transition takes place between levels
being populated according to NLTE, and according to LTE.

In all cases, the gas dissipatively settles into an irregular, central
configuration which has a filamentary and knotty appearance. The fluid regions
with the highest densities are the first to undergo runaway collapse due
to gravitational instability, and to form clumps with initial
masses $\sim 10^{3}M_{\odot}$, close to the characteristic Jeans scale.
These results suggest
that the first stars might have been quite massive, possibly even very massive
with $M_{\star}\ga 100\:M_{\odot}$.

After a gas element has undergone runaway collapse,
and has reached densities in excess of $10^{8}$ cm$^{-3}$, a sink particle
is created. This procedure allows us to follow the evolution of the overall
system beyond the point where the first nonlinear region would otherwise
force the calculation to a halt.
These later evolutionary stages,
during which the clumps grow in mass due to
accretion and merging with other clumps, are quite sensitive
to the initial conditions.
The key process in building up very massive clumps, with masses up to a few 
times $10^{4}M_{\odot}$, is merging between clumps. Since the merging rate
sensitively depends on the density of the gas, halos with the highest degree
of central concentration are able to assemble the most massive clumps.
Among these are halos with a low spin ($\lambda \simeq 0.01$), and with
DM fluctuations imprinted according to a white-noise spectrum.

\end{abstract}

\footnotetext[1]{Present address: Institute of Astronomy, University
of Cambridge, Madingley Road, Cambridge CB3 0HA, UK}
\section{INTRODUCTION}

This paper is concerned with the initial stages in the formation of
cosmic structure.
How did the universe evolve from the
extreme uniformity of the big bang fireball into its highly organized
and clustered present state? An increasing wealth of observational data
has become available to guide theoretical thinking. The study of anisotropies
in the cosmic microwave background (CMB) provides a window into the earliest
phases of structure formation, when the primordial density fluctuations
were still linear, and therefore amenable to an exact physical description
(e.g., White, Scott, \& Silk 1994; Lawrence, Scott, \& White 1999).
Thus, we have a powerful probe of the last scattering surface at
$z\simeq 1000$, corresponding to $\sim 10^{6}$ yr after the big bang.
Complementary to the CMB studies are observations of high-redshift
quasars and galaxies (e.g., Hu, Cowie, \& McMahon 1998; Chen, Lanzetta, \&
Pascarelle 1999; Fan et al. 2000).
The light from these objects originated at $z\simeq 5$, 
when the universe was $\sim 10^{9}$ yr old. Very little is known
about the crucial era in between, at $z=1000-5$, which 
has been termed the ``dark ages'' (e.g., Loeb 1999, Rees 1999).
This serene epoch ended when the first luminous objects lit up
the universe again. In the context of hierarchical scenarios of structure
formation, as specified by a variant of the cold dark matter (CDM) model,
the collapse of the first baryonic objects is expected at redshifts
$ z\simeq 50-10$, involving dark matter halos of mass $\sim 10^{6}M_{\odot}$
(Tegmark et al. 1997).

The importance of the first stars and quasars derives from the crucial
feedback they exert on the intergalactic medium (IGM). A generation of
stars which formed out of primordial, pure H/He gas (the so-called
Population III) must have existed, since heavy elements can only be
synthesized in the interior of stars. Population III stars, then, were
responsible for the initial enrichment of the IGM with heavy elements. From
the absence of Gunn-Peterson absorption in the spectra of high-redshift
quasars, we know that the universe has undergone a reionization event
at $z > 5$ (Gunn \& Peterson 1965). UV photons from the first stars, perhaps
together with an early population of quasars, are expected to have contributed
to the reionization of the IGM (Haiman \& Loeb 1997; Ferrara 1998;
Miralda-Escud\'{e}, Haehnelt, \& Rees 2000). The energy input from the first
stars might have left a measurable imprint on the CMB on very small scales
(Vishniac 1987; Dodelson \& Jubas 1995).

Despite an intense observational effort, the discovery of a true Population~III
star remains elusive. Surveys of the metal-poor population in the halo
of our Galaxy have resulted in stars with metallicities $Z \ga 10^{-4} Z_{\odot}$ 
(Beers 2000). Spectroscopic studies of high-redshift Lyman $\alpha$ clouds,
the supposedly most pristine objects in the universe, find a minimum
metallicity $Z_{min}\sim 10^{-3}Z_{\odot}$ (Cowie \& Songaila 1998). Therefore,
wherever we look, we find contaminated material. To probe the time
when star formation first started, consequently entails observing at even 
higher redshift. This is the main purpose of the {\it Next Generation
Space Telescope} (NGST) which is designed to reach $\sim$ nJy sensitivity
at near-infrared wavelengths (Loeb 1998). In preparation for this upcoming
observational revolution, the study of the first stars is very timely,
providing a theoretical framework for the interpretation of what NGST
might discover, less than a decade from now.

The question arises how one can make any progress in understanding 
primordial star formation, given the lack of direct observational
constraints. The physics of the first stars, however, is characterized 
by some important simplifications, as compared to the extreme complexity
of present-day star formation (Larson 1998; Loeb 1998). The absence of metals,
and consequently of dust, leaves atomic and molecular hydrogen as the main
agent of radiative cooling and the source of opacity. Magnetic fields were
likely to be dynamically insignificant, prior to the onset of efficient
(stellar) dynamo amplification (Kulsrud 1997). The chemistry and heating
of the primordial gas was not yet complicated by the presence of a UV
radiation background. The IGM must have been a rather quiescent place,
with no source to sustain turbulent motion, as long as the
density perturbations remain in their linear stage. Only after the
explosion of the first supernovae, and the associated input of mechanical
and thermal energy, is this state of primordial tranquility bound to change
 (Loeb \& Haiman 1997; Miralda-Escud\'{e} \& Rees 1997). Therefore, the
physics of primordial star formation is mainly governed by gravity, thermal
pressure, and angular momentum. This situation renders the problem
theoretically more straightforward and tractable than the highly complex
present-day case which continues to defy attempts to formulate a 
fundamental theory of star formation.
Finally, the initial conditions for the collapse of a primordial star 
forming cloud are given by the adopted model of cosmological structure
formation. The initial abundances of the chemical species are predicted
by standard Big-Bang nucleosynthesis (e.g., Copi, Schramm, \& Turner 1995).

In this paper, we investigate the question: {\it How do the first primordial
star forming clouds evolve in the context of a hierarchical model of
structure formation?} The collapse of the clouds, having masses close to
the cosmological Jeans mass ($\sim 10^{6}M_{\odot}$), results in the formation
of high density clumps. These clumps are the immediate progenitor of
primordial protostars. This second stage in the primordial star formation
process will be discussed in a subsequent paper (henceforth Paper II).

The subject of the formation of the first stars has a long
and venerable history (e.g., Yoneyama 1972; Hutchins 1976; Silk 1977, 1983;
Carlberg 1981; Kashlinsky \& Rees 1983; Palla, Salpeter, \& Stahler 1983;
Carr, Bond, \& Arnett 1984;
Couchman \& Rees 1986; Haiman, Thoul, \& Loeb 1996; Uehara et al. 1996;
Tegmark et al. 1997; Omukai \& Nishi 1998; Nakamura \& Umemura 1999).
Recently, it has become possible to address this problem in the context
of full cosmological simulations, due to dramatic improvements in
numerical resolution, and in the modelling of the relevant gas physics (Anninos
\& Norman 1996; Ostriker \& Gnedin 1996; Abel et al. 1998; Abel, Bryan, \&
Norman 2000, ABN2000 henceforth; Fuller \& Couchman 2000).

Our approach is complementary to these last studies in that we focus
on the collapse of an isolated overdensity. This choice has its advantages
and shortcomings. Paramount among the latter, we ignore the tidal effects
of the large scale matter distribution which are responsible for generating
the angular momentum in the collapsing structures. The amount and
distribution of angular momentum therefore are free parameters in our
simulations. In prescribing them, however, we can draw on the insight from
cosmological numerical simulations. These result in a statistical description
of the angular momentum (spin) which a given dark matter halo is expected
to acquire (e.g., Barnes \& Efstathiou 1987). The primary advantage of
our method, on the other hand, is that it allows us to comprehensively
explore the behavior of the primordial gas under a variety of initial 
conditions. In doing so, we hope to single out the essential physics, and
to test its robustness. In a previous publication, we have already presented
first results (Bromm, Coppi, \& Larson 1999).

The organization of this paper is as follows. In \S 2, we discuss the relevant
physical ingredients, both for the dark matter and baryonic components. A
description of our numerical method is given in \S 3, whereas \S 4 presents
the results of our exploratory survey. Finally, \S 5 contains our conclusions
and debates avenues for further progress.

\section{THE PHYSICAL INGREDIENTS}
\subsection{Dark Matter}

Current scenarios of cosmological structure formation assume the dark
matter (DM) to be `cold', in the sense of having negligible velocity dispersion.
Candidates include the lightest supersymmetric particle, the photino,
of estimated mass $\sim$ 100 GeV, and the invisible axion (e.g., Peacock
1999). Primordial density fluctuations in the cold dark matter have
consequently survived on all scales due to the absence of free-streaming.
Provided these fluctuations obey Gaussian statistics, as predicted by
inflation, they are fully described by the power spectrum $P(k)$, with
$k$ denoting the wavenumber.
The standard CDM model (e.g., Blumenthal et al. 1984) predicts that the 
fluctuations
decrease with mass, leading to a hierarchical
(bottom-up) picture of structure formation. Variants of CDM agree qualitatively
with each other, and we take the standard CDM model as a convenient template
for our discussion. In the critical Einstein-de Sitter model, the rms
density contrast on a mass scale $M$ grows 
in time according to $\sigma(M)=\sigma_{0}(M)/(1 + z)$,
as long as the fluctuation remains in the linear stage. $\sigma_{0}(M)$
is the linear extrapolation to the present epoch. In a model
with a cosmological constant, growth is suppressed at recent epochs ($z < 10$),
but at earlier times, the simple Einstein-de Sitter behavior still applies.
For a DM halo of given mass $M$, corresponding to a $\nu \sigma$-peak in
the Gaussian random field, the redshift of collapse can be estimated as
\begin{equation}
1+z_{vir}=\frac{\nu \sigma_{0}(M)}{\delta_{c}}
\mbox{\ \ \ .}
\end{equation}
The threshold overdensity for collapse is often taken to be $\delta_{c}=1.69$
(Peacock 1999). We discuss in Section 2.3 that the first stars are
expected to form in DM halos of mass $\sim 10^{6}M_{\odot}$, corresponding
to $3-4\sigma$ peaks. At these
small scales, halos of increasing mass collapse in rapid succession. This
(`cross-talk') behavior is characteristic of the CDM model (Rees 2000).
The CDM power spectrum approaches $P(k)\propto k^{-3}$ asymptotically
on small scales. This is characteristic for the DM fluctuations within
a Population III star forming region. Since the baryonic mass is smaller
than the initial Jeans mass, all perturbations in the gas have been wiped
out by pressure forces. 
The presence of the small-scale DM fluctuations, therefore,
might play an important role in shaping the fate of the collapsing gas.

A collapsing DM halo is expected to acquire angular momentum via tidal
interactions with neighboring overdensities. The outcome of this
process can be conveniently described by the dimensionless spin parameter
\begin{equation}
\lambda=\frac{L|E|^{1/2}}{G M^{5/2}}
\mbox{\ \ \ ,}
\end{equation}
where $L,E,$ and $M$ are the total angular momentum, energy, and mass,
respectively, with $G$ being Newton's constant.
Numerical simulations result in an average spin parameter of
$\lambda\simeq 0.05$ (e.g., Barnes \& Efstathiou 1987).

Once a perturbation on a given mass scale
reaches overdensities $\delta=(\rho_{halo}-
\rho_{b})/ \rho_{b}$ of close to unity, with $\rho_{b}$ being the density of the
unperturbed background universe, the linear description breaks down, and the
halo is undergoing nonlinear collapse. A convenient analytical model for
the nonlinear evolution of DM halos is given by the spherical top-hat
model (e.g., Padmanabhan 1993). 
At redshifts close to $z_{vir}$, the DM particles
reach a state of
virial equilibrium.
The approach to virial equilibrium occurs through
the process often called violent relaxation in the literature
(Lynden-Bell 1967). This process operates
on a dynamical timescale, as opposed to the slow two-body relaxation.
The halo density after virialization is estimated to be
\begin{equation}
\rho_{vir}\simeq 18\pi^{2} \rho_{b}=18\pi^{2}\rho_{0}(1+z_{vir})^{3}
\mbox{\ \ \ ,}
\end{equation}
where $\rho_{0}=1.88\times 10^{-29}$g cm$^{-3}$ $\Omega_{m}h^{2}$ is
the density of the present-day universe.
The baryons partake in the DM collapse, and acquire, through shocks
and adiabatic compression, the virial temperature
\begin{equation}
T_{vir}\simeq \frac{G M m_{\mbox{\scriptsize H}}}{2 R_{vir} k_{\mbox{\scriptsize
 B}}}\simeq 100 \mbox{K} h^{2/3} \left(\frac{M}{10^{6}M_{\odot}}\right)^{2/3}
(1 + z_{vir})
\mbox{\ \ \ .}
\end{equation}
Here, $R_{vir}$ denotes the virial radius, $k_{\mbox{\scriptsize B}}$
is Boltzmann's constant,
and $m_{\mbox{\scriptsize H}}$ is the mass of a hydrogen atom.
Although realistic DM halos are much more complicated than this simple
model, the top-hat results give an intuitive way to understand the 
physics of complex situations to within factors of a few.
It also allows us to specify the initial conditions for our numerical
simulations (see Section 4).

In Figure 1, we show the density evolution of a top-hat perturbation,
and compare the analytical prediction with the result of one of our
numerical simulations. It can be seen that the prediction of $\rho_{vir}$
in equ. (3) is nicely reproduced in the simulation, after an epoch
of settling down to the equilibrium state.

\setcounter{figure}{0}
\begin{center} 
\psfig{file=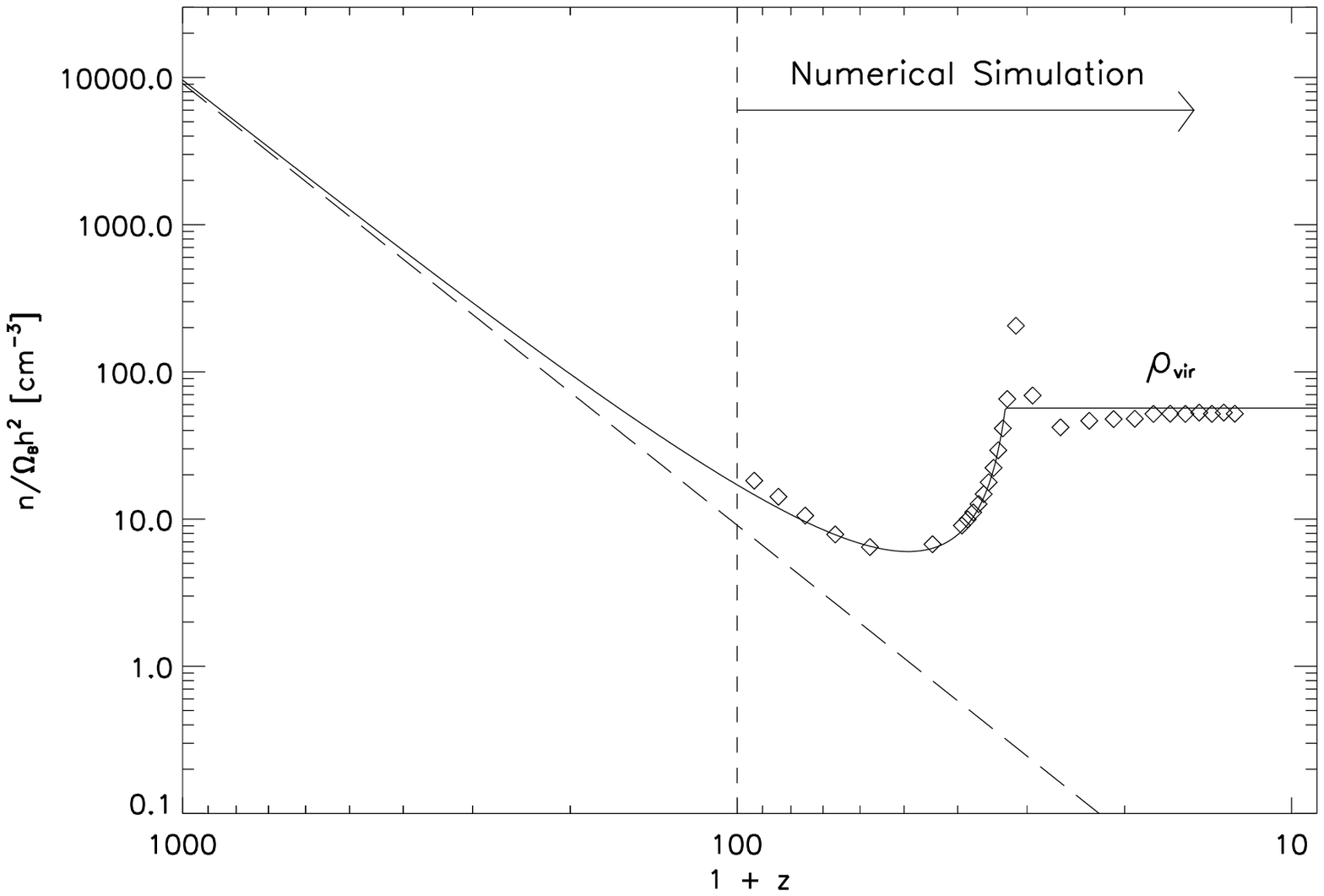,width=8.cm,height=6.3cm}
\figcaption{
Top-hat model for an overdensity collapsing at $z_{vir}\simeq 30$.
Shown is the number density of hydrogen atoms vs. redshift.
{\it Solid line:} Density evolution according to top-hat solution.
{\it Long-dashed line:} Density evolution in the expanding background (E-dS)
universe. $\rho_{vir}$ denotes the ``virial plateau'', the estimated density
after violent relaxation has established virial equilibrium.
{\it Diamond-shaped symbols:} Evolution of the average gas density
in adiabatic test case. The simulation is initialized at $z=100$ to reproduce
the top-hat collapse. As can be seen, the agreement with the analytical
solution is good. Notice also that
the average gas density settles eventually to a value
close to $\rho_{vir}$.
\label{fig1}}
\end{center}

\subsection{Baryons}
The fate of the baryons is determined by the strife 
between gravity and opposing pressure, and it is, therefore, essential
to understand 
the thermal history of the gas, a topic which we address in the following
subsection.

\subsubsection{\it Cooling and Heating}
The thermal evolution of the gas is governed by the equation:
\begin{equation}
\frac{\mbox{D}u}{\mbox{D}t}=\frac{P}{\rho^{2}}\frac{\mbox{D}\rho}{\mbox{D}t}
+\frac{\Gamma - \Lambda}{\rho}
\end{equation}
Here, $P$ and $\rho$ are the gas pressure and density, $u$ is the specific
internal energy (in erg g$^{-1}$),  
and $\Gamma$ and $\Lambda$ are the contributions
from radiative heating and cooling, respectively.
The first term on the right-hand side describes adiabatic cooling due to
expansion or heating due to compression.
We now discuss the relevant
processes of radiative cooling.

In the absence of metals, H$_{2}$ is the main coolant below $\sim 10^{4}$ K,
which is the temperature range typically encountered in collapsing
Population III objects. Being homonuclear, H$_{2}$ possesses no permanent
dipole moment. Rotational transitions, therefore, can only occur via
electric quadrupole radiation with the corresponding very small transition
probabilities ($A\sim 10^{-11}$ s$^{-1}$ for the lowest-lying transition).
The H$_{2}$ cooling function
has been investigated by a number of authors whose results, however, differed
by up to a factor of 100 (Hollenbach \& McKee 1979; Lepp \& Shull 1983). A
crucial uncertainty concerned the low temperature (a few 100 K) regime, where
quantum-mechanical effects were not yet taken into proper account. Recently,
the cooling rates have been thoroughly recomputed. We have implemented
H$_{2}$ cooling using the new parameterization of Galli \& Palla (1998).
These authors have included improved collisional rates for $ T > 600$ K
(Martin, Schwarz, \& Mandy 1996), as well as for $T<600$ K (Forrey et al. 1997),
and have assumed ortho- and para-H$_{2}$ to be present in their equilibrium
ratio of 3:1. In general, the cooling rate (in erg s$^{-1}$ cm$^{-3}$)
can be written as
\begin{equation}
\Lambda=\sum_{u \rightarrow l} n_{u}A_{ul}\Delta E_{ul} \mbox{\ \ \ ,}
\end{equation}
where the sum extends over all possible transitions $u \rightarrow l$.
$A_{ul}$ is the Einstein coefficient for spontaneous emission,
$\Delta E_{ul}$ the respective energy difference, and $n_{u}$ the
density of H$_{2}$ in the (upper) level $u$. In the low-density
regime ($n\rightarrow 0$), each collisional excitation is almost
instantaneously followed by spontaneous emission. The level populations
are then determined by the rate of H-H$_{2}$ collisions:
\begin{equation}
n_{u}^{(n \rightarrow 0)} \propto 
n_{\mbox{\scriptsize H$_{2}$}}
n_{\mbox{\scriptsize H}}\gamma_{lu}\propto n_{\mbox{\scriptsize H}}^{2}
\mbox{\ \ \ ,}
\end{equation}
with $\gamma_{lu}$ being the coefficient of collisional excitation.
At high density, the levels are populated according to LTE:
\begin{equation}
n_{u}^{(\mbox{\scriptsize LTE})} \propto 
n_{\mbox{\scriptsize H}}\mbox{exp}(-\Delta E_{ul}/k_{B}T)
\propto n_{\mbox{\scriptsize H}}
\mbox{\ \ \ .}
\end{equation}
Summarizing the limiting behavior, one has:
\begin{equation}
\Lambda \rightarrow \left\{
\begin{array}{ll}
\Lambda^{(n \rightarrow 0)} 
\propto  n_{\mbox{\scriptsize H}}^{2} & \mbox{for \ } n \ll n_{crit} \\
\Lambda^{(\mbox{\scriptsize LTE})} 
\propto  n_{\mbox{\scriptsize H}} & \mbox{for \ } n \gg n_{crit}
\\
\end{array}
\right.
\end{equation}
The critical density $n_{crit}$, above which de-excitation
due to collisions dominates over the competing radiative decay,
is a function of temperature only,
and is defined as
\begin{equation}
n_{crit}\equiv \frac{\Lambda^{(\mbox{\scriptsize LTE})}} {\Lambda^{(n \rightarrow 0)}}
n_{\mbox{\scriptsize H}}
\mbox{\ \ \ .}
\end{equation}
The cooling function can finally be expressed in the form
\begin{equation}
\Lambda=
\frac{\Lambda^{(\mbox{\scriptsize LTE})}} {1 +n_{crit}/
n_{\mbox{\scriptsize H}}}
\mbox{\ \ \ ,}
\end{equation}
smoothly joining the low-density with the high-density case.
In Figure 2, we show this cooling function for various densities.
At a given temperature, the cooling per molecule ($\Lambda /
n_{\mbox{\scriptsize H$_{2}$}}$) first increases with density, but then
approaches (for $n > n_{crit}$) the LTE value which constitutes the maximum
possible cooling rate. Since H$_{2}$ has a comparatively small
moment of inertia, $I\simeq m_{\mbox{\scriptsize H}} a_{0}^{2}$, the 
rotational energy, $E_{rot}=L^{2}/2 I$, is correspondingly large
for a given amount of angular momentum $L=\hbar\sqrt{J(J+1)}$.
Estimating the internuclear seperation to be of order the
Bohr radius, $a_{0}\simeq 0.5$ \AA, one then finds
for the lowest-lying ($2\rightarrow 0$)
rotational transition the equivalent minimum temperature
\begin{equation}
T_{min}=\frac{\Delta E_{20}}{k_{B}}=\frac{6\hbar^{2}}
{2 m_{\mbox{\scriptsize H}}a_{0}^{2}k_{B}}
\mbox{\ \ \ ,}
\end{equation}
resulting in $T_{min}\sim 500$ K. The high-energy tail of the Maxwellian
velocity distribution will allow the gas to attain somewhat lower 
temperatures, but it is one of the crucial aspects of the primordial
gas physics that cooling down to temperatures below $T \sim 100$ K is
not possible in the absence of coolants other than molecular hydrogen.
The corresponding critical density for $T\sim 100-300$ K is then
$n_{crit}\simeq 10^{3}-10^{4}$ cm$^{-3}$. These characteristic values
of temperature and density,
based on the microphysics of H$_{2}$ cooling,
leave their imprint on the thermal evolution of the primordial
gas, as will be discussed below.

\begin{center} 
\psfig{file=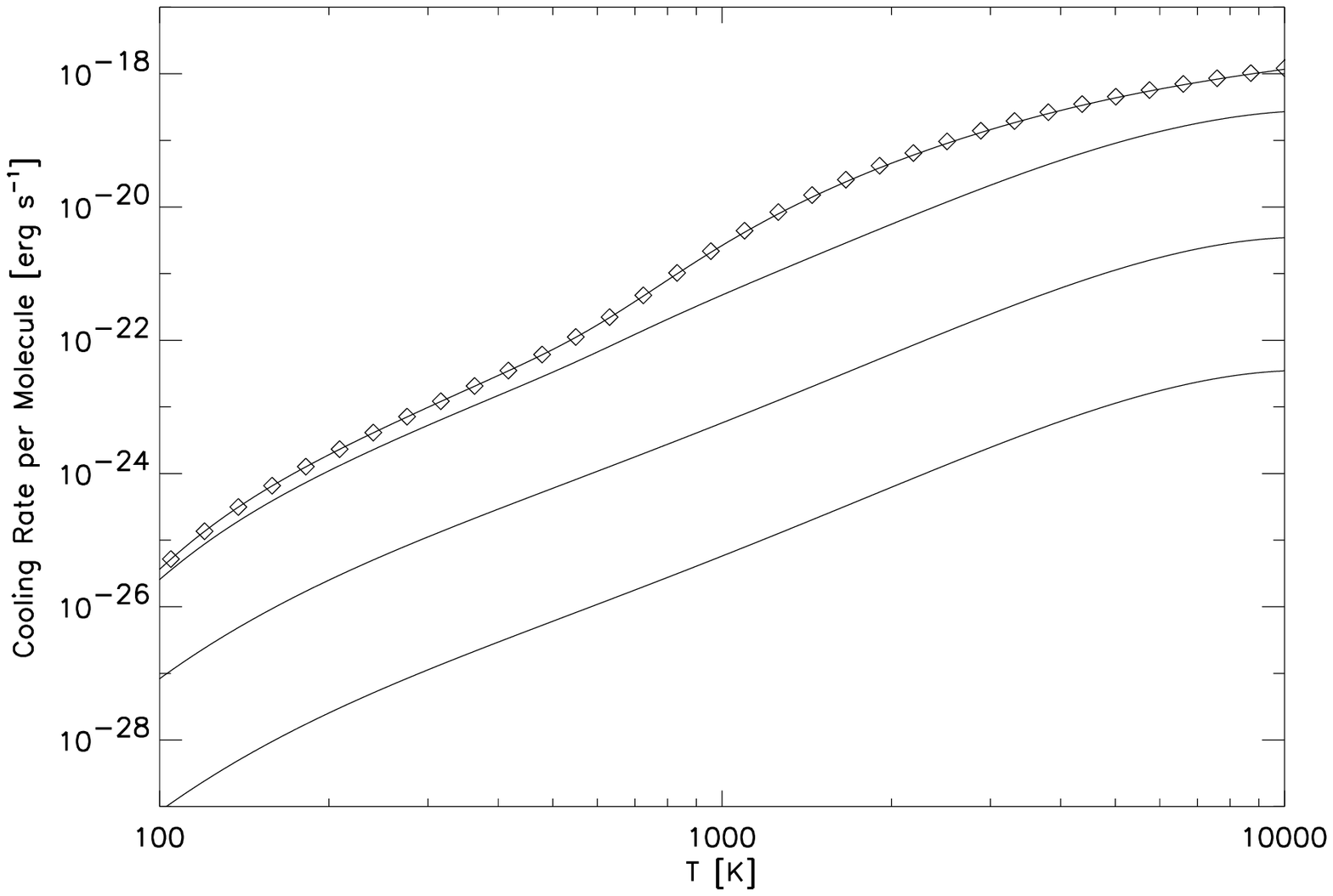,width=8.cm,height=6.3cm}
\figcaption{
H$_{2}$ cooling function \rm (Galli \& Palla 1998).
{\it Solid lines:} Cooling rates per H$_{2}$ molecule for various 
densities. {\it From bottom to top:} $n = 10^{-1},10^{1},10^{3},10^{5}$
 cm$^{-3}$.
{\it Diamond-shaped symbols:} Cooling rate in LTE. This latter rate
gives the maximum possible cooling per molecule at a given temperature.
Notice the saturation in the cooling rate due to the transition
from NLTE to LTE level populations.
\label{fig2}}
\end{center} 

Another possible coolant in the primordial gas is HD, deuterium
hydride. The low abundance, $n_{\mbox{\scriptsize D}}\simeq
10^{-5}n_{\mbox{\scriptsize H}}$, is partially offset by the
fact that HD does possess a permanent electric dipole moment
with the correspondingly larger radiative transition probabilities
($A \sim 10^{-8}$ s$^{-1}$ for the lowest-lying rotational level,
a factor $\sim 1000$ enhancement over H$_{2}$). In addition, since the
rotational transition $1\rightarrow 0$ is allowed, HD cooling
can reach lower temperatures than H$_{2}$: $T_{min}\simeq \Delta E_{10}/
k_{\mbox{\scriptsize B}}\simeq 160$ K. Consequently, cooling due to HD
might become important at temperatures $\sim 100-200$ K. Our treatment
of HD cooling relies on recent work of Flower et al. (2000), who have
carefully computed the relevant collisional rates, and provide an
analytical fit to their results. 

Since radiative cooling to temperatures below that of the CMB
\begin{equation}
T_{\mbox{\scriptsize CMB}}=2.7 \mbox{K} (1 + z)
\end{equation}
is thermodynamically not possible, we write for the cooling due to H$_{2}$
and HD:
\begin{equation}
\Lambda = \Lambda (T) - \Lambda (T_{\mbox{\scriptsize CMB}})
\end{equation}
This approximate treatment ensures that $T \geq T_{\mbox{\scriptsize CMB}}$,
unless cooling proceeds via adiabatic expansion.

Finally, we have included two cooling mechanisms of lesser importance.
Firstly, for temperatures approaching $\sim 10^{4}$ K, cooling due to collisionally
excited lines of atomic hydrogen becomes effective (Cen 1992):
\begin{equation}
\Lambda_{\mbox{\scriptsize H}}=7.5\times 10^{-19}(1 + T_{5}^{1/2})^{-1}
\mbox{exp}(-118,348/T) n_{e} n \mbox{\scriptsize [H]}
\end{equation}
Only a small fraction of the gas in a Population III DM halo, however,
reaches that high a temperature, and atomic hydrogen could not have
facilitated the collapse and fragmentation of the primordial gas.
Secondly, as long as there remains a residual degree of ionization,
energy is exchanged between the photons of the CMB and the electrons
of the gas by (normal and inverse) Compton scattering:
\begin{equation}
\Lambda_{Comp}=5.4\times 10^{-36}(1 + z)^{4} n_{e} 
 (T - T_{\mbox{\scriptsize CMB}})
\end{equation}
At redshifts below $z\sim 200$, this coupling becomes increasingly
weak, and does not significantly contribute to the cooling of the gas.

In Figure 3, we show the relative importance of the various heating and
cooling processes for a fluid element, whose time evolution is 
representative of what we find in our simulations.
At early times, while the cloud is still expanding, cooling due to
adiabatic expansion dominates. After turnaround, the gas heats up
by adiabatic compression. Only close to the moment of virialization
does H$_{2}$ cooling exceed the adiabatic heating, allowing the gas
to cool upon further contraction. In these early stages, cooling due
to HD is never important. After virialization, when the gas has settled into 
a cold (a few 100 K) central configuration, HD cooling becomes somewhat
more important, without ever changing the thermal evolution dramatically.

\begin{center} 
\psfig{file=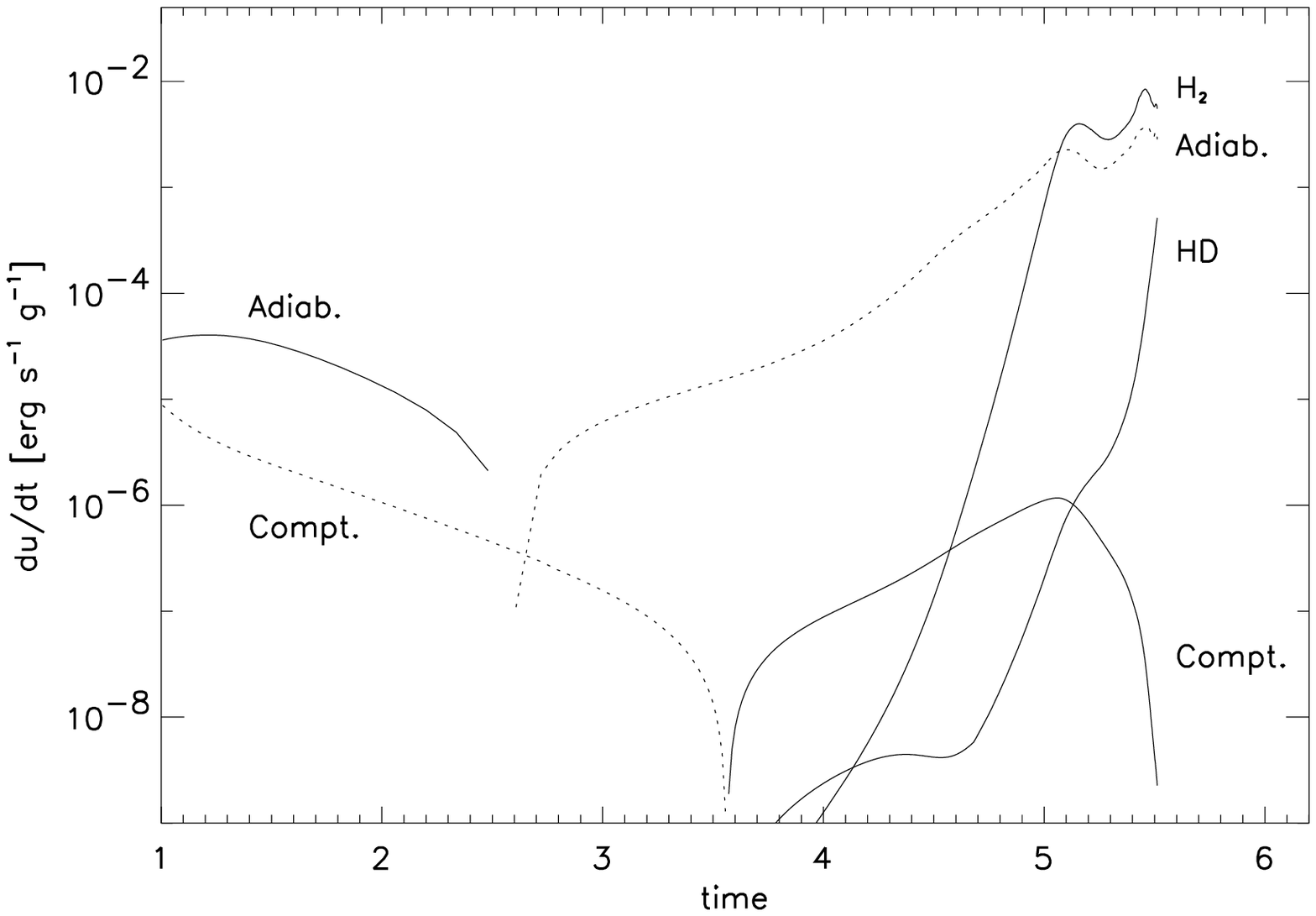,width=8.cm,height=6.3cm}
\figcaption{
Importance of cooling and heating terms during collapse.
{\it Solid lines:} Cooling (in erg s$^{-1}$ g$^{-1}$) vs. time.
{\it Dotted lines:} Heating (in erg s$^{-1}$ g$^{-1}$) vs. time.
Time is measured in units of the initial free-fall time,
$\sim 2\times 10^{7}$ yr. 
Adiabatic cooling changes into adiabatic heating after turnaround.
Compton heating (for $T < T_{\mbox{\scriptsize CMB}}$), on the other hand,
turns into Compton cooling 
(for $T > T_{\mbox{\scriptsize CMB}}$).
H$_{2}$ is the dominant coolant, but can compete with the heating due to 
adiabatic compression only close to the moment of virialization.
\label{fig3}}
\end{center} 

The treatment of radiative cooling, therefore, necessitates knowing
the abundances of H$_{2}$ and HD. This is the subject we take up next.

\subsubsection{\it Chemistry}
As shown in the previous section, radiative cooling is mainly due to
H$_{2}$. Cooling due to HD is less important, but we have included it 
for the sake of completeness. It is consequently essential to predict the
respective abundances of these molecules. The appropriate primordial 
chemistry has the following simplifying features. Helium is assumed to
be always completely neutral, and we neglect all reactions involving
He, He$^{+}$, He$^{++}$. This assumption is justified since the
temperature is typically low enough ($T < 10^{4}$ K) to render the He
chemistry inert. In addition, we also ignore all processes of photoionization
and -destruction. Although photoreactions due to the CMB are very important
at redshifts $ z > 100$, they can be safely neglected in our simulations.
Finally, there does not yet exist a UV background, prior to the onset
of Population III star formation.
Since H$_{2}$ and HD are formed in nonequilibrium, it is necessary to
consider the respective reaction networks. First, we discuss the chemistry
of H$_{2}$ formation and destruction.

Our H$_{2}$ network is based on the compilation of Haiman et al. (1996), and
is presented in Table 1. 
In the absence of dust grains, the main 
route for the production of H$_{2}$ is given by the H$^{-}$ channel 
(McDowell 1961):
\begin{displaymath}
\mbox{H}+ e^{-} \rightarrow \mbox{H}^{-} + h\nu 
\end{displaymath}
\begin{displaymath}
\mbox{H}^{-} + \mbox{H} \rightarrow \mbox{H}_{2} + e^{-} 
\end{displaymath}
An alternative formation channel relies on the intermediary H$_{2}^{+}$.
Since the H$^{-}$ channel always dominates in our simulations, we ignore
reactions involving H$^{+}_{2}$. The validity of this assumption has been
tested by comparison calculations with the full network.
At redshifts $z > 100$, on the other hand, the H$_{2}^{+}$ channel becomes
important due to the ready destruction of H$^{-}$, which has a binding
energy of only 0.75 eV, by energetic CMB photons. Thus, hydrogen molecules
are produced as long as there is a sufficient abundance of free electrons
which act as catalysts in the H$^{-}$ channel. For the low temperatures
and weak shocks in the typical collapse of a primordial cloud, there is no
efficient way to ionize the gas. The residual abundance of free
electrons $x=n_{e}/n \sim 10^{-4}$ recombines on a timescale
$t_{rec}\simeq (k_{rec} x n)^{-1}\simeq 10^{6}-10^{7}$ yr. The recombination
coefficient $k_{rec}\sim 10^{-12}$ cm$^{3}$ s$^{-1}$ is evaluated for
typical temperatures, $T \sim 5000$ K, and densities, $n \sim 10^{2}$ cm$^{-3}$,
as found in our simulations. One can then estimate the maximal abundance
of hydrogen molecules,
$f=n_{\mbox{\footnotesize H}_2}/n$, according to
\begin{displaymath}
\frac{f}{t_{rec}}\sim \frac{\mbox{d}f}{\mbox{d}t}\simeq n k_{3} x \mbox{\ .}
\end{displaymath}
The rate for the radiative attachment of H$^{-}$ and $e^{-}$ (see reaction
(3) in Table 1) is approximately $k_{3} \sim 10^{-15}$ cm$^{3}$ s$^{-1}$. The
asymptotic abundance of H$_{2}$, produced through the H$^{-}$ channel, is then
$f\simeq k_{3} x n t_{rec}\simeq 10^{-3}$. This prediction is borne out
in our simulations under a wide range of circumstances. H$_{2}$ abundances
of $10^{-4}-10^{-3}$, albeit low, nevertheless allow the primordial gas
to efficiently cool, thus enabling the condensation into high density 
structures. Once the gas reaches densities $n > 10^{8}$ cm$^{-3}$, 
three-body
reactions (Palla et al. 1983)
convert the gas rapidly into fully molecular form. 
These reactions are included in Paper II, where we investigate the
further collapse of a high density clump, but are not important here.

The chemistry responsible for the formation and destruction of HD has
recently been discussed by Galli \& Palla (1998). We adopt their
minimum model, as listed in Table 2.

\subsection{Properties of the Star Forming Region}
To determine the properties of the primordial star forming region, two
ingredients have to be considered. Firstly, one needs to know how the dark
matter evolves. This history of hierarchical gravitational clustering,
resulting in the collapse of increasingly massive DM halos, is specified
by the adopted variant of the CDM model. Secondly, one has to address the 
question of whether the baryons are able to fall together with the dark
matter. 
The 
opposing effect of gas pressure in collapsing DM
halos has been studied
by Tegmark et al. (1997). These authors argue that 
the primordial gas can only undergo continued
collapse and fragmentation, and consequently star formation, if it 
manages to efficiently radiate away the gravitational binding energy which is
released in this process. The efficiency of cooling can be quantified by
the cooling timescale, $t_{cool}\simeq n k_{\mbox{\scriptsize B}}T/ \Lambda$,
and is to be compared with the freefall time, $t_{ff}\simeq 1/ \sqrt{G\rho}$.
One then has the classical Rees-Ostriker criterion
for fragmentation and continued collapse (Rees \& Ostriker 1977):
$t_{cool} < t_{ff}$. In the early stages of the collapse, temperatures
are low enough ($\sim 1000$ K) for cooling to proceed mainly via
the lowest-lying rotational transition of H$_{2}$. 
In evaluating
the criterion $t_{cool}<t_{ff}$, Tegmark et al. (1997) find the minimum virial temperature
necessary for sufficient cooling, and by applying equ. (4) the
corresponding minimum halo mass as a function of collapse redshift.
The result of this calculation is
that a $3\sigma$-peak of mass $\sim 10^{6}M_{\odot}$, and collapsing
at a redshift $z_{vir}\sim 30$, does meet the requirement for efficient
cooling. Halos with these characteristic masses and collapse redshifts
are therefore predicted to be the sites for the formation of the first
stars. 

To understand the
complicated physics of gas fragmentation, we now turn to a description
of our numerical simulations.

\section{NUMERICAL METHODOLOGY}
In this section, we describe the elements of our numerical approach.
The evolution of the dark matter and gas components is calculated with
a version of TREESPH (Hernquist \& Katz 1989), combining
the smoothed particle hydrodynamics (SPH) method with a hierarchical (tree)
gravity solver. 
The details
of the gravitational N-body solver and the treatment of the 
hydrodynamics are discussed in the Appendix. We here
present our additions to the code which are necessary
for the investigation of the primordial gas.
These are a method to solve the primordial chemistry network, and
 a technique to create sink particles.

\subsection{\it Solving the Reaction Network}

Following the abundance evolution of the 8 species H, H$^{+}$, H$^{-}$,
H$_{2}$, D, D$^{+}$, HD, and e$^{-}$, entails solving the corresponding
coupled set of kinetic equations. Due to the very short reaction
timescales, compared to the dynamical time, the adopted method of solution
has to be implicit, as in the case of the thermal energy equation.
Traditional matrix-based techniques such as the STIFBS routine (Press et
al. 1992), or the Livermore LSODAR solver (Hindmarsh 1983), prove to be
computationally too expensive for three-dimensional applications. We
therefore have implemented a fast method of solution which is based on
the approximate backwards differencing formula (BDF). This
approach has been pioneered by Anninos et al. (1997). Consider the rate
equation for a given species $i$, which can be expressed as
\begin{equation}
\frac{\mbox{d}n_{i}}{\mbox{d}t}=
-D n_{i} + C
\mbox{\ \ \ .}
\end{equation}
All reactions contributing to the destruction of species $i$ are contained
in $D$, while all reactions leading to the production of that species
are summarized by the symbol $C$. Both $D$ and $C$ are functions of
temperature and the abundances of the other species. Evaluating
the right-hand side of equation (17) at the new timestep, results in the
discretization
\begin{equation}
n_{i}^{new}=\frac{C^{new}\Delta t + n_{i}^{old}}{1 + D^{new}\Delta t}
\mbox{\ \ \ .}
\end{equation}
We update the hydrogen and deuterium species in the following order, which
has been found by experimentation to be both stable and accurate:
e$^{-}$, H$^{+}$, H$^{-}$, H$_{2}$, H, D$^{+}$, HD, D. An estimate for
$C^{new}$ and $D^{new}$ is obtained by first inserting the abundances at
the old timestep, and then successively replacing them with the updated ones,
as the algorithm proceeds from species to species. In determining the
abundances of the normal hydrogen species, we do not include the reactions
of the deuterium network. This is justified by the very low abundance
of deuterium ($n_{\mbox{\scriptsize D}}\sim 10^{-5} n_{\mbox{\scriptsize H}}$).
The evolution of the deuterium network, on the other hand, crucially
depends on the abundances of the normal hydrogen species.

For increased accuracy, a subcycling procedure is adopted: The algorithm
loops over the species repeatedly, each time with a small timestep
$\Delta t_{sub}$ given by
\begin{equation}
\Delta t_{sub}=\epsilon\frac{n_{e}}{\dot{n}_{e}}
\mbox{\ \ \ ,}
\end{equation}
where $n_{e}$ is the abundance of free electrons, and $\epsilon=0.1$ is an
accuracy parameter.

The reactions determining the abundance of the intermediary species H$^{-}$
proceed much more rapidly than the other reactions. Therefore, one
can safely assume that H$^{-}$ is always present with its equilibrium
value:
\begin{equation}
n_{\mbox{\scriptsize H$^{-}$}}=\frac{k_{3}n_{\mbox{\scriptsize H}}
n_{e}+k_{6}n_{\mbox{\scriptsize H$_{2}$}}}{(k_{4}+k_{11})n_{\mbox{\scriptsize
 H}}+(k_{5}+k_{12})n_{\mbox{\scriptsize H$^{+}$}} + k_{10}n_{e}}
\mbox{\ \ \ .}
\end{equation}
The identification of the reaction rates can be found in Table 1.

To test our chemistry solver, we compare the analytically determined 
equilibrium abundances as a function of temperature, obtained by
setting $\mbox{d}n_{i}/\mbox{d}t=0$ for all $i$, with those given by the
chemistry solver, after the integration has converged. The analytical
estimates are derived as follows. For H$^{+}$, we assume $n_{e}=n_{\mbox{
\scriptsize H$^{+}$}}$ and consider the collisional balance:
\begin{eqnarray}
\mbox{H}+ e^{-} &\rightarrow& \mbox{H}^{+} + 2 e^{-}\nonumber\\
\mbox{H}^{+} + e^{-} &\rightarrow& \mbox{H} + h\nu \nonumber
\end{eqnarray}
With the reaction rates in Table 1, one finds
\begin{equation}
n_{\mbox{\scriptsize H$^{+}$}}=\frac{k_{1}}{k_{2}}n_{\mbox{\scriptsize H}}
\mbox{\ \ \ .}
\end{equation}
The abundance of atomic hydrogen is assumed to be $n_{\mbox{\scriptsize H}}=
n_{\mbox{\scriptsize H}, tot} - n_{\mbox{\scriptsize H$^{+}$}}$, where
$n_{\mbox{\scriptsize H}, tot}$ is the total density of normal hydrogen.
For H$_{2}$, we obtain (see Table 1):
\begin{equation}
n_{\mbox{\scriptsize H$_{2}$}}=\frac{k_{4}n_{\mbox{\scriptsize H}}
n_{\mbox{\scriptsize H$^{-}$}}}{(k_{6}+k_{8}+k_{9})n_{\mbox{\scriptsize
 H}}+k_{7}n_{\mbox{\scriptsize H}}}
\mbox{\ \ \ .}
\end{equation}
The abundance of the intermediary species H$^{-}$ is given by equation (20),
evaluated with $n_{\mbox{\scriptsize H$_{2}$}}=0$.
For the deuterium species, D$^{+}$ and HD, the corresponding expressions are:
\begin{equation}
n_{\mbox{\scriptsize D$^{+}$}}=\frac{k_{D2}n_{\mbox{\scriptsize D}}
n_{\mbox{\scriptsize H$^{+}$}}}{k_{D1}n_{e}+
k_{D3}n_{\mbox{\scriptsize H}}}
\mbox{\ \ \ ,}
\end{equation}
and
\begin{equation}
n_{\mbox{\scriptsize HD}}=\frac{k_{D4}n_{\mbox{\scriptsize D$^{+}$}}
n_{\mbox{\scriptsize H$_{2}$}}}{k_{D5}
n_{\mbox{\scriptsize H$^{+}$}}}
\mbox{\ \ \ .}
\end{equation}
Here, $k_{D1}, k_{D2},\ldots $ correspond to the reactions in Table 2, and
we assume
$n_{\mbox{\scriptsize D}}=
n_{\mbox{\scriptsize D}, tot} - n_{\mbox{\scriptsize D$^{+}$}}$, with
$n_{\mbox{\scriptsize D}, tot}$ being the total density of heavy hydrogen.
The result of this comparison is presented in Figure 4. As can be seen,
the chemistry solver nicely reproduces the analytical expectation.

\begin{center} 
\psfig{file=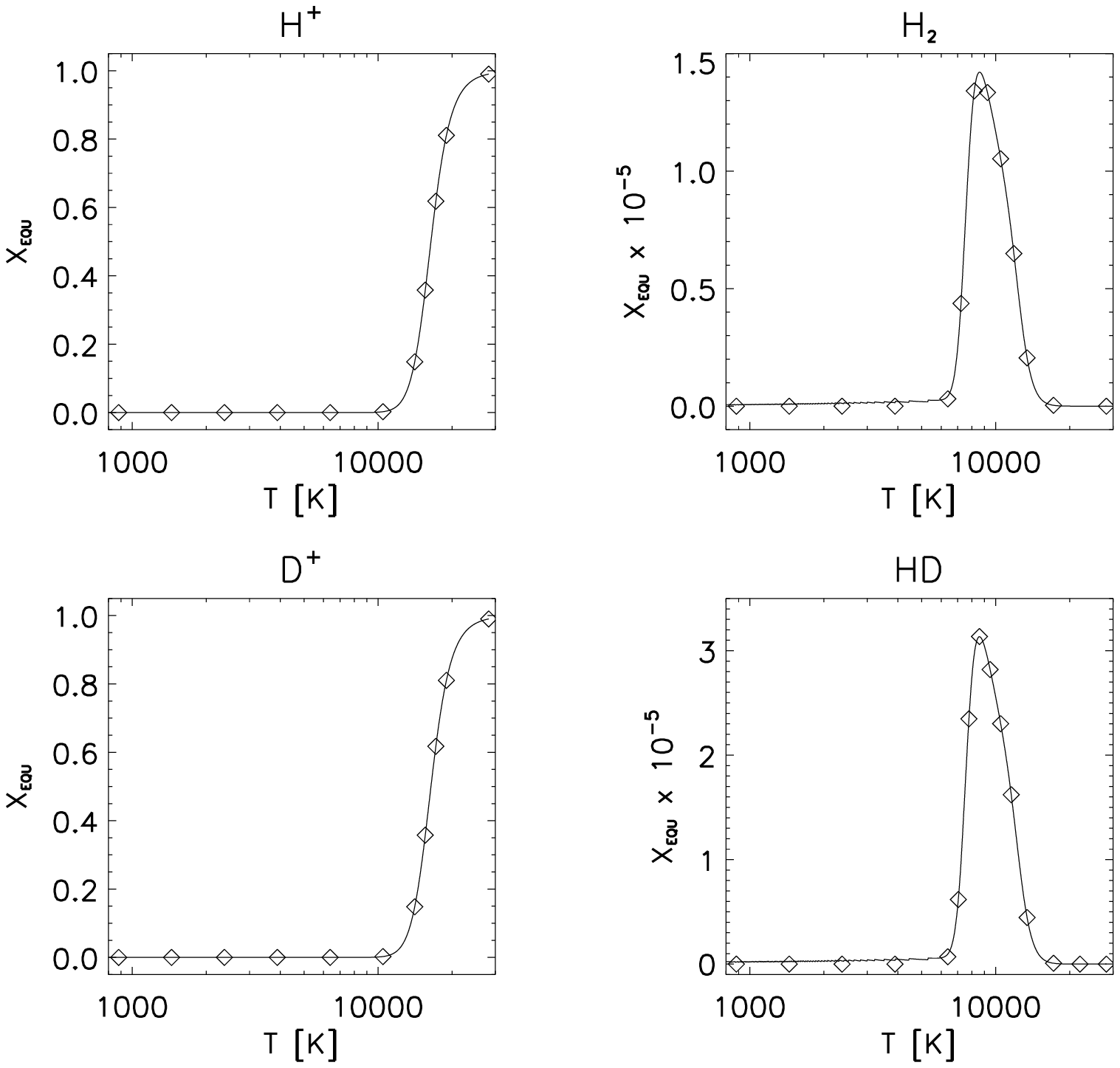,width=8.4cm,height=7.56cm}
\figcaption{
Testing the chemistry solver.
Shown are the equilibrium abundances
for the species H$^{+}$, H$_{2}$, D$^{+}$, and HD. In each case, fractional
abundance is plotted vs. temperature.
{\it Solid lines:} Abundances from allowing the chemistry solver to reach
convergence (to within $10^{-10}$).
{\it Diamond-shaped symbols:} Analytic estimate for the equilibrium value.
It can be seen that the equilibrium abundances are nicely reproduced by the
solver. Notice also the extreme similarity between the normal hydrogen
and the deuterium species.
\label{fig4}}
\end{center} 

\subsection{\it Creation of Sink Particles}

When in the course of a simulation the gas attains increasingly high
density, the SPH smoothing length decreases according to $h\simeq N_{neigh}^{1/3}
n^{-1/3}$. The Courant condition then enforces the adoption of smaller and
smaller timesteps $\Delta t\simeq h/c_{s}$, with $c_{s}$ being the sound speed.
Upon the onset of gravitational
instability and the resulting runaway collapse, the simulation therefore
effectively grinds to a halt. To overcome this fundamental numerical 
limitation, and to be able to follow the evolution of the overall system for
a few dynamical times, a method of creating sink particles is necessary.
Here, we describe our strategy in doing so. Recently, Bate, Bonnell, and
Price (1995) have developed another such technique in the context of SPH
and applied it to the study of protobinary accretion. 

In developing the algorithm, one has to address the following questions.
(1) {\it When a sink particle is created during a simulation, would the
incorporated particles really continue to collapse, or would they
escape again from each other, if one were to follow their evolution further?}
To ensure that only gravitationally bound particles are merged to form
a sink particle, we utilize the runaway nature of the Jeans instability.
Consider how the density does evolve with time for the SPH particles
close to the first collapsing region. Densities are initially close to
$n\sim 10^{4} - 10^{5}$ cm$^{-3}$, and subsequently either experience only
a modest rise in density, or a rapid increase by many orders of magnitude.
This dichotomy allows for the unambiguous identification of the collapsing
gas particles. The first, and physically most important, criterion for
a particle to be eligible for merging therefore is: (a) $n > n_{th}$.
The value of $n_{th}$ has to be adjusted to the physical characteristics of
the problem. We find that $n_{th}=10^{8}$ cm$^{-3}$ works well in our
case of primordial gas which collapses and fragments in the center of
a dark matter halo. Another reason for this choice lies in the fact that
beyond a density of $10^{8}$ cm$^{-3}$, additional physics has to
be considered: Three-body reactions convert the gas into fully molecular
form, and opacity effects start to become important. We address this
high-density regime in Paper II,
but otherwise assume that $n < 10^{8}$ cm$^{-3}$. 
A second test a particle has to pass is: (b) $\nabla\cdot\vec{v}<0$, i.e., that
it is part of a converging flow.
 
To further examine whether a given collection of particles is gravitationally
bound, one has to determine whether the binding energy of the system is
negative (criterion (c)):
\begin{equation}
E_{total}=E_{grav}+E_{kin}+E_{th} < 0
\mbox{\ \ \ ,}
\end{equation}
where $E_{grav}, E_{kin},$ and $E_{th}$ are the total gravitational, kinetic,
and thermal energies, respectively. A system will continue to collapse only
if gravity overwhelms the combined opposing efforts of thermal pressure and
rotation. Defining the usual parameters $\alpha=E_{th}/|E_{grav}|$ and
$\beta=E_{rot}/|E_{grav}|$, with $E_{rot}$ being the total rotational energy
of the system, the requirement of continued collapse can be written as
(criterion (d)):
\begin{equation}
\alpha + \beta < 1
\mbox{\ \ \ .}
\end{equation}
In Figure 5, we show these quantities as a function of distance from the
density maximum, at the instant of creating the sink particle.

\begin{center} 
\psfig{file=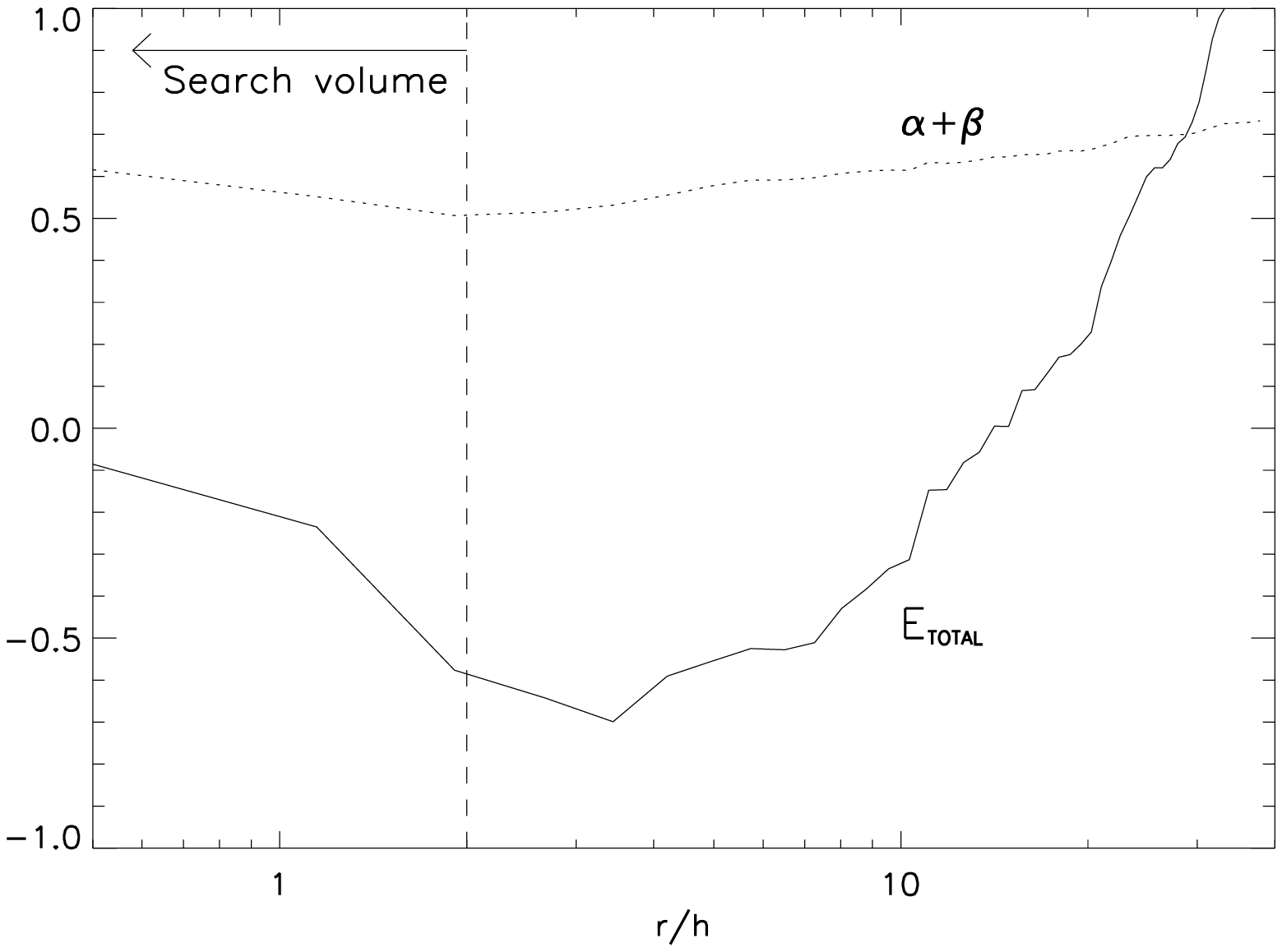,width=8.0cm,height=6.3cm}
\figcaption{
Search radius for inclusion into sink particle.
{\it Solid line:} Total energy vs. distance from density maximum. The total
energy of all particles within a given radius is defined as $E_{total}=
E_{grav}+E_{kin}+E_{th}$, and plotted with an arbitrary normalization.
Distances are in units of $h(r_{max})$, the smoothing length of the particle
with the highest density.
{\it Dotted line:} Sum of the ratios $\alpha = E_{th}/|E_{grav}|$ and
$\beta = E_{rot}/|E_{grav}|$ vs. $r/h$.
Only particles within $\Delta r < 2 h$ ({\it dashed line}) are considered
to become part of the sink particle. It is evident that this is a very
conservative choice, safely meeting the criteria $\alpha + \beta < 1$
and $E_{total}<0$. A more aggressive merging would search within
$\Delta r < 10 h$, but this would entail a more elaborate testing of whether
a candidate particle would really end up in the clump, or escape again.
\label{fig5}}
\end{center} 

Our merging algorithm now proceeds by sorting the eligible particles according
to density. Subsequently, a search is performed within a range of
$r_{search}=2 h_{max}$ around the particle with the highest density.
All particles within this volume are merged to form the sink particle, provided
they fulfill criteria (a) and (b). This procedure is repeated until all 
eligible particles are assigned to a sink particle. We assume that passing
criteria (a) and (b) implies $E_{total}<0$ and $\alpha + \beta <1$, without
explicitly testing for it. This assumption is justified by the following
physical argument. As can be seen in Figure 5, the choice of $r_{search}
=2 h_{max}$ is very conservative. All particles within this search volume
are safely bound, and far from pressure or rotational support. Surrounding
the search volume, there is a massive, collapsing envelope which is part
of the Jeans unstable flow. Consequently, a more aggressive merging could
be adopted with $r_{search}\simeq 10 h_{max}$. To reliably determine
whether one of the less bound, outer particles is to be merged, the explicit
testing for criteria (c) and (d) would be necessary in this case (Bate et
al. 1995). We have carefully verified the reliability of our merging
procedure by performing test calculations with and without criteria (c) and
(d), and find that in each case our assumption, (a)+(b)$\Rightarrow$
(c)+(d), is valid.

The next conceptual step concerns the properties of the newly
created sink particle, and how to treat the boundary to the neighboring
region. This leads to the second question: (2) {\it Does the presence
of a sink particle unphysically affect the SPH particles in its vicinity}?
Upon its formation, the sink particle (or clump) has mass
\begin{equation}
M_{Cl}=\sum_{i}m_{i}
\mbox{\ \ \ ,}
\end{equation}
a mass-weighted position vector
\begin{equation}
\vec{r}_{Cl}=\frac{\sum_{i}m_{i}\vec{r}_{i}}{M_{Cl}}
\mbox{\ \ \ ,}
\end{equation}
and a mass-weighted velocity
\begin{equation}
\vec{v}_{Cl}=\frac{\sum_{i}m_{i}\vec{v}_{i}}{M_{Cl}}
\mbox{\ \ \ ,}
\end{equation}
where the summation extends over all contributing particles.
We treat the sink particle as being still gaseous, and as consequently
still participating in the SPH interactions and smoothing procedure.
A constant gas density, $n_{Cl}=n_{th}$, and temperature, $T_{Cl}=
\sum_{i}m_{i}T_{i}/M_{Cl}$, are assigned to the clump. The resulting
(constant) pressure is higher than in the surroundings which is slightly
less dense and hot. This prescription approximately models a negative
pressure gradient. The unphysical infall of neighboring particles onto the
clump is therefore prevented. Such a situation would occur if the sink
particle were assumed to be collisionless, unless special care is taken
to formulate appropriate boundary conditions (see Bate et al. (1995)).
With continuing mass accretion, gravity becomes increasingly dominant,
and the sink particle asymptotically assumes a fully collisionless
character. In Figure 6, we demonstrate that this procedure leads to
a satisfactory treatment of the boundary between the clump and its 
neighboring particles.

\begin{center} 
\hspace{0.5cm}
\psfig{file=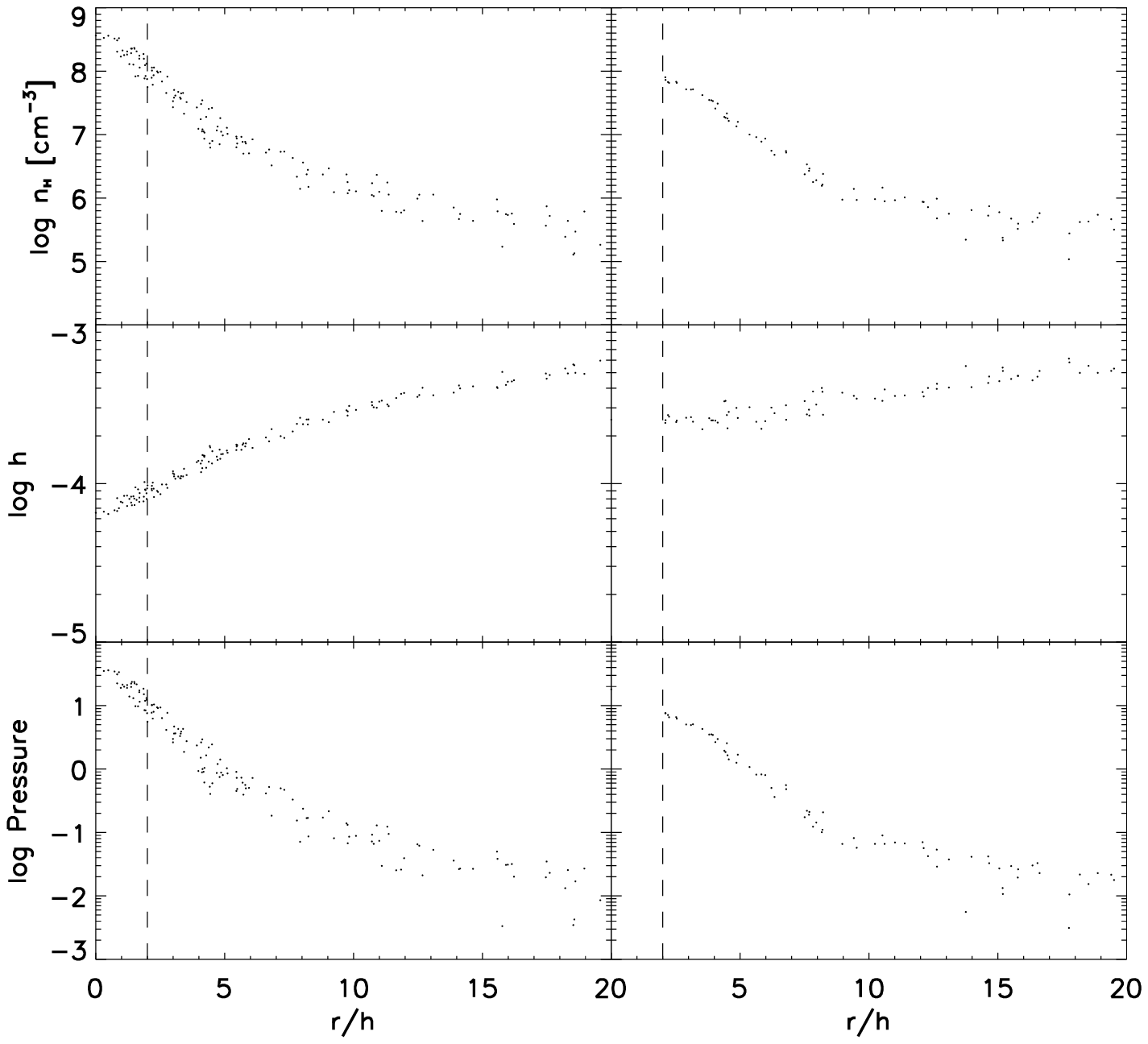,width=7.32cm,height=6.552cm}
\vspace{12pt}
\figcaption{
Effect of merging on neighboring particles.
{\it Right column:} Properties of SPH particles in the vicinity of the
sink particle, briefly after its creation.
{\it Left column:} Particle properties for the comparison case, where no
merging takes place, and the gas evolution is followed to increasingly
higher density. {\it Top panels:} Number density vs. distance from density
maximum. Distances are in units of $h(r_{max})$, the smoothing length
of the particle with the largest density, evaluated at the instant of
merging. {\it Middle panels:} Smoothing length vs. $r/h$. {\it Bottom
panels:} Gas pressure vs. $r/h$. The vertical lines at $r/h=2$ delineate
the effective size of the sink particle. It can be seen that the
thermodynamic properties, together with the radial gradients of
density and pressure, are very similar in the two cases. The smoothing
lengths, on the other hand, are noticeably larger in the presence of a sink 
particle. This is to be expected, since removal of particles entails a
degrading spatial resolution.
\label{fig6}}
\end{center} 

The final element of the algorithm handles the subsequent accretion
of gas, and addresses the question: (3) {\it Is the accretion
of a given gas particle physically justified, or does the particle venture
only temporarily into the vicinity of the clump}? A particle is accreted
onto a pre-existing clump if criteria (a) and (b) are fulfilled, and if
it approaches to within the accretion radius $r_{A}=2 h_{Cl}$, where
$h_{Cl}$ is the smoothing length of the accreting clump. This
procedure proves to be very reliable, again due to the fact that in order
to reach a density of $10^{8}$ cm$^{-3}$, the particle has to be part
of a Jeans unstable gas flow which has already proceeded well into its
runaway collapse. The position and velocity of the merged clump and
accreted particle are again taken to be the mass-weighted averages.
The density and temperature of the clump retain their initial values, 
as mentioned above.

This algorithm allows us to investigate the complex dynamics
of clump formation, subsequent gas accretion, and the occasional
merger of clumps.

\section{THE SIMULATIONS}

With all the ingredients in hand, we now turn
to the description of our simulations, and 
present an exploratory survey of how the primordial gas behaves under a
wide range of initial conditions.
We begin by describing our procedure of initializing the numerical
simulations, and explaining our choice of parameters.
We then turn to the description of the simulations, focusing first
on the early evolutionary stages, and subsequently on the later ones.

\subsection{Initial Conditions}

We model the site of primordial star formation as an isolated overdensity,
corresponding to a high$-\sigma$ peak in the random field
of density perturbations.
The numerical simulations are initialized at $z_{i}=100$ by endowing a
spherical region, containing dark matter and gas, with a uniform density
and a Hubble expansion according to the top-hat solution (see Section 2.1).
The top-hat overdensity is embedded in an Einstein-deSitter universe with
a Hubble constant $h = H_{0}/(100 \mbox{\,km\,/\,s\,/\,Mpc}) = 0.5$. We consider
halos collapsing at $z_{vir}\simeq 20$ and at $z_{vir}\simeq 30$, corresponding
to $\sim 2 \sigma$ and $\sim 3 \sigma$ peaks, respectively, with total masses
of $2\times 10^{5}M_{\odot}$ and $2\times 10^{6}M_{\odot}$. The baryonic
mass fraction is usually taken to be $\Omega_{B}=0.05$, but we also
consider the case $\Omega_{B}=0.20$.

While this is clearly a rather idealized initial configuration, the subsequent
evolution of the halo quickly departs from a simple, monolithic collapse.
In response to the imprinted small-scale DM fluctuations (see below),
the halo develops a very lumpy, nonspherical morphology.
This model, which allows for the
formation of DM substructure, and the hierarchical merger of DM clumps,
embedded in a collapsing background medium, does qualitatively describe
the dynamical behavior of the cold dark matter on very small scales.

The major shortcoming of this approach lies in the inability to 
self-consistently incorporate the angular momentum of the halo. In
a realistic cosmological setting,
angular momentum is generated by the tidal interaction of neighboring
overdensities which are, in general, nonspherical. For an isolated halo, one
therefore has to specify the distribution and magnitude of the angular
momentum in an explicit way. We here assume that the halo is initially in
rigid rotation with a given angular velocity $\omega$.
We prescribe $\omega$ in
accordance with the prediction for the spin parameter $\lambda$,
as found in cosmological N-body simulations (see Section 2.1).
In computational units with $G=M=R=1$,
one approximately has the relation: $\omega \simeq 6.7 \lambda$.
The initial angular velocity has values $\omega=0.1, 0.2,$ and 0.4, 
corresponding to $\lambda = 0.015, 0.03,$ and 0.06, respectively.

We now describe our procedure of initializing the positions and velocities of
the DM and gas particles.
To imprint small-scale density fluctuations on the dark matter, we carry out
the following steps. After setting up the DM particles on a Cartesian grid,
we impose periodic boundary conditions, and perturb the particles according
to the Zeldovich (1970) approximation. The random density field $\delta(
\vec{x})$ has the Fourier decomposition $\delta=\sum \delta_{\vec{k}}\mbox{
exp}(i\vec{k}\cdot\vec{x})$ with $\delta_{\vec{k}}=A_{\vec{k}}\mbox{exp}(
i\varphi_{\vec{k}})$. A given mode with wavevector $\vec{k}$ is specified
by the random phase $\varphi_{\vec{k}}$, distributed uniformly in the interval
$[0,2\pi]$, and the amplitude $A_{\vec{k}}=\nu\sqrt{P(k)}$, where $\nu$
is drawn from a Rayleigh distribution (see Padmanabhan 1993).
The particles, having undisturbed position $\vec{q}$, are displaced in such
a way as to reproduce the desired density field. In the Zeldovich approach, one
finds for the displaced particle position: $\vec{x}=\vec{q}+\vec{f}(\vec{q})$.
The displacement field $\vec{f}$ is related to the density via
$\delta=-\nabla\cdot\vec{f}$. It is then straightforward to show that
$\vec{f}=\sum \vec{f}_{\vec{k}}\mbox{exp}(i\vec{k}\cdot\vec{q})$ with
$\vec{f}_{\vec{k}}=i \delta_{\vec{k}}\vec{k}/k^{2}$. The Zeldovich approximation
also allows one to self-consistently assign peculiar velocities:
 $\vec{v}_{pec, i}=H_{i}\vec{f}$, where $H_{i}$ is the Hubble parameter
at $z_{i}=100$. Adding the peculiar velocity to the Hubble expansion and
the solid-body rotation, gives the resulting initial velocity for each
DM particle: $\vec{v}_{i}=\vec{v}_{H,i}+\vec{v}_{rot,i}+\vec{v}_{pec,i}$.

For Gaussian fluctuations, the power spectrum $P(k)=|\delta_{\vec{k}}|^{2}
=A k^{n}$ fully describes
the random density field. On small scales ($M < 10^{7} M_{\odot}$), the 
standard CDM model predicts an asymptotic behavior of $P(k)\propto k^{-3}$,
and we take $n=-3$ as our standard value. To investigate the dependence
of the gas fragmentation on the character of the DM substructure, we
also consider the case $P(k)\propto k^{0}$, corresponding to
white-noise perturbations. To finally fix the amplitude $A$, we
specify the initial variance of the fluctuations
\begin{equation}
\sigma^{2}_{i}=A \sum k^{n}
\mbox{\ \ \ .}
\end{equation}
The summation is over all contributing modes, where the minimum wavenumber
is given by the overall size of the Cartesian box, and $k_{max}$ by
the Nyquist frequency. We typically have $\sigma_{i}^{2}\simeq 0.1$.

The rms fluctuation at the moment
of collapse is then 
\begin{displaymath}
\sigma(z=30)=\left(\frac{1 + z_{i}}{1 + z}\right) \sigma_{i}\simeq 1
\mbox{\ \ \ .}
\end{displaymath}
This choice ensures
that the substructure develops on a similar timescale as the overall
collapse of the background medium.
A potential problem with this initialization procedure is the poor
sampling of ${\bf k}$-space for the longest wavelength modes.
Since
the overall structure of the halo is predominantly determined by these
modes (for a $k^{-3}$ spectrum),
the DM morphology will change significantly from one random
realization to the next, and we can therefore not claim to
have simulated the DM morphology in a representative way. This is
true in particular for the early stages of the collapse, whereas at
later times, different initial configurations approach similar equilibrium
states.
To remedy this shortcoming, we have studied cases with different
random realizations of the DM fluctuations,
keeping all other parameters constant.
Finally, all particles
within a given radius are selected for the simulation. The typical
number of DM particles is $N_{\mbox{\scriptsize DM}}\simeq 17,000$, but we have also
performed test calculations with $N_{\mbox{\scriptsize DM}}\simeq 130,000$.

The SPH particles, representing the baryonic component, are placed 
randomly within the given spherical volume. This procedure inevitably
introduces unphysical shot-noise. As opposed to the dark matter, however,
these numerically induced perturbations are quickly erased by pressure forces,
since the initial gas mass is close
to the Jeans mass, $M_{J}\sim 10^{5}M_{\odot}$.
The gas particles have velocities, consisting of Hubble expansion and
rigid rotation:
$\vec{v}_{i}=\vec{v}_{H,i}+\vec{v}_{rot,i}$.
The fractional
free electron and hydrogen molecule abundances are taken to be, respectively,
$x_{i}=4.6\times 10^{-4}$ and $f_{i}=2\times 10^{-6}$ (Anninos \& Norman 1996).
We assume a deuterium abundance of $n_{\mbox{\scriptsize D}}=4\times
10^{-5} n_{\mbox{\scriptsize H}}$ (Galli \& Palla 1998), and initialize
the density of
D$^{+}$ and HD according to $n_{\mbox{\scriptsize D$^{+}$}}=
10^{-8}n_{\mbox{\scriptsize H}}$, and
$n_{\mbox{\scriptsize HD}}=
10^{-3}n_{\mbox{\scriptsize H$_{2}$}}$, respectively. The gas 
temperature is finally chosen to be $T_{i}=200$ K.
The typical
number of SPH particles is $N_{\mbox{\scriptsize SPH}}\simeq 16,384$, but we have also
performed simulations with $N_{\mbox{\scriptsize SPH}}\simeq 65,536$ and
$N_{SPH}\simeq 131,072$, to test for convergence.

Initializing our simulations at $z_{i}=100$ poses the question whether
this is early enough to investigate objects that are collapsing at $z\sim 30$.
The crucial initial values for the chemical abundances,
the gas temperature and density are actually obtained by integrating
the respective governing equations beginning at
$z=1100$, along the line of Tegmark et al. (1997).
Although the use of the Zeldovich approximation in advancing
the density perturbations to $z_{i}=100$ will introduce an error, the
overall nature of the DM collapse would not change in a
significant way if the simulations were started at somewhat higher
redshift. 

In Table 3, we summarize the parameters of the different simulations.
Run A constitutes the fiducial case, which we will describe first, and
against which we subsequently compare the other runs.

\subsection{Early Evolution}

Given the almost complete absence of observations to constrain the theoretical
study of primordial star formation, it is important to
single out the essential physical processes.
We here present evidence for the existence of characteristic values for the
temperature and density of the primordial gas, which in turn translate into
a characteristic Jeans scale for fragmentation. To illustrate this 
characteristic behavior, we first describe a fiducial simulation (Run A)
in greater detail, and then investigate the effect of varying the initial
conditions.
 
\subsubsection{\it The Characteristic Mass Scale for Fragmentation}

The parameters of Run A, a halo of total mass $2\times 10^{6}M_{\odot}$ with
$10^{5}M_{\odot}$ in baryons, and virializing at $z_{vir}\simeq 30$, are
chosen to safely satisfy the Rees-Ostriker criterion for continued collapse
and fragmentation, $t_{cool} < t_{ff}$ (see Section 2.3). 
In a previous publication, we have
described a similar case (Bromm, Coppi, \& Larson 1999). The
simulation presented here has improved on that earlier treatment by
including cooling due to HD, and a more realistic initial H$_{2}$ abundance
of $f_{i}=2\times 10^{-6}$ (instead of $f_{i}=10^{-4}$). Furthermore,
Run A is initialized with a different random realization of the DM
fluctuation field.

\begin{center} 
\psfig{file=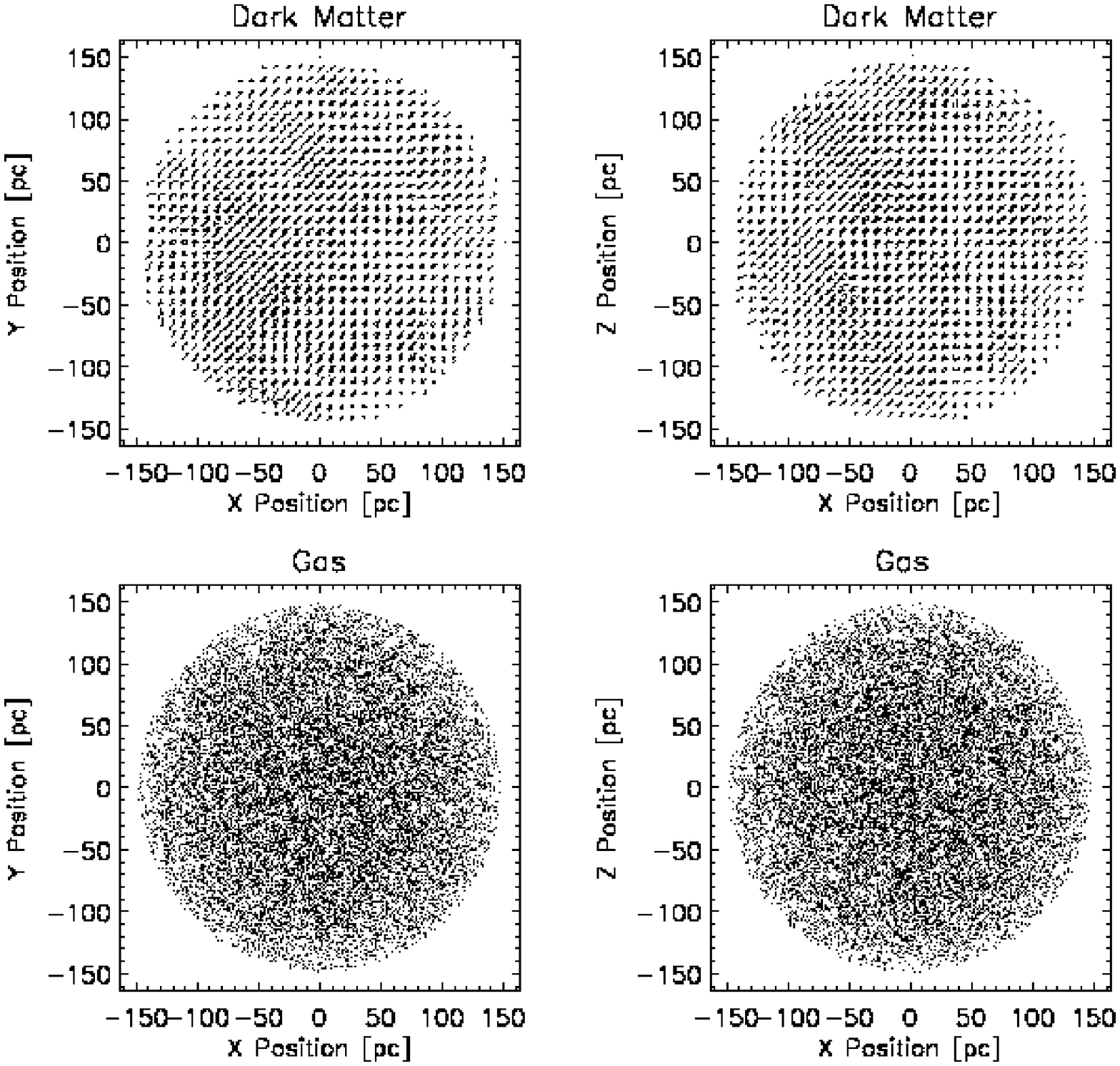,width=8.4cm,height=7.56cm}
\figcaption{
Run A: Initial configuration for top-hat collapse.
The halo has a total mass of
$2\times 10^{6}M_{\odot}$, and is endowed with a Hubble expansion such
that virialization occurs at $z_{vir}\simeq 30$.
{\it Top row:} The DM particles are perturbed from a regular grid
according to $P(k)\propto k^{-3}$.
{\it Bottom row:} The gas particles are placed at random, and comprise a
mass fraction of $\Omega_{B}=0.05$.
Both components are initially in solid body rotation with the angular
momentum vector pointing in the $z$-direction.
{\it Left panels:} Face-on view.
{\it Right panels:} Edge-on view.
\label{fig7}}
\end{center} 

\begin{center} 
\psfig{file=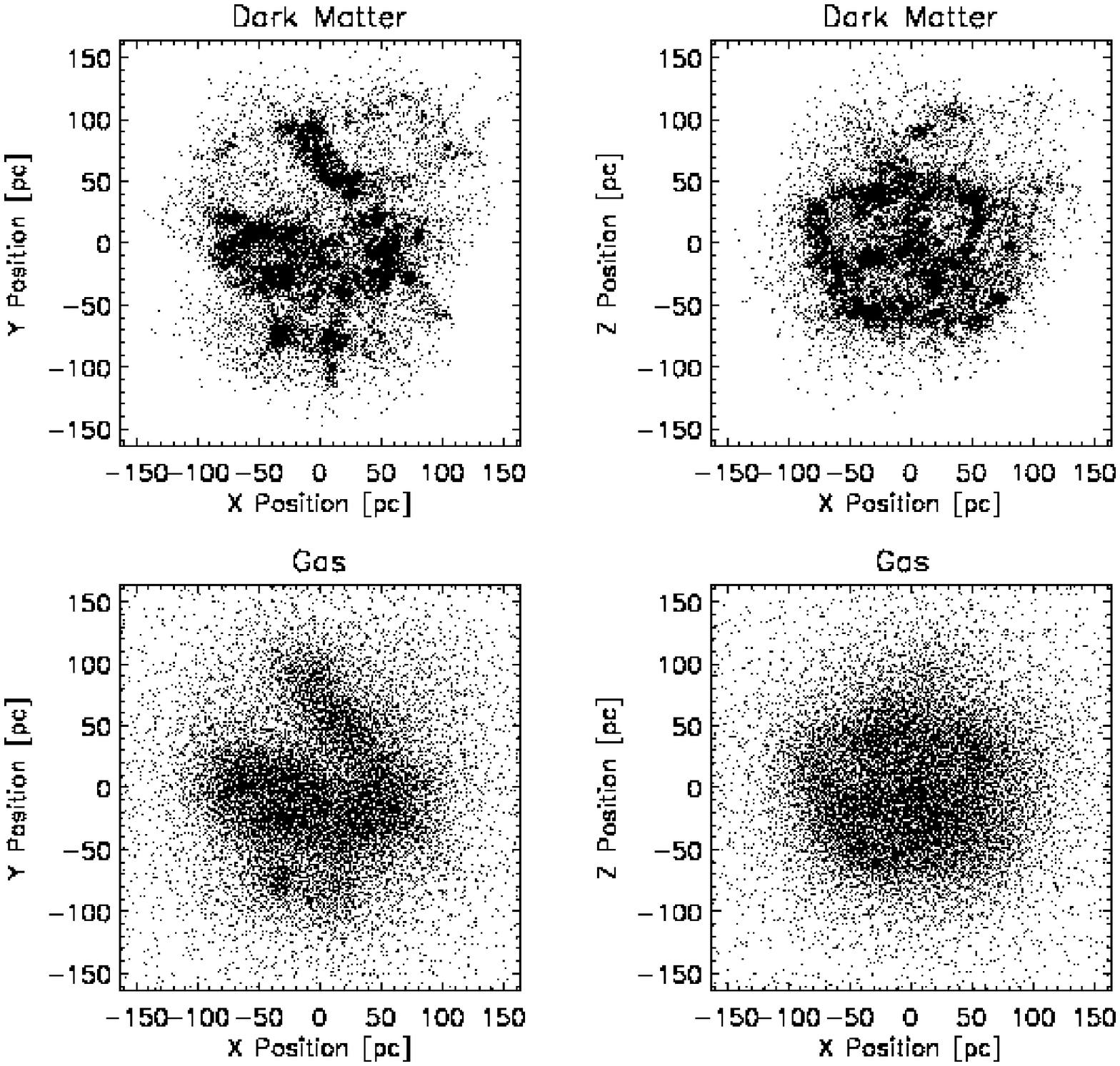,width=8.4cm,height=7.56cm}
\figcaption{
Run A: Morphology at $z=33.5$.
The manner of presentation is the same as
in Figure 7.
The DM has developed significant substructure, and the baryons are just
beginning to fall into the corresponding potential wells.
\label{fig8}}
\end{center} 

\begin{center} 
\psfig{file=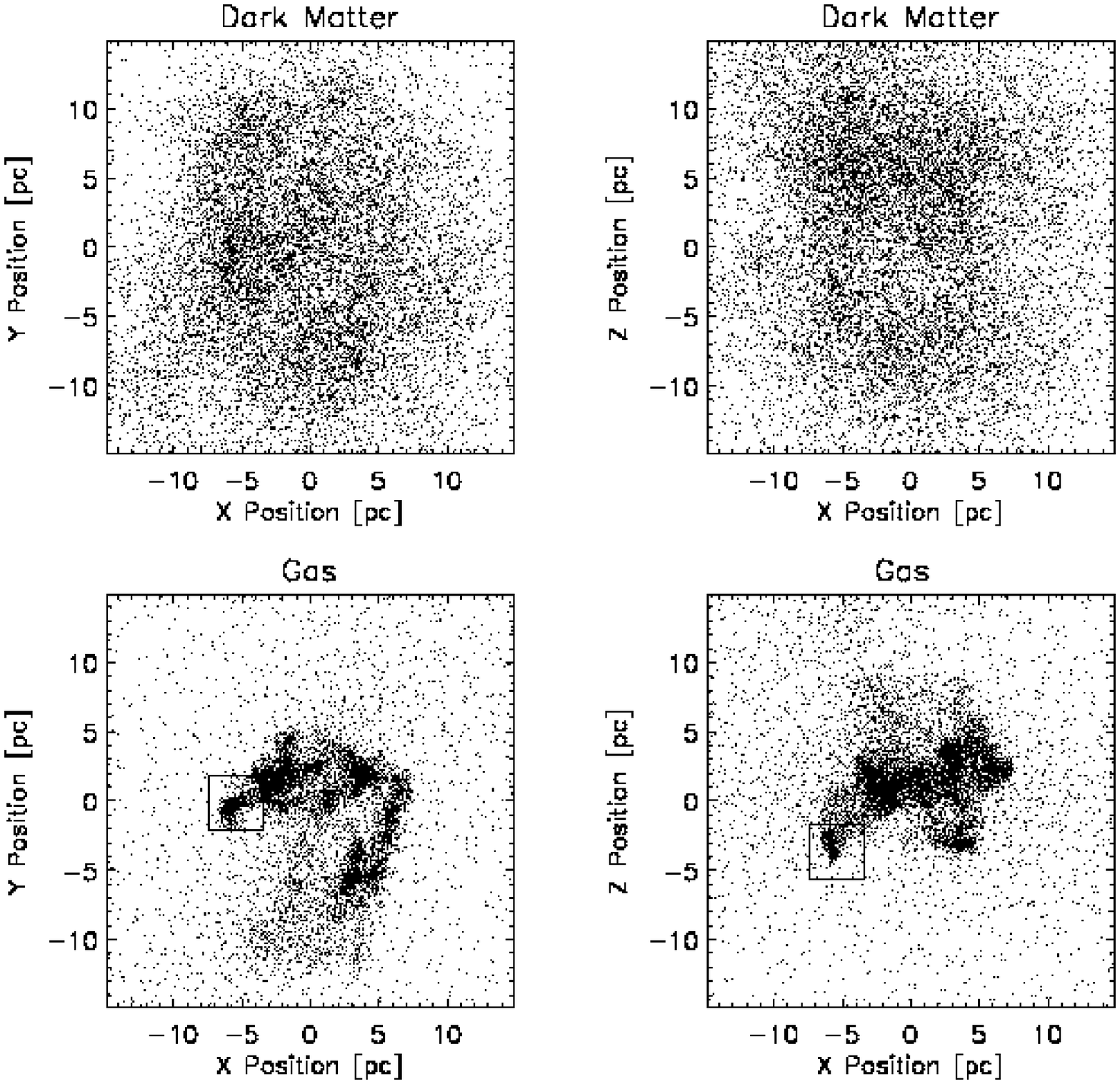,width=8.4cm,height=7.56cm}
\figcaption{
Run A: Morphology at $z=31.2$.
The convention in Fig. 7 is adopted for the rows and columns.
The box size is 30 pc.
The DM is in the process of undergoing violent relaxation with the concurrent
smoothing out of the substructure. Having developed a lumpy and elongated
morphology, the
gas has settled into the center
of the DM potential well. 
Shown is the situation briefly after the formation of the first clump of
mass $1400 M_{\odot}$ ({\it small box}).
\label{fig9}}
\end{center} 

\begin{center} 
\psfig{file=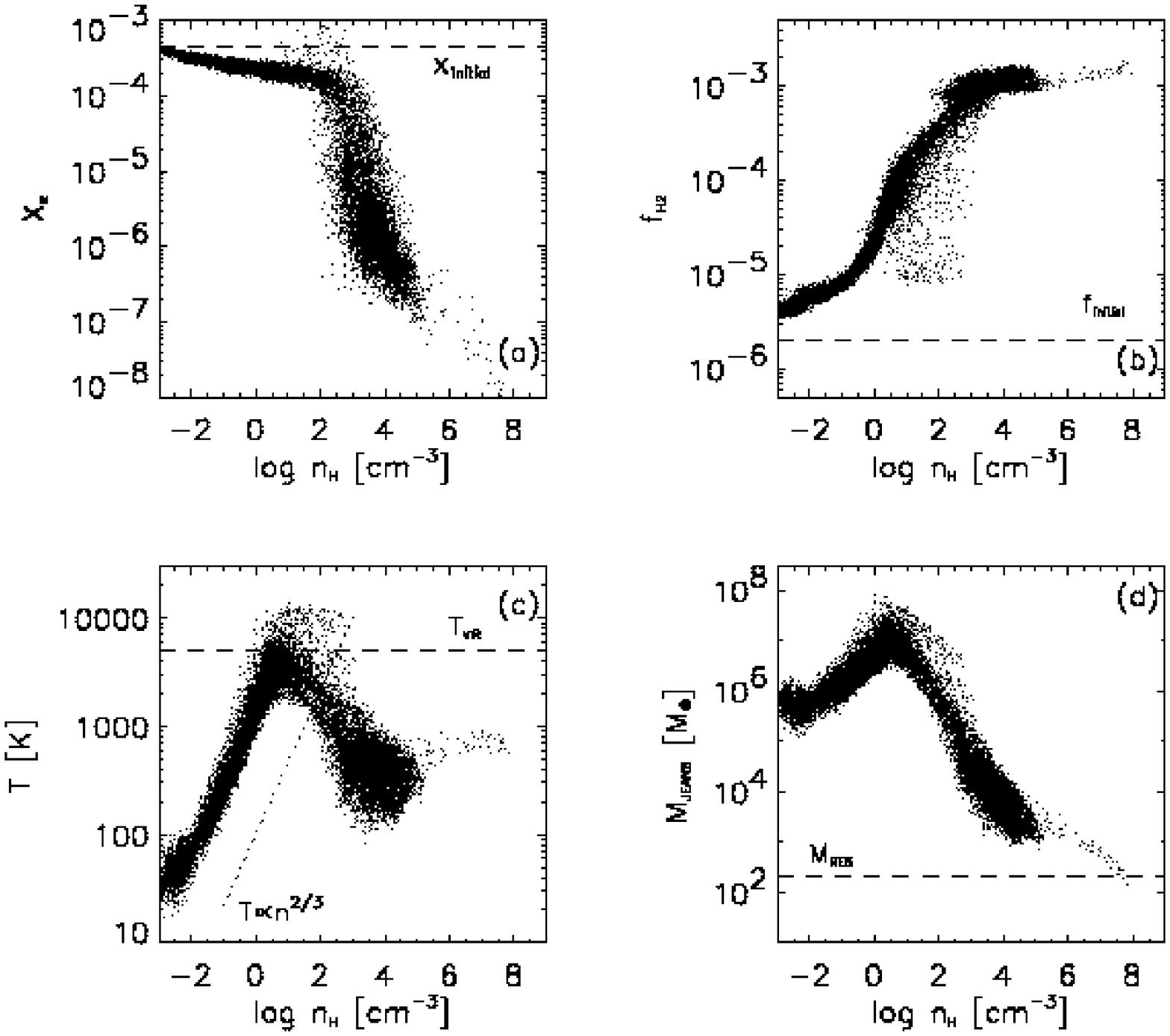,width=8.4cm,height=7.56cm}
\figcaption{
Run A: Gas properties at $z=31.2$. 
{\bf (a)} Free electron abundance vs. hydrogen number density (in cm$^{-3}$).
At densities exceeding $n\sim 10^{3}$ cm$^{-3}$, recombination is very
efficient, and the gas becomes almost neutral.
{\bf (b)} Hydrogen molecule abundance vs. number density. After a quick
initial rise, the H$_{2}$ abundance
approaches the asymptotic value of $f\sim 10^{-3}$, due to the operation
of the H$^{-}$ channel.
{\bf (c)} Gas temperature vs. number density. At densities below $\sim 1$ cm$^
{-3}$, the gas temperature rises because of adiabatic compression until
it reaches the virial value of $T_{vir}\simeq 5000$ K.
At higher densities, cooling due to H$_{2}$
drives the temperature down again, until the gas settles into a quasi-
hydrostatic state at $T\sim 500$ K and $n\sim 10^{4}$ cm$^{-3}$.
Upon further compression due to the onset of the gravitational
instability, the temperature experiences a modest rise again.
{\bf (d)} Jeans mass (in $M_{\odot}$) vs. number density. The Jeans mass
reaches a value of $M_{J}\sim 10^{3}M_{\odot}$ for the quasi-hydrostatic
gas in the center of the DM potential well, and reaches the
resolution limit of the simulation, $M_{res}\simeq 200 M_{\odot}$, for
densities close to the merging threshold of $n=10^{8}$ cm$^{-3}$.
\label{fig10}}
\end{center} 

In Figure 7, we show the initial configuration of Run A at $z_{i}=100$.
Comparing the dark matter and gaseous components, the different way of
initializing them, as described above, is clearly visible. Initially, the
halo is still expanding, until the moment of turnaround at $z_{ta}\simeq 50$.
The subsequent collapse of the dark matter, ocurring on a dynamical
timescale of $t_{ff}\sim 1/\sqrt{G\rho(z=100)}\sim 5\times 10^{7}$ yr, leads
to the eventual establishment of virial equilibrium at $z_{vir}\sim 30$.
Since the initial gas pressure is dynamically unimportant, the baryons freely
fall together with the dark matter. Upon compression, the gas temperature
rises adiabatically until it reaches the virial value
\begin{equation}
T_{vir}\simeq\frac{G M m_{\mbox{\scriptsize H}}}
{2 k_{\mbox{\scriptsize B}}R_{vir}}\sim 5000\mbox{\ K}
\mbox{\ \ \ ,}
\end{equation}
where $R_{vir}\simeq 100$ pc is the virial radius. At this point, enough
H$_{2}$ molecules have been formed ($f\sim \mbox{\,a few \,}10^{-4}$) to
provide an efficient cooling mechanism. Consequently, the temperature
decreases again with further compression. Figure 8 shows the situation
at $z=33.5$, briefly before the virialization of the dark matter. In 
response to the initially imprinted $k^{-3}$-noise, the dark matter has
developed a pronounced substructure. Evidently, the collapse proceeds in a
very inhomogeneous manner, far from the monolithic, spherically symmetric
evolution of the analytic top-hat model.
The baryons have just begun to fall into the potential wells which
are created by the DM substructure. Thus, the DM imparts a `gravitational
head-start' to certain regions of the gas, which subsequently act as
the nucleization centers for the formation of high-density clumps.

At the end of the free-fall phase, shown in Figure 9, the gas has developed
a very lumpy, filamentary structure in the center of the DM potential.
By now, the dark matter has lost the memory of the primordial perturbations,
but only after having imprinted its signature on the gas. This almost
complete erasure of the DM substructure might be due to insufficient
numerical resolution, as has been recently suggested by, e.g.,
Moore et al. (1999). We have performed a test calculation with eight times
as many DM particles ($N_{\mbox{\scriptsize DM}}\sim 130,000$ instead
of 17,000), and find no qualitative change, although a still larger
number of particles may be required ($10^{6}-10^{7}$) to see the
survival of the DM lumps.

The corresponding thermodynamic and chemical state of the gas is
summarized in Figure 10. Since the abundances, temperature and density
are plotted for every SPH particle, this mode of presentation has an
additional dimension of information: Particles accumulate (`pile up') in those
regions of the diagram where the evolutionary timescale is slow.
In panel (c) of Figure 10, one can clearly discern such a preferred state
at temperatures of a few 100 K, and densities of $10^{3}-10^{4}$ cm$^{-3}$.
These characteristic values have a straightforward physical explanation,
along the line of argument presented in Section 2.2 regarding the microphysics
of H$_{2}$ cooling. A temperature of $T\sim 100-200$ K is the minimum one
attainable via H$_{2}$ cooling. The corresponding critical density, beyond
which the H$_{2}$ rotational levels are populated according to LTE,
is then $n_{crit}\simeq 10^{3}-10^{4}$ cm$^{-3}$. 
Due to the now inefficient cooling, the gas `loiters'
and passes through a phase of quasi-hydrostatic, slow contraction.

\begin{center} 
\psfig{file=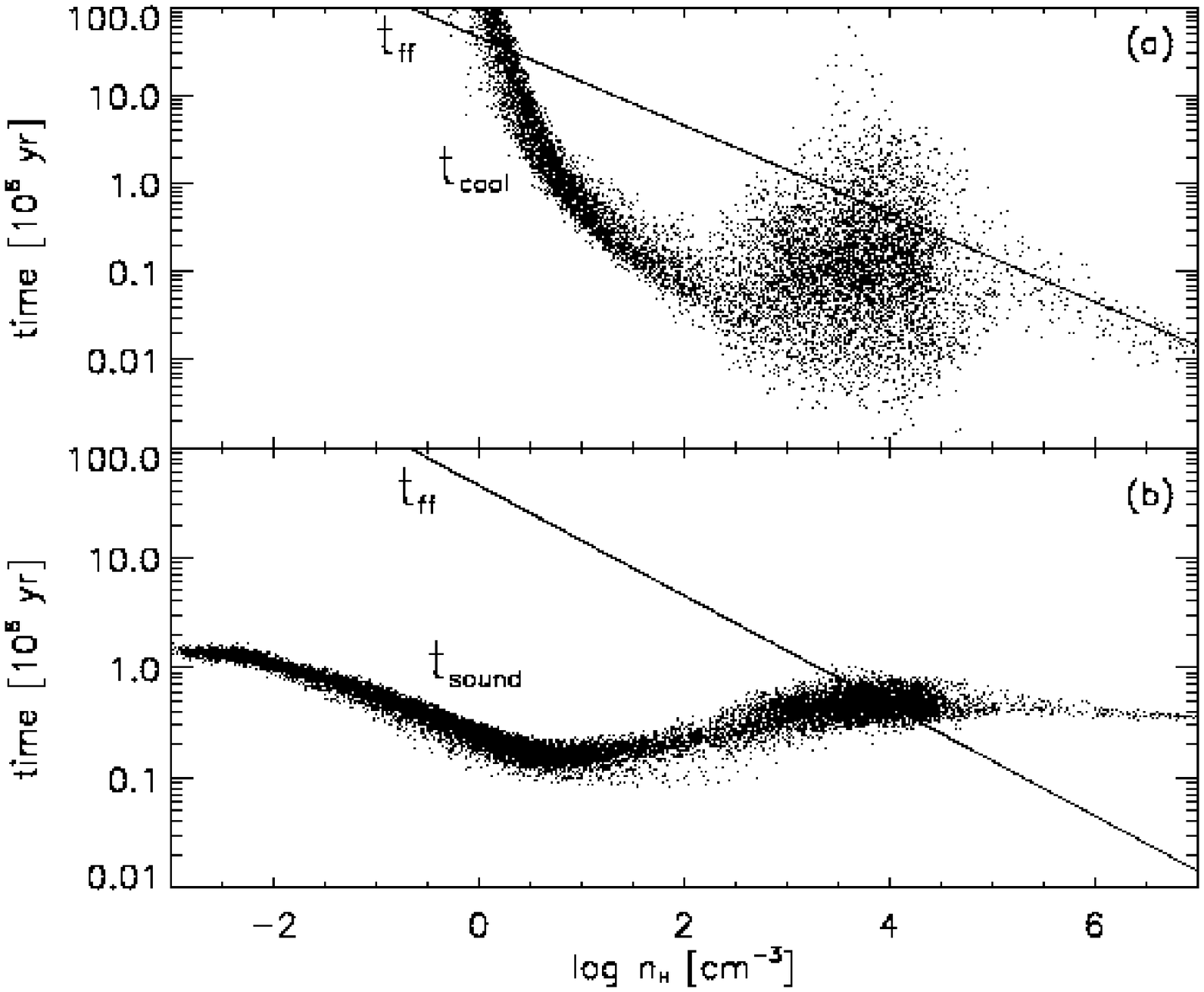,width=8.cm,height=6.3cm}
\figcaption{
Run A: Important timescales at $z=31.3$. 
Shown is the situation briefly before the formation of the first clump.
{\bf (a)} Free-fall timescale ({\it solid line}) and cooling timescale 
({\it dotted symbols}) 
 vs. number density (in cm$^{-3}$).
Timescales are in units of $10^{6}$ yr. The gas particles pile up
at a density of $n\sim 10^{4}$ cm$^{-3}$, where $t_{cool}\simeq t_{ff}$.
{\bf (b)} Free-fall timescale ({\it solid line}) and sound-crossing timescale 
({\it dotted symbols}) 
 vs. number density (in cm$^{-3}$).
The onset of the Jeans instability (i.e., $t_{sound} > t_{ff}$) at 
$n\sim 10^{4}$ cm$^{-3}$ coincides with the condition $t_{cool}\simeq
t_{ff}$ in panel (a).
\label{fig11}}
\end{center} 

Further insight can be gained by considering the characteristic timescales
of the problem, which are displayed in Figure 11. We consider the free-fall
time $t_{ff}$, the cooling time $t_{cool}\simeq n k_{\mbox{\scriptsize B}}
T/\Lambda_{\mbox{\scriptsize H$_{2}$}}$, and the sound-crossing time
$t_{sound}\simeq L_{char}/c_{s}$. Here, $L_{char}\simeq 1$ pc is the
characteristic size of the filamentary gas in Figure 9. The comparison
of $t_{cool}$ and $t_{ff}$ explains the evolutionary behavior of the gas,
as described above. At densities exceeding $n\sim 10^{0}$ cm$^{-3}$, the
gas can cool efficiently (i.e., $t_{cool} < t_{ff}$), until it reaches the
quasi-hydrostatic phase, where $t_{cool}\sim t_{ff}$. To move away from
this loitering regime, and to attain higher densities, the gas has to 
become gravitationally unstable. The condition for the onset of instability,
$t_{sound} > t_{ff}$, is shown in panel (b) of Figure 11. The gas is
Jeans unstable for $n > 10^{4}$ cm$^{-3}$. Alternatively, we can evaluate
the Jeans mass for the characteristic values $T\sim 200$ K and
$n\sim 10^{3}-10^{4}$ cm$^{-3}$, resulting in $M_{J}\sim 10^{3}M_{\odot}$.
When enough gas has accumulated in a given region to satisfy $M > M_{J}$,
runaway collapse of that fluid region ensues. We find
that the gas becomes self-gravitating ($\rho_{\mbox{\scriptsize B}} >
\rho_{\mbox{\scriptsize DM}}$) coincident with the onset of the Jeans
instability. Although the dark matter has played an important role in
determining where most of the gas ends up, it henceforth ceases to influence
the primordial gas on its further course to stardom.

\subsubsection{\it The Onset of Gravitational Instability}

After the onset of instability, the temperature rises again, up to
$T\sim 1000$ K at $n\simeq 10^{8}$ cm$^{-3}$, but not sufficiently to halt the
collapse. Once the gas has reached a density of $n_{th}=10^{8}$ cm$^{-3}$,
a clump (or sink particle) is formed, and we do not follow the evolution
to increasingly higher density.
The initial mass of the clump, as shown
in Figure 9, is $1400 M_{\odot}$, close to the characteristic Jeans mass
of $M_{J}\sim 10^{3}M_{\odot}$. 

\begin{center} 
\psfig{file=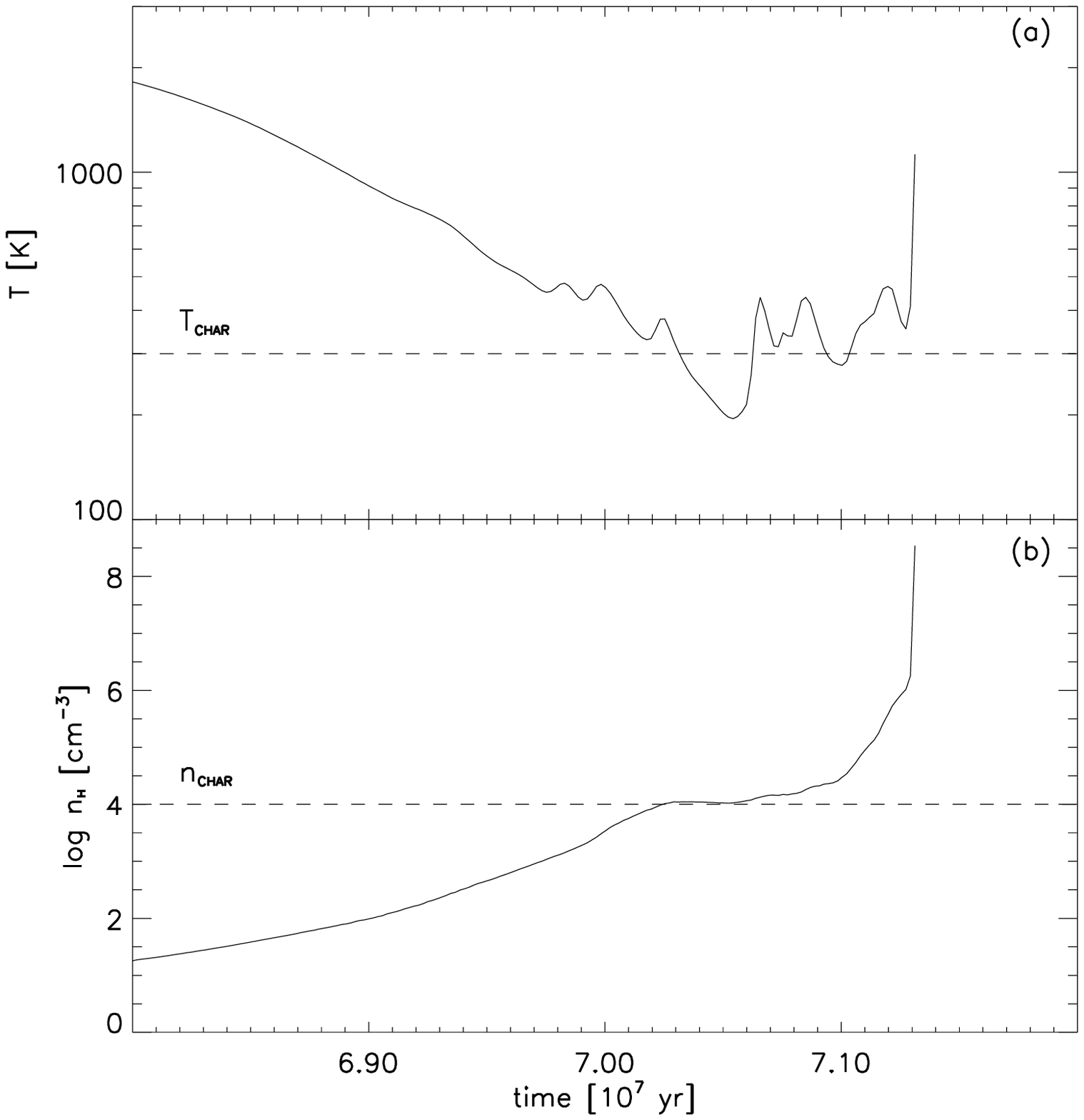,width=8.cm,height=6.3cm}
\vspace{12pt}
\figcaption{
History of first runaway fluid element. 
{\bf (a)} Gas temperature (in K) vs. cosmic time (in $10^{7}$ yr).
{\it Dashed line:} Characteristic temperature $T_{char}\simeq 300$ K.
{\bf (b)} Hydrogen number density (in cm$^{-3}$) vs. cosmic time.
{\it Dashed line:} Characteristic density $n_{char}\simeq 10^{4}$ cm$^{-3}$.
The fluid element spends $\sim 10^{6}$ yr at temperatures and densities
close to the characteristic values. After this period of ``loitering'',
the runaway collapse sets in, operating on a timescale $\sim 10^{5}$ yr.
\label{fig12}}
\end{center} 

In Paper II, we take up the question of how the further collapse of a clump
does proceed, up to densities of $n \sim 10^{14}$ cm$^{-3}$. In doing so,
we find no indication for further subfragmentation. Despite the additional
boost in the cooling due to the action of three-body reactions, which
convert the gas into almost fully molecular form at $n> 10^{8}$ cm$^{-3}$,
no runaway cooling occurs. At the end of the simulation presented 
in Paper II, a central core of $\sim 100 M_{\odot}$ is in a state of 
free-fall, surrounded by an extended envelope with an approximately
isothermal density profile, $\rho \propto r^{-2}$.

Another way to understand the nature of the runaway collapse is shown
in Figure 12, where we plot the history of the first runaway SPH
particle. This particle marks the center of the fluid region which is first
to become gravitationally unstable. It is evident that the temperature
reaches $T\sim 300$ K, and subsequently oscillates around that value for
approximately $10^{6}$ yr, to finally experience a very rapid rise
to $\sim 1000$ K. The oscillatory behavior is due to the negative feedback
of the H$_{2}$ cooling, with increased cooling for higher temperature,
and vice versa. Similarly, the density stays at $n\sim 10^{4}$ cm$^{-3}$
for $\sim 10^{6}$ yr, before the onset of the runaway collapse.

It has been pointed out by Bate \& Burkert (1997) that to avoid
numerical fragmentation, one has to
resolve the Jeans scale, $M_{J} > M_{res}$. One can estimate the
resolution limit, $M_{res}$, of the simulation as
\begin{equation}
M_{res}\simeq\left(\frac{N_{neigh}}{N_{\mbox{\scriptsize SPH}}}\right)
M_{\mbox{\scriptsize B}}
\mbox{\ \ \ .}
\end{equation}
For Run A, where $N_{\mbox{\scriptsize SPH}}=16,384$, this results in
$M_{res}\sim 200 M_{\odot}$, and the criterion above is reasonably well
satisfied. To test for numerical convergence, we have performed a simulation
with $N_{\mbox{\scriptsize SPH}}=131,072$ (Run B), and a corresponding mass
resolution of $M_{res}\sim 25 M_{\odot}$. This high-resolution run 
confirms that clumps initially form with masses close to $\sim 10^{3}
M_{\odot}$.

It is a longstanding question whether the
Jeans mass is relevant for the understanding of the characteristic mass
of present-day star formation. In the primordial case, however, where one
can argue that magnetic and turbulent pressures are initially unimportant,
one is left with the classic battle between gravity and thermal pressure,
as originally envisioned by Jeans (1902). Primordial star formation,
therefore, might be the most clear-cut setting for the application
of the Jeans criterion.

\subsubsection{\it Exploring Parameter Space}

We now harness the key advantage of our method, the ability to perform
controlled experiments, and ask how sensitive the results, obtained in 
Run A, are to variations in the initial conditions. Again, we here discuss
the simulations up to the formation of the first clumps, and turn to
the further clump evolution later.

\vspace{4pt}
\noindent
{\bf (i) Spectral index}
\vspace{4pt}

To investigate the role of the DM substructure, we consider in Run E the
case of a white-noise spectrum, $P(k)\propto k^{0}$. Admittedly, such a
spectrum is physically ad hoc, in contrast to the $k^{-3}$ case which
is ultimately motivated by the theory of inflation. With the rms
fluctuation on a given mass scale being $\sigma(M)\propto M^{-(n+3)/6}$
for spectral index $n$, one finds
\begin{equation}
\sigma(M)\propto\left\{
\begin{array}{ll}
\mbox{const.}
& \mbox{for \ } k^{-3} \\
M^{-1/2} &  \mbox{for \ } k^{0} \\
\end{array}
\right. 
\mbox{\ \ \ .}
\end{equation}
The $k^{-3}$ case, therefore, has approximately equal power on all mass scales,
whereas the smallest resolvable scale (given by the Nyquist frequency)
dominates the white-noise realization.

In Figures 13 and 14, we present the evolution of Run E up to the onset
of gravitational instability. This time sequence is to be compared to the
corresponding Figures 7 - 9 for Run A.
At redshift $z=33.5$, briefly before virialization, the baryons have not
yet begun to fall into the shallow DM potential wells, in contrast to
Run A. Towards the end of the free-fall phase, at $z=31.2$, the gas has
settled into a ring-like central configuration with a morphology which
is somewhat more regular than in Run A. This distribution derives from the
rather homogeneous character of the overall collapse which deviates much
less from spherical symmetry than the $k^{-3}$ case. High-density clumps are
formed again with initial masses close to $\sim 10^{3}M_{\odot}$, and
the full difference to Run A becomes manifest only during the later
evolutionary stages, as will be discussed below.

The occurrence of the regular ring structure in Figure 14 might be
an artefact of the initial conditions in Run E which are highly
symmetric. Such a ring-like configuration, however, is not a typical
result. In general, the resulting morphology is a somewhat accidental
feature of our results, and is not important for the main 
conclusions of this study.

\begin{center} 
\psfig{file=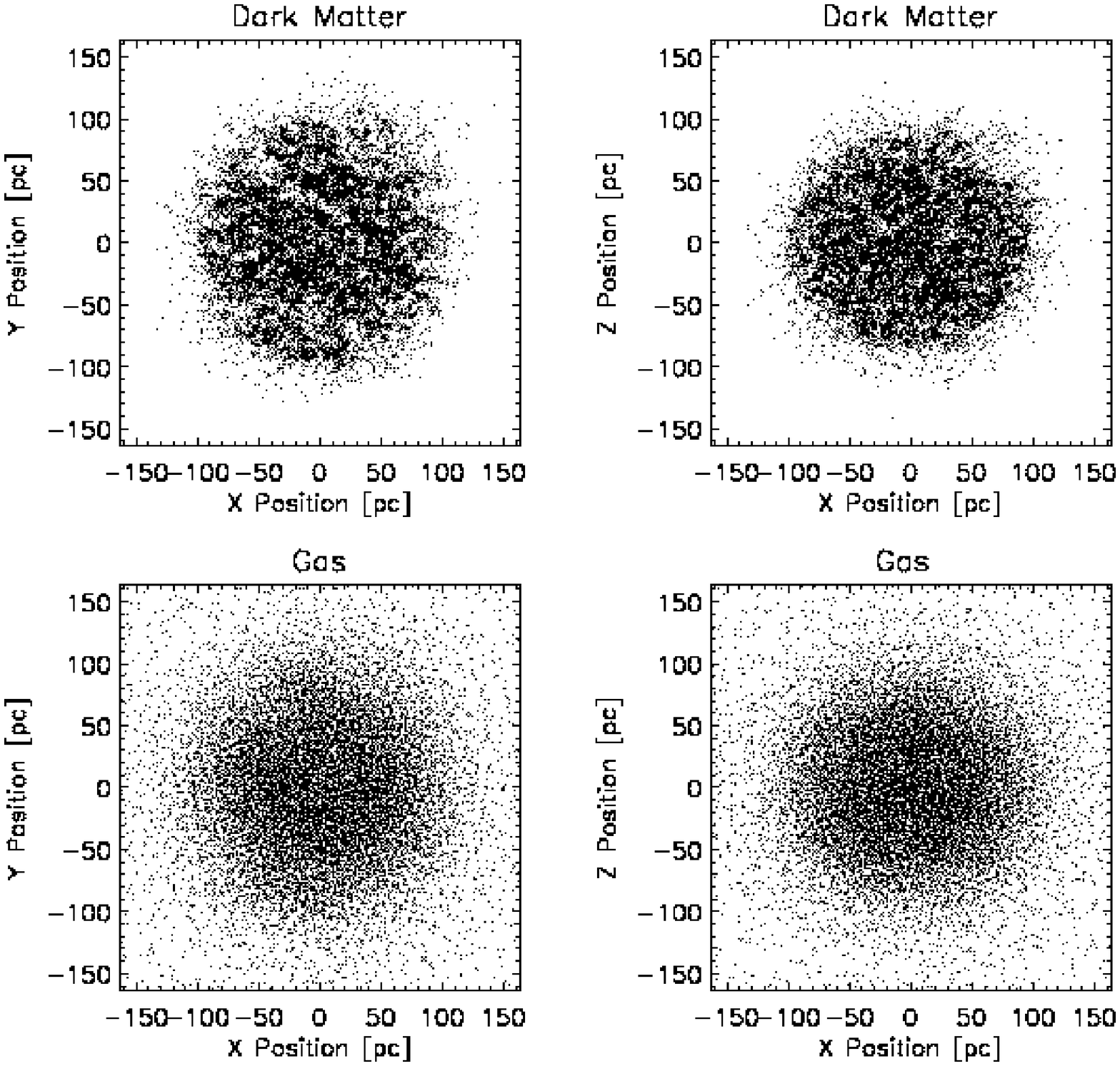,width=8.4cm,height=7.56cm}
\figcaption{
Run E: Morphology at $z=33.5$.
The convention in Fig. 7 is adopted for the rows and columns.
Compared to Run A in Figure 8,
there is relatively more DM substructure
on the smallest resolvable scales, and the overall collapse
proceeds in a more regular way.
The baryons do not yet fall into the shallow DM 
potential wells.
\label{fig13}}
\end{center} 
\begin{center} 
\psfig{file=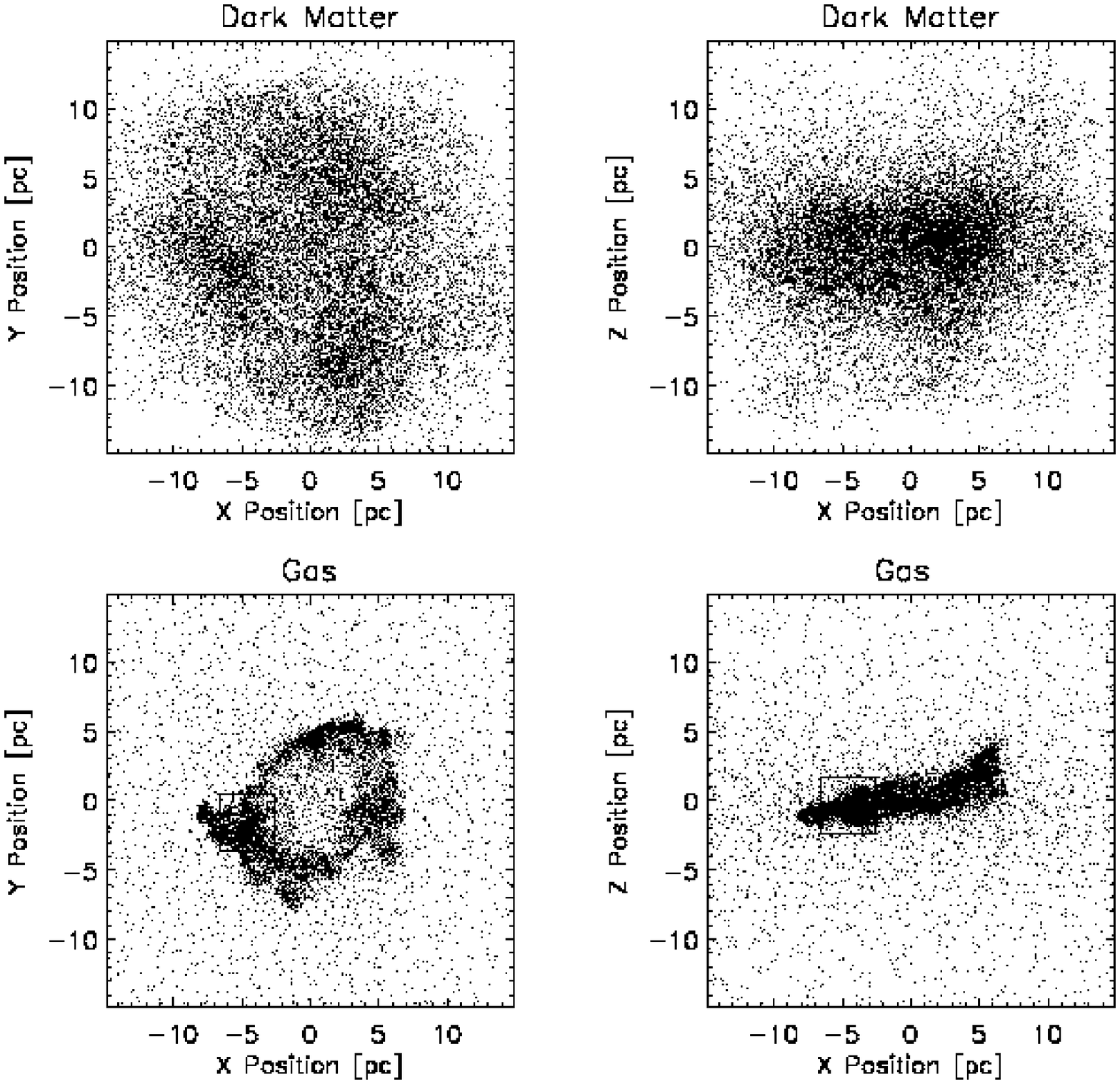,width=8.4cm,height=7.56cm}
\figcaption{
Run E: Morphology at $z=31.2$.
The convention in Fig. 7 is adopted for the rows and columns. 
The boxsize is 30 pc.
Shown is the situation briefly before the formation of the first clump 
(enclosed in the small box).
Again, the dark matter is loosing its substructure in the process
of virialization, whereas the gas has settled into a ring-like, central
configuration which has a very knotty appearance.
\label{fig14}}
\end{center} 

\vspace{4pt}
\noindent
{\bf (ii) Random realization of DM fluctuations}
\vspace{4pt}

To address the problem of poor {\bf $k$}-space sampling for the longest
wavelength modes, and the resulting morphological variations in the dark
matter, we now compare Runs A and K, and the corresponding Figures 9
and 15. Run K has the same parameters as Run A, but a dark matter 
component which is perturbed according to a different realization of
the Gaussian random process. The appearance of the central gas configuration
at $z=31.2$ is quite different indeed, with a much more extended, filamentary
distribution of gas in Run K. Also, the first two clumps to form have
masses close to $\sim 500M_{\odot}$, compared to the one $\sim 1500M_{\odot}$
clump in Run A. We will see
below, however, that these two cases later on converge
to a rather similar state, despite the differences during the early
evolutionary stages.

\begin{center} 
\psfig{file=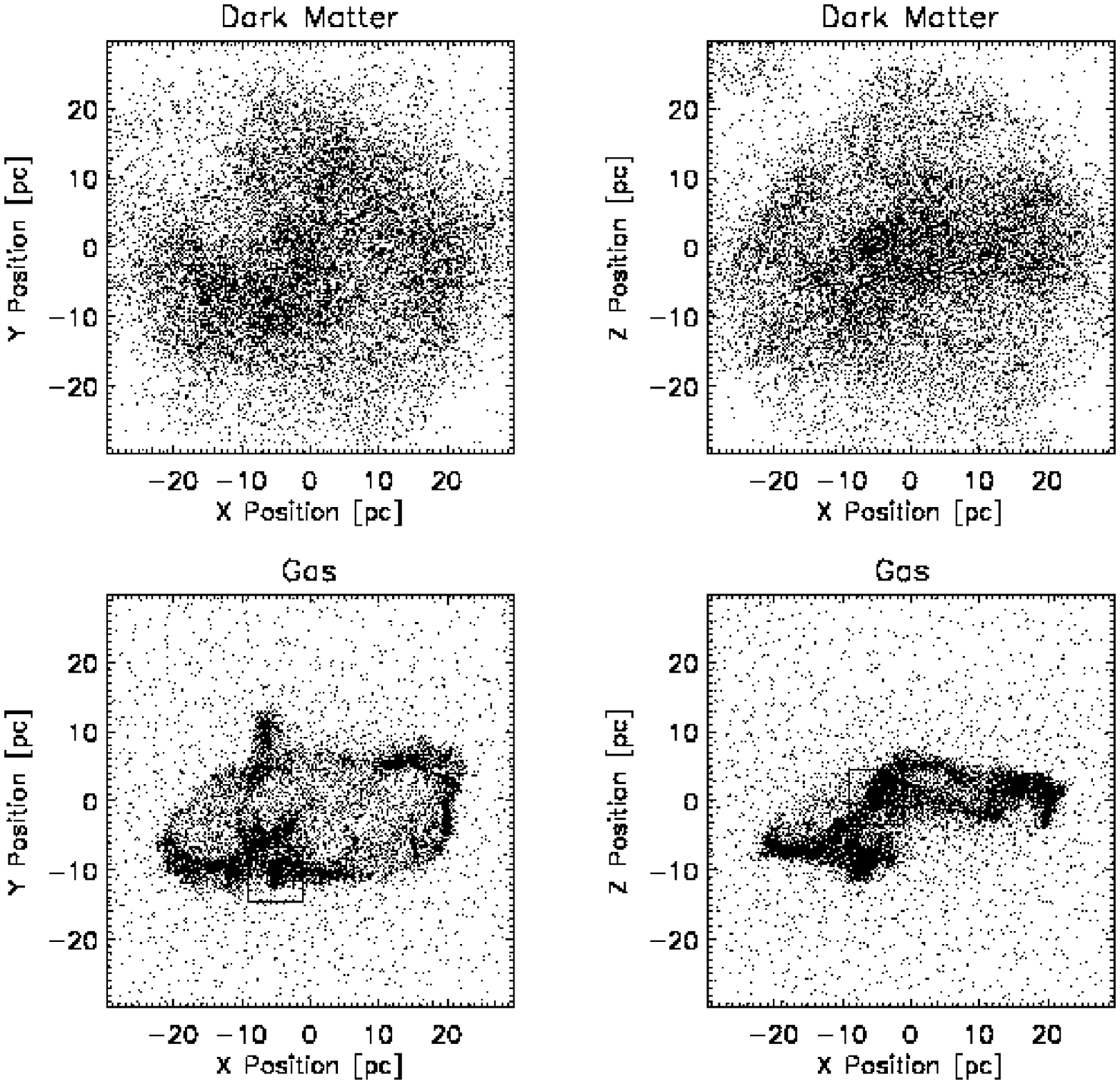,width=8.4cm,height=7.56cm}
\figcaption{
Run K: Case with different realization of $P(k)\propto k^{-3}$ noise.
Shown is the morphology at $z=31.2$, briefly after the formation of the
first two clumps with masses $360$, and $520 M_{\odot}$ 
({\it small box}).
The convention in Fig. 7 is adopted for the rows and columns.
The boxsize is 60 pc.
This case is to be compared to Run A in Figure 9. Both cases have
the same initial conditions with the exception of the random realization
of the DM fluctuation field. It can be seen that the resulting morphology
is very different here, with a much more extended central gas configuration.
\label{fig15}}
\end{center} 
\begin{center} 
\psfig{file=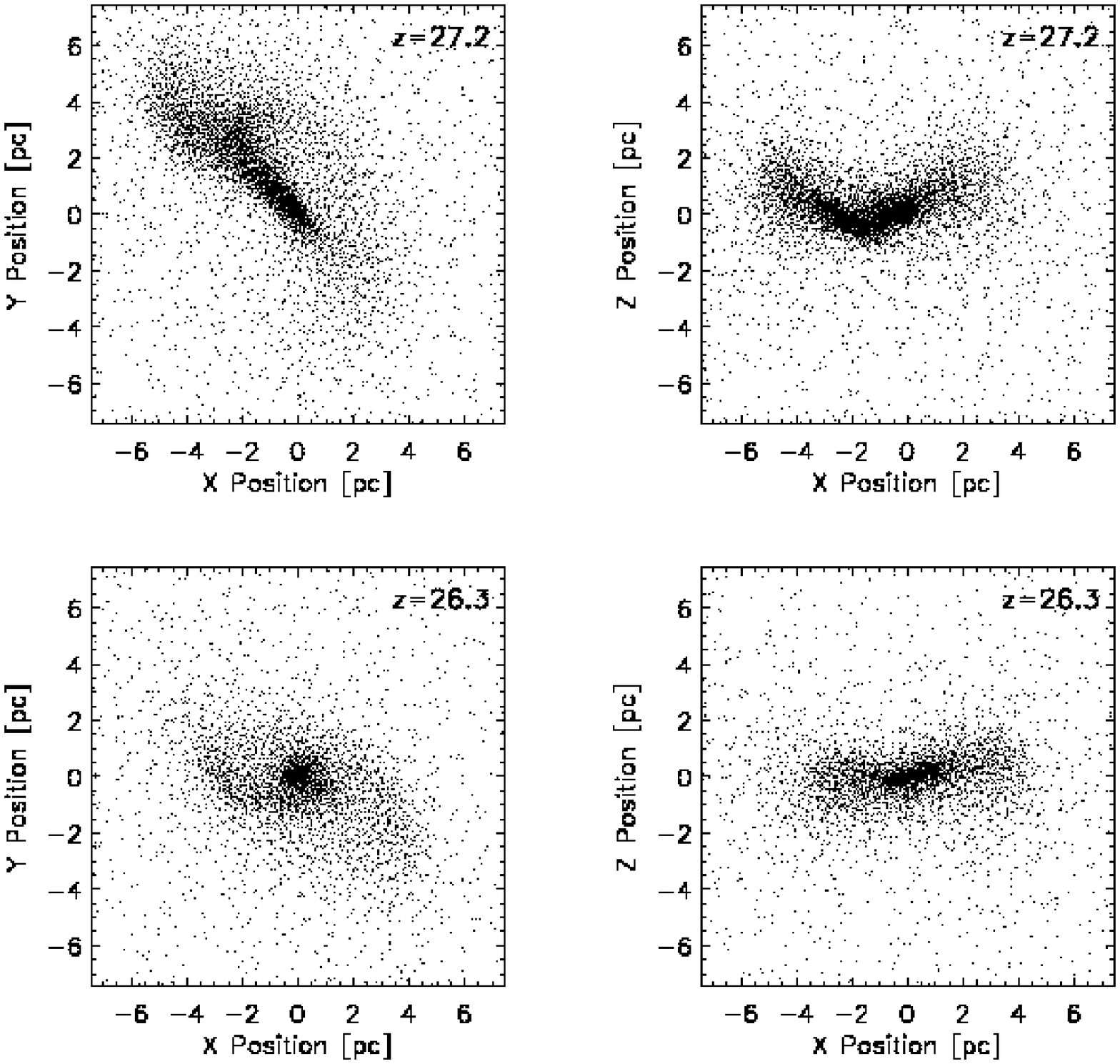,width=8.4cm,height=7.56cm}
\figcaption{
Run H: Case of halo with total mass of $2\times 10^{5}M_{\odot}$.
Shown is the morphology of the gas for two subsequent times.
{\it Top row:} Gas distribution at $z=27.2$, briefly after the formation
of a clump with $M\sim 400 M_{\odot}$. 
{\it Bottom row:} Gas distribution at $z=26.3$. By now, the clump has grown
in mass to 
$M\sim 1800 M_{\odot}$. 
{\it Left panels:} Face-on view.
{\it Right panels:} Edge-on view.
The box has a linear size of 15 pc.
In the case of this low-mass halo, where the requirement for efficient
cooling, $t_{cool} < t_{ff}$, is only marginally satisfied,
one clump forms in the center of the DM potential, and reaches a final
mass of $\sim 2000 M_{\odot}$.
\label{fig16}}
\end{center} 

\vspace{4pt}
\noindent
{\bf (iii) Angular momentum}
\vspace{4pt}

In Runs C, A, and D with initial angular velocities of $\omega=0.1$, 0.2,
and 0.4, respectively, we investigate the influence of angular momentum
(or spin) on the evolution of the primordial gas. Increasing the amount of
spin has two main effects. First, the moment of virialization and
of the onset of gravitational instability is delayed, leading to the
sequence of collapse redshifts: $z_{vir}\simeq 31.7$, 31.2, and 29.8 for
Runs C, A, and D. Second, the gas is less centrally concentrated, resulting
in a reduced rate for the merging of clumps, as will be discussed in the
following section. 

The first effect can be understood in terms of a straightforward
modification of the analytical top-hat model, adding to it the presence
of angular momentum. By considering the energy balance at turnaround, an
estimate for the turnaround radius can be obtained as follows (in
dimensionless units where $G=M=R=1$):
\begin{equation}
\frac{1}{R_{ta}}\simeq 1-\frac{1}{3}\omega_{i}^{2}-\frac{1}{2}H_{i}^{2}
\mbox{\ \ \ ,}
\end{equation}
where $\omega_{i}$ and $H_{i}$ are the initial angular velocity and Hubble
parameter, respectively. With the virial radius given by $R_{vir}\simeq
1/2 R_{ta}$, the redshift of virialization is approximately
\begin{equation}
1+z_{vir}\simeq (1+z_{vir,nr})(1-20.5\lambda^{2})
\mbox{\ \ \ .}
\end{equation}
Choosing the redshift of virialization in the absence of rotation as
$z_{vir,nr}=31.8$, nicely reproduces the numerical results. The presence
of angular momentum, therefore, delays the collapse by reducing the
binding energy of the halo.

\vspace{4pt}
\noindent
{\bf (iv) Halo mass}
\vspace{4pt}

In Run H, we study the collapse of a less massive halo of total mass
$2\times 10^{5}M_{\odot}$. This case only marginally satisfies
the Rees-Ostriker criterion (see Section 2.3). The smaller halo mass
translates into a lower virial temperature, $T_{vir}\sim 2000$ K, which
in turn leads to a reduced efficiency of H$_{2}$ cooling. Consequently,
the condition for the termination of the free-fall phase, $t_{cool}\sim
t_{ff}$, is reached already at lower densities ($10^{2}-10^{3}$ cm$^{-3}$).
The gas, therefore, goes through a prolonged phase of quasi-hydrostatic
contraction, and attains a roughly spherical configuration.
In Figure 16, we show the central gas distribution at two successive
redshifts. Only one clump forms in the center of the cloud with an initial
mass of $\sim 400 M_{\odot}$. Subsequently, this clump grows in mass up
to $\sim 2000 M_{\odot}$. 

The main difference to Run A, besides the formation of only one clump
instead of a few, is that here the Jeans instability proceeds less violent,
since the opposing effect of pressure is non-negligible in this case.

\vspace{4pt}
\noindent
{\bf (v) Baryon fraction}
\vspace{4pt}

In a universe with a significant contribution to the critical density
in the form of vacuum energy, a larger fraction of the matter resides
in the baryonic component. For $\Omega_{m}=1-\Omega_{\Lambda}=0.3$, one
has $\rho_{\mbox{\scriptsize B}}\simeq 0.20 \rho_{m}$, where $\rho_{m}$
is the total density of matter. In Run L, we consider a halo with 20\%
of the mass in gaseous form. From the outset,  this case behaves very
different from the simulations with a low baryon fraction.
As can be seen in Figure 17, the gravitational instability is already
triggered during the free-fall phase in the three dominating
DM condensations. Since the thermal properties of the gas are not very
different from those in Run A, and there is a four times larger amount
of available gas, gravity more easily overwhelms thermal pressure.
The peculiar character of Run L continues into the later evolutionary 
stages, where an altogether larger fraction of the gas is able to
condense into high-density clumps.

\begin{center} 
\psfig{file=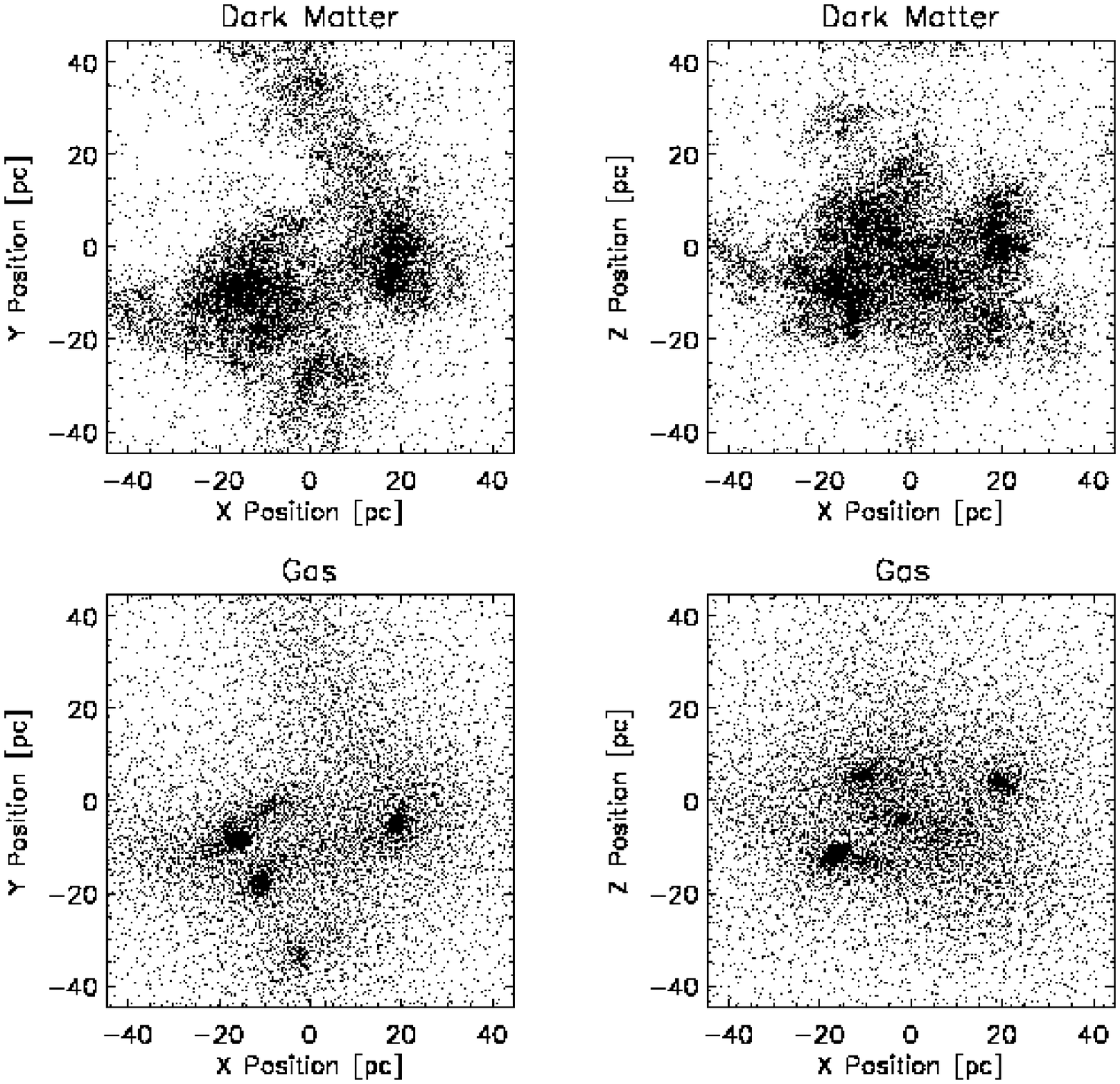,width=8.4cm,height=7.56cm}
\figcaption{
Run L: Morphology of case with high $\Omega_{B}$ at $z=31.7$.
The convention in Fig. 7 is adopted for the rows and columns.
The boxsize is 90 pc.
Shown is the situation briefly after the formation of the first clumps 
with masses of $750$, $850$, and $1440 M_{\odot}$.
The halo is still in its initial free-fall collapse, and the DM has not
yet virialized. Already in this early dynamical stage,
the gas in the deepest DM potential wells has undergone runaway collapse.
This behavior is in marked contrast to the cases with a low baryon
fraction ($\Omega_{B}=0.05$).
\label{fig17}}
\end{center} 
\begin{center} 
\psfig{file=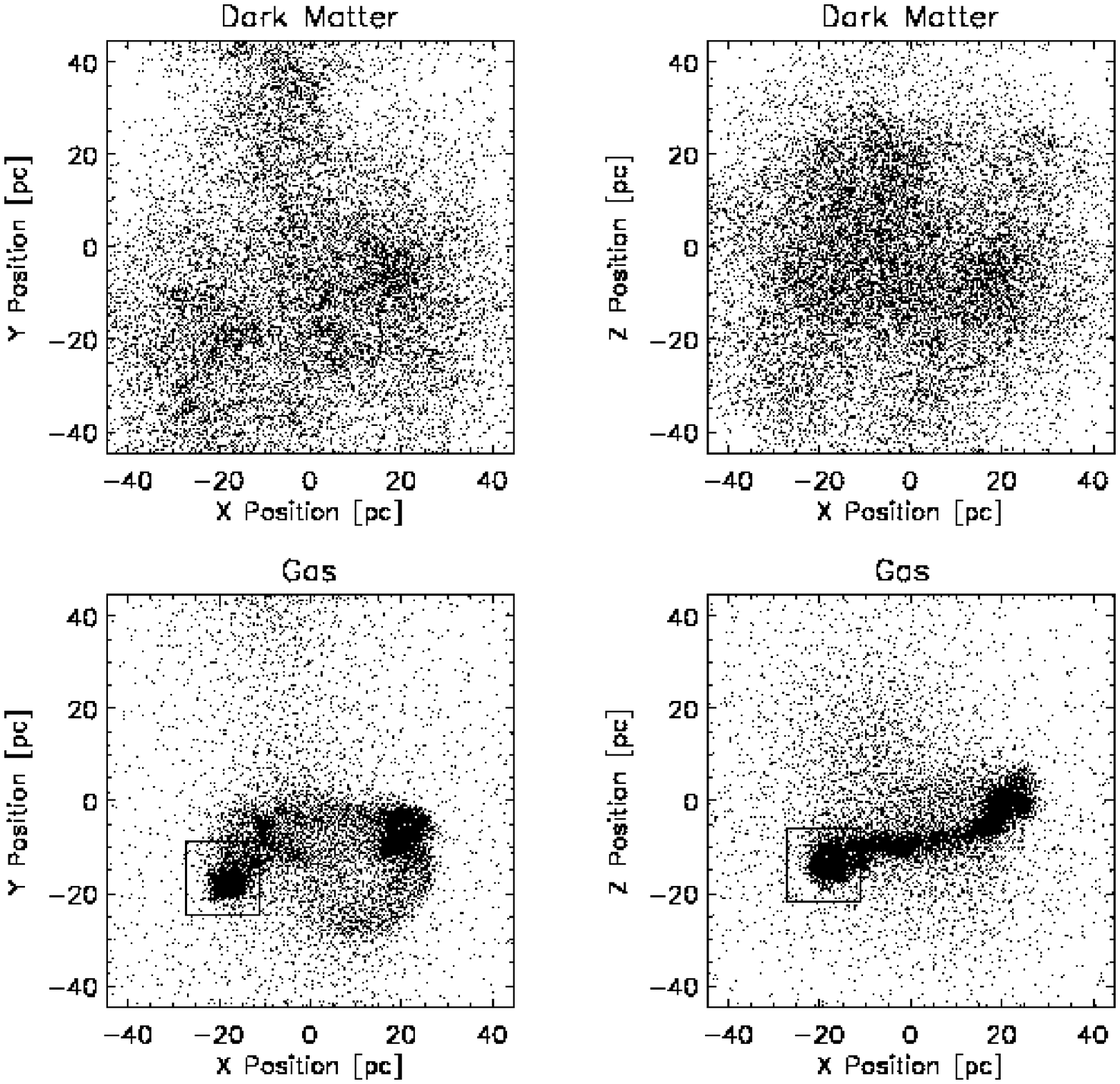,width=8.4cm,height=7.56cm}
\figcaption{
Run G: Case of top-hat with $z_{vir}\simeq 20$.
Shown is the morphology at $z=20.6$, briefly after the first
clump has formed with a mass of $920 M_{\odot}$ (in the region marked
by the small box).
The convention in Fig. 7 is adopted for the rows and columns.
The boxsize is 90 pc.
Compared to Run A (a top-hat of the same mass collapsing at $z_{vir}\simeq 30$)
in Figure 9, the resulting gas configuration is much more extended.
\label{fig18}}
\end{center} 
\begin{center} 
\psfig{file=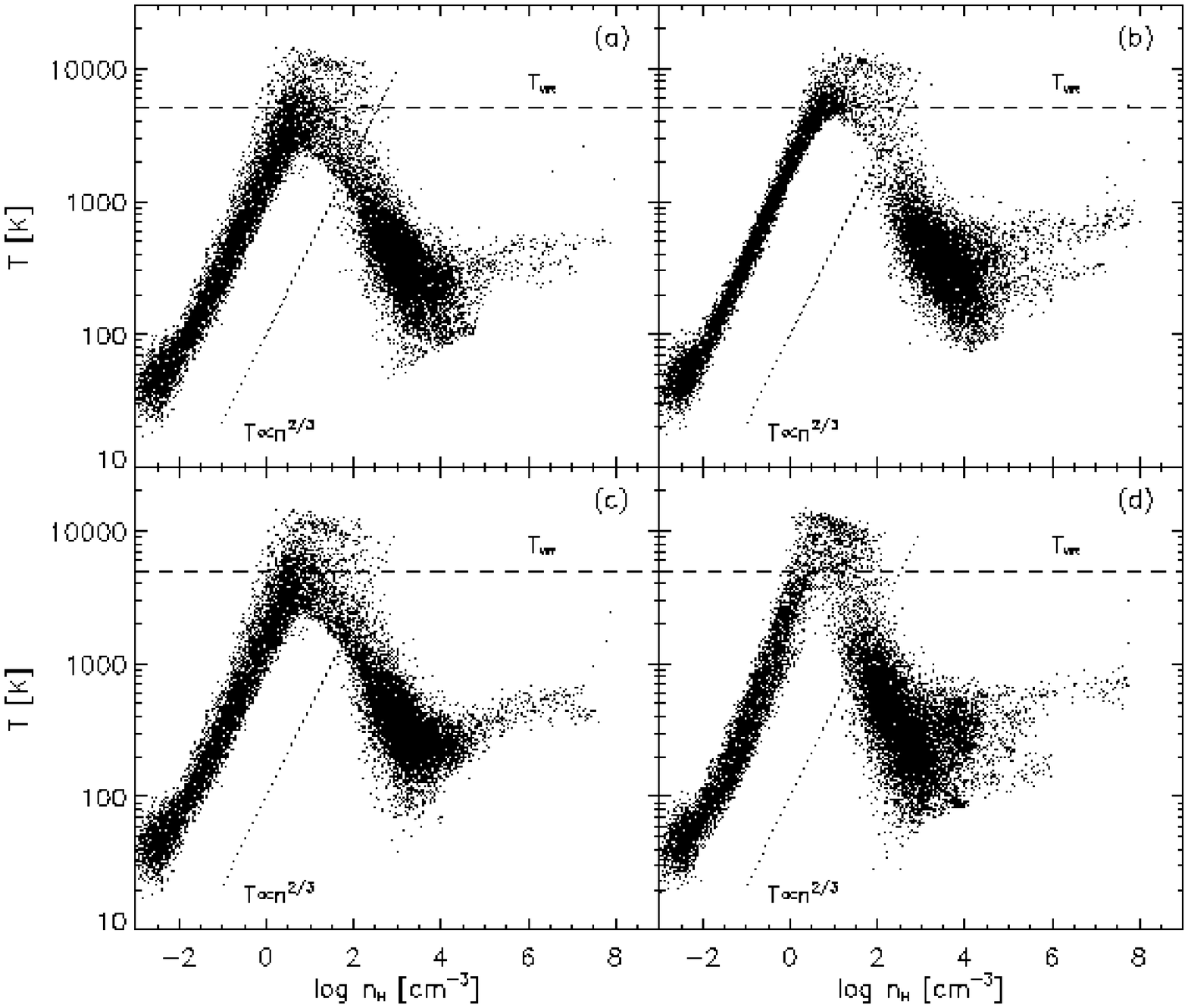,width=8.4cm,height=7.56cm}
\figcaption{
Gas properties
in simulations with different initial conditions I.
Temperature vs. hydrogen
number density. All runs are shown at very nearly the same instant, 
at $z=31.0$.
{\bf (a)} Fiducial case (Run A): Halo of total mass $2\times 10^{6}M_{\odot}$,
collapsing at $z_{vir} = 30$, initialized with
$P(k)\propto k^{-3}$, and including HD cooling.
{\bf (b)} Varying the power spectrum (Run E): Same as (a), but $P(k)\propto k^{0}$.
{\bf (c)} Varying the cooling (Run F): Same as (a), but no HD cooling.
{\bf (d)} Same as (a), but different realization of the random density field
 (Run K).
\label{fig19}}
\end{center} 
\begin{center} 
\psfig{file=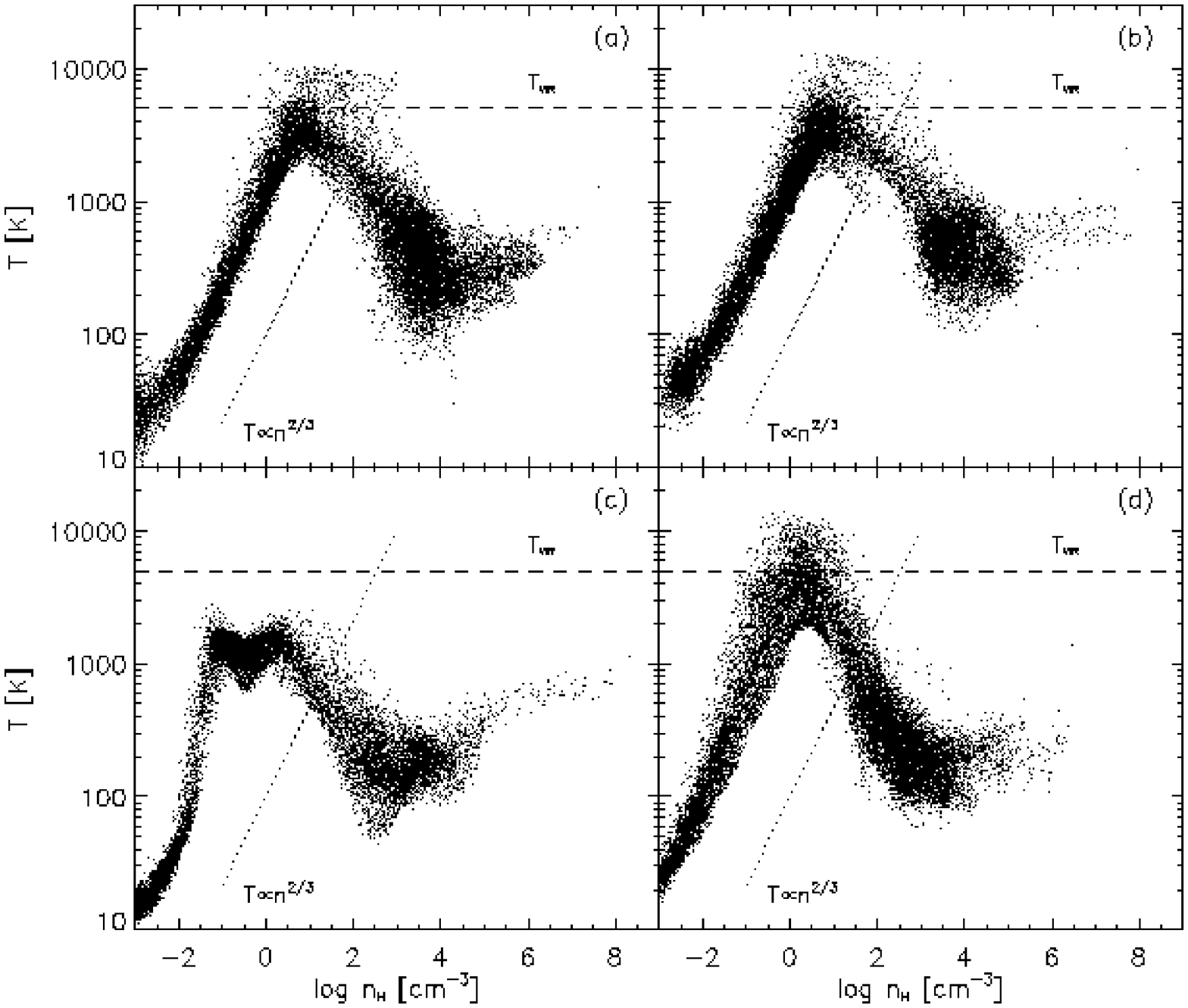,width=8.4cm,height=7.56cm}
\figcaption{
Gas properties
in simulations with different initial conditions II.
Temperature vs. hydrogen
number density.
{\bf (a)} Varying the number of particles (Run B): Simulation with large
number of SPH particles, shown at $z=31.0$. For clarity and ease of 
comparison, only every eighth particle is plotted.
{\bf (b)} Varying the angular momentum (Run C): Low-spin case, shown at $z=31.7$.
{\bf (c)} Varying the halo mass (Run H): Less-massive halo, shown at $z=27.2$.
{\bf (d)} Varying the collapse redshift (Run G): Halo with $z_{vir}\simeq 20$, shown at $z=20.6$.
The thermodynamic behavior, displayed in Figures 23 and 24, is very robust
under variation of the initial conditions, and in each case, the gas
attains characteristic values of temperature and density close to
$T\sim 200$ K and $n\sim 10^{3}-10^{4}$ cm$^{-3}$, respectively.
\label{fig20}}
\end{center} 

\vspace{4pt}
\noindent
{\bf (vi) Collapse redshift}
\vspace{4pt}
 
In Run G, we consider the collapse of a halo with total mass $2\times 10^{6}
M_{\odot}$, collapsing at $z_{vir}\simeq 20$. Run A and G correspond
to a $\sim 3 \sigma$ and $\sim 2 \sigma$ peak in the Gaussian random
field, respectively. We show the situation briefly after the formation
of the first clump in Figure 18, which should be compared to the
equivalent stage for Run A in Figure 9. The initial clump mass is
again $\sim 10^{3}M_{\odot}$, and the main difference between the two
cases lies in the much more extended morphology of the gas in Run G, with
a linear size of $\sim 40$ pc compared to $\sim 10$ pc in Run A. The
larger extension of the central gas configuration is simply due to
the smaller binding energy of the halo in Run G, with almost the same
amount of initial rotational energy as in Run A. Otherwise, the evolution
of the two simulations is very similar.
\newline

Summarizing the results from our exploratory survey up to the onset
of gravitational instability, two important lessons can be learned.
First, the morphology of the collapse varies significantly among the different
cases, depending on the initial conditions. Second, the thermodynamic behavior
of the priomordial gas is very similar for all the cases studied, despite the
morphological diversity. To demonstrate this, we show in Figures 19 and 
20 the location of the individual SPH particles in the temperature vs. density
plane for eight different runs. As can be seen, in each case, the gas does
attain the characteristic values of the temperature and density, $T\sim\mbox{
a few 100 K}$ and $n\sim 10^{3}-10^{4}$ cm$^{-3}$. In this preferred region
of $T-n$ space, the evolution of the system slows down, allowing to imprint
the corresponding characteristic Jeans scale of $M_{J}\sim 10^{3}M_{\odot}$
onto the gas. From examining Figures 19 and 20, it is also evident that
the gas particles begin their runaway collapse with these characteristic
values, to swiftly attain a temperature of $T\sim 1000$ K at the theshold
density of $n=10^{8}$ cm$^{-3}$, at which point they are incorporated into
a sink particle (i.e., a clump).

We turn next to the discussion of how the high-density clumps subsequently
evolve, and of the processes through which they grow in mass, the accretion
of surrounding diffuse material and the merging of clumps.

\subsection{Later Evolution}

At the end of the free-fall phase, the primordial cloud has fragmented
into clumps with initial masses of $M_{Cl}\sim 10^{2}-10^{3}M_{\odot}$.
These clumps are the basic elements in the building-up of a spectrum
of masses through the processes of accretion and merging. To address
the complex dynamics which is involved in the shaping of the mass function,
it is essential to be able to follow the systems's evolution over a few
dynamical timescales. The technique of creating sink particles allows us to
do so, which is another important advantage of our approach.

In the following, we first discuss the physics of accretion and merging, and
then the resulting distribution of clump masses.

\subsubsection{\it Accretion and Merging of Clumps}

The next question to ask is: {\it What determines the fraction of gas which
ends up in high-density clumps, and how does the difference in this quantity
between the 
various simulations arise?} With $M_{B}=10^{5}M_{\odot}$ being the total
amount of gas, and $t_{dyn}\sim 10^{7}$ yr the initial dynamical timescale
of the DM halo, the overall rate of conversion between diffuse gas and clumps
can be estimated to be $\dot{M}_{conv}\simeq M_{B}/t_{dyn}\simeq 10^{-2}
M_{\odot}$ yr$^{-1}$. The conversion rate, $\dot{M}_{conv}$, sets an upper
limit to the star formation rate (SFR). Feedback effects from the developing
protostars are expected to limit the SFR to a somewhat smaller value which
at present is not known with any certainty.

\begin{center} 
\hspace{0.5cm}
\psfig{file=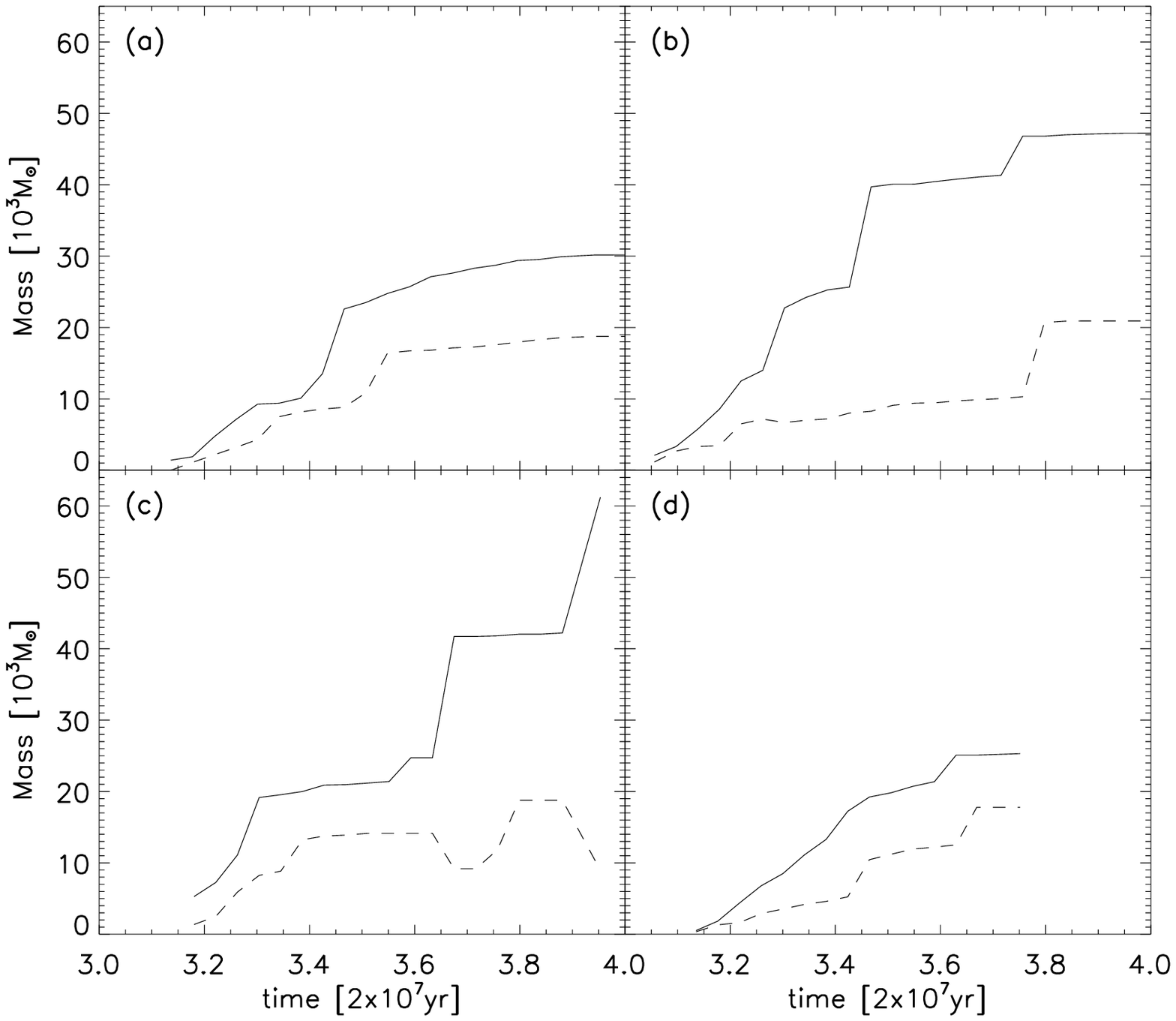,width=7.56cm,height=6.804cm}
\vspace{15pt}
\figcaption{
Growth of clumps in different runs. 
{\it Solid line:} Mass of most massive clump 
vs. time.
{\it Dashed line:} Mass of second most massive clump vs. time.
Mass is plotted in units of $10^{3}M_{\odot}$, and time in $2\times 10^{7}$yr,
which corresponds to the initial dynamical timescale.
{\bf (a)} Run A: Fiducial case. {\bf (b)} Run C: Low-spin
case. {\bf (c)} Run E: $P(k)\propto k^{0}$. {\bf (d)} Run K: Different
realization of $P(k)\propto k^{-3}$.
A rapid rise in
mass (a `step') corresponds to a merger with another clump, whereas a phase
of slow and steady growth is due to accretion of diffuse gas.
Notice the similarity in
panels (a) and (d), with two dominant clumps surviving without further merging.
The simulations in panels (b) and (c), corresponding to runs which result in
a more centrally concentrated morphology, 
on the other
hand, build up one very massive clump by successive merger events.
\label{fig21}}
\end{center} 

The clumps are comprised of gas which at some point has become Jeans unstable,
and which has been drawn from the reservoir of cooled gas at the characteristic
values of temperature and density, $T\sim$ a few 100 K and $n\simeq 10^{3}-
10^{4}$ cm$^{-3}$. In approximate pressure equilibrium with the cooled gas,
a second, hot phase has formed at $T\sim 10^{4}$ K and $n\sim 10^{1}$ cm$^{-3}$.
Whatever material resides in this hot phase is not available for the
formation of clumps. The total amount of cooled gas is typically 
$M_{cool}\sim 8 - 9 \times 10^{4}M_{\odot}$.
Once there is not enough gas left to overwhelm the opposing pressure, 
i.e., $M < M_{J}$, the Jeans instability ceases. If $\overline{M}_{J}$ is the
average Jeans mass at the moment of virialization, where the average includes
all SPH particles which have been able to cool, the baryonic mass fraction in
clumps is approximately given by
\begin{equation}
f_{Cl}\simeq \frac{M_{cool}-\overline{M}_{J}}{M_{B}}
\mbox{\ \ \ .}
\end{equation}
In Table 4, we summarize the resulting conversion efficiency for three 
different simulations. The two simulations (Runs C and E) with a more
concentrated gas morphology and correspondingly higher gas density are
characterized by $f_{Cl}\sim 0.7$, as compared to $f_{Cl}\sim 0.5$ in the 
case of the less centrally concentrated gas configuration in Runs A and K.
As can be seen, equation (36) nicely describes the numerical results. All
runs have 
a similar average temperature $\overline{T}\sim 300$ K, but the average gas
density is an order of magnitude larger in Runs C and E. For these latter
two simulations, therefore, the corresponding Jeans mass is smaller, and
less of the cooled gas is left behind in a pressure supported state.
In general, the conversion efficiency $f_{Cl}$ increases with the central
concentration and density of the gas. The larger central concentration is
due to the low degree of angular momentum in Run C, and to the absence of
a significant deviation from spherical symmetry in Run E.
The fraction $f_{Cl}$
constitutes an upper limit for the star formation efficiency (SFE), with
the same degree of uncertainty as in the case of $\dot{M}_{conv}$ and the
SFR.

In Figure 21, we show how the two most massive clumps grow in mass for
four different simulations, corresponding to the cases in Table 4.
The clumps are formed with initial masses
close to $M\sim 10^{3}M_{\odot}$, and then gain in mass by the slow accretion
of surrounding gas, and by merging with other clumps. Both mechanisms can be
clearly discerned in the figure, where the merging events correspond to the
step-like, sudden increase in mass. It is evident that there is significant
merging activity in the high-density simulations (Runs C and E),
whereas in Runs A and K, the clumps at late times only grow by steady accretion.
This difference can be understood by considering the timescale for the
collision of clumps, $t_{coll}$, and the corresponding collision rate
(e.g., Bonnell et al. 1998)
\begin{equation}
\frac{1}{t_{coll}}=16\sqrt{\pi}n_{Cl}v r_{acc}^{2}\left[
1 + \frac{G M_{Cl}}{2 v^{2} r_{acc}}\right]
\mbox{\ \ \ .}
\end{equation}
Here, $n_{Cl}$ is the number density of clumps, $v$ the velocity dispersion,
$r_{acc}$ the accretion radius, and $M_{Cl}$ the mass of a clump. The second
term in the brackets is the Safronov number, describing the effect of
gravitational focusing. Estimating the accretion radius as $r_{acc}\simeq
2 h_{Cl}\simeq 0.1$ pc (see Section 3.2), and taking the clump mass to be
$M_{Cl}\simeq 20,000M_{\odot}$, the result of evaluating equation (37) is
summarized in Table 5. We accordingly expect of order 5 merger events in 
the high-density simulations (Runs C and E), as opposed to only 1 event in
the low-density cases (Runs A and K). As can be seen in Figure 21, this
prediction is borne out in the numerical simulations. A clump can become
very massive by undergoing multiple mergers, up to $\sim 50,000M_{\odot}$ in
Run C, and $\sim 60,000M_{\odot}$ in Run E. This runaway growth of one
central clump is analogous to the evolution of a supergiant cD galaxy in the
center of rich clusters of galaxies.
From Figure 21, it is evident that at late evolutionary stages, clumps grow
in mass mainly by merging with other clumps. In comparison, accretion of
surrounding gas is a rather inefficient process.

\subsubsection{\it Distribution of Clump Masses}

In this section, we discuss the clump mass spectrum which results from
the complex dynamics of merging and accretion described above.
Although the initial masses of the clumps are close to $\sim 10^{3}M_{\odot}$,
rather independent of the initial conditions, the subsequent evolution of the
clumps proceeds differently from case to case. As we have seen, the efficiency
of merging is very sensitive to the central density of the post-virialization
cloud. Despite the existence of a characteristic mass scale for Population III
star formation, the mass spectrum, being determined by the merging history
of the clumps, is therefore expected to vary significantly.
In the following, we investigate
how the distribution of clump masses depends on the initial conditions.
For each simulation, the gas and clump morphology is shown at a late
evolutionary stage, where most of the available, cooled gas has already been
incorporated into clumps.

\begin{center} 
\psfig{file=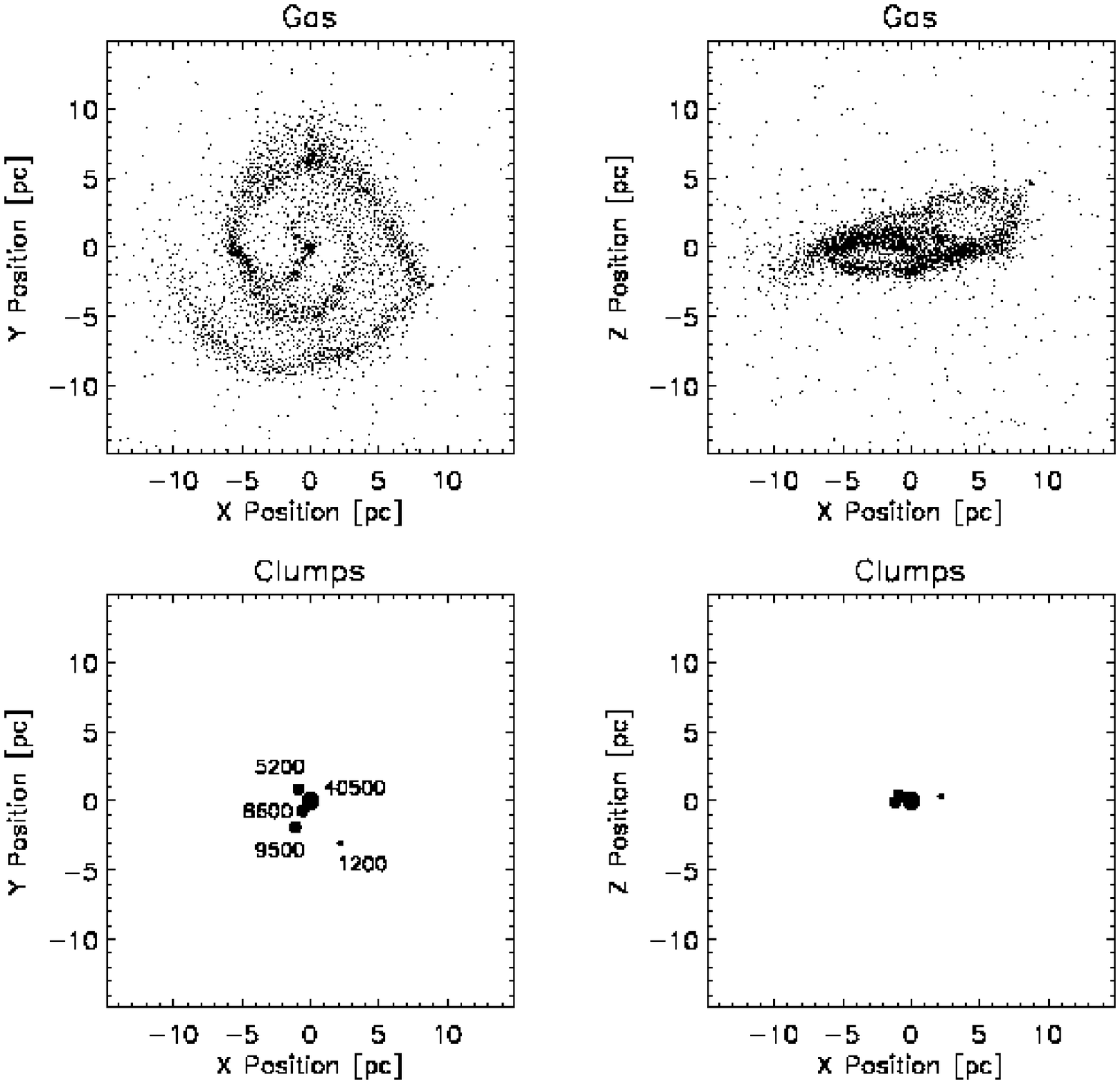,width=8.4cm,height=7.56cm}
\figcaption{
Run C: Low-spin case at $z=28.9$.
{\it Top row:} The remaining gas in the diffuse phase.
{\it Bottom row:} Distribution of clumps. The numbers next to the dots
denote clump mass in units of $M_{\odot}$.
{\it Left panels:} Face-on view.
{\it Right panels:} Edge-on view.
The length of the box is 30 pc.
Dominated by a massive clump of $\sim 40,000 M_{\odot}$, comprising
$\sim$40\% of the initially present gas, a compact, disk-like feature
has formed in the center of the DM potential.
\label{fig22}}
\end{center} 
\begin{center} 
\psfig{file=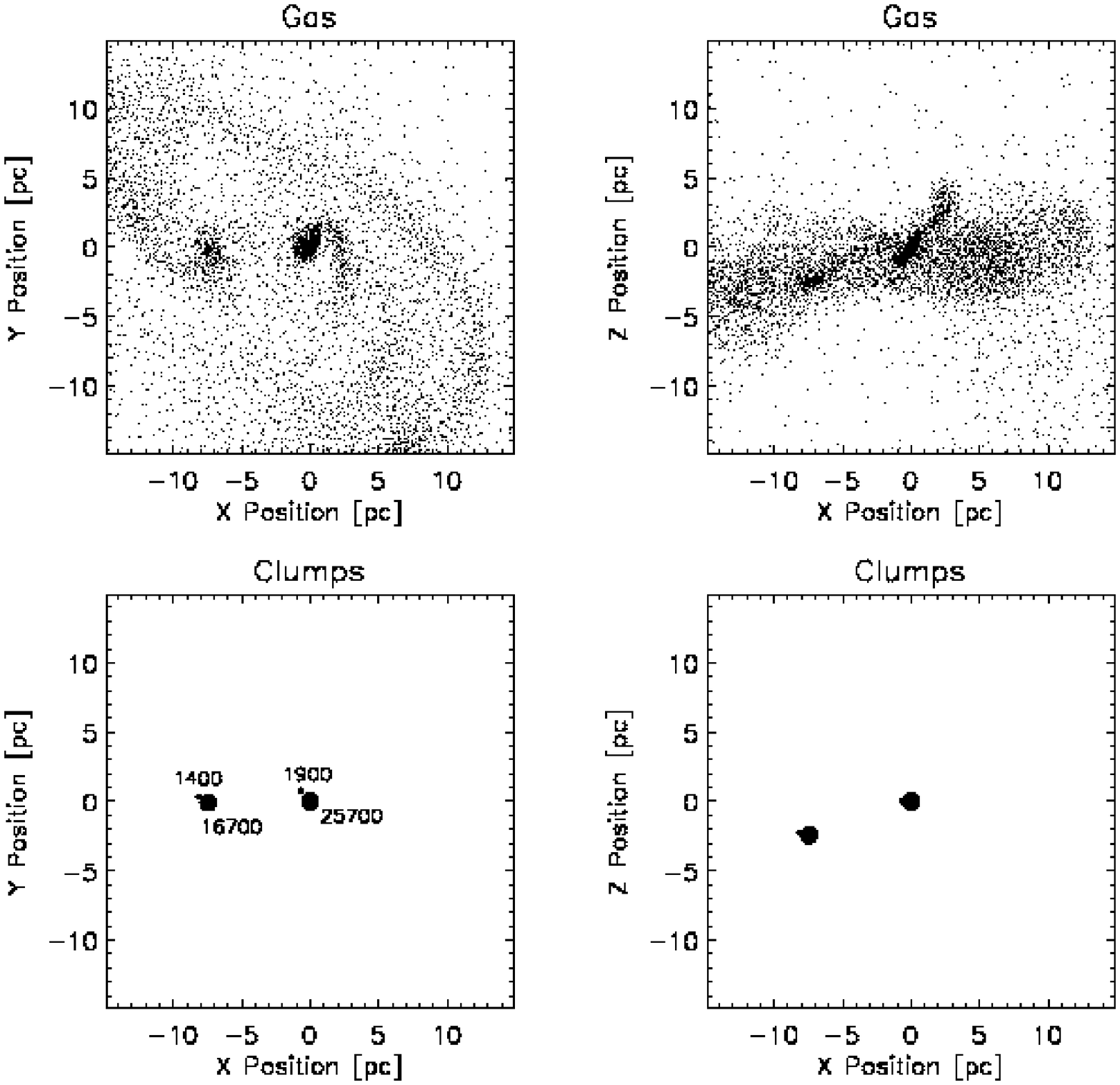,width=8.4cm,height=7.56cm}
\figcaption{
Run A: Intermediate-spin case at $z=28.9$.
The manner of presentation is the same as in Figure 22.
Compared to the low-spin case, the gas has settled into a less regular, more
extended configuration with two dominant clumps of mass close to 
$20,000 M_{\odot}$. During the subsequent evolution, the clumps survive
without merging, and grow in mass only slightly by accretion of surrounding
gas.
\label{fig23}}
\end{center} 
\begin{center} 
\psfig{file=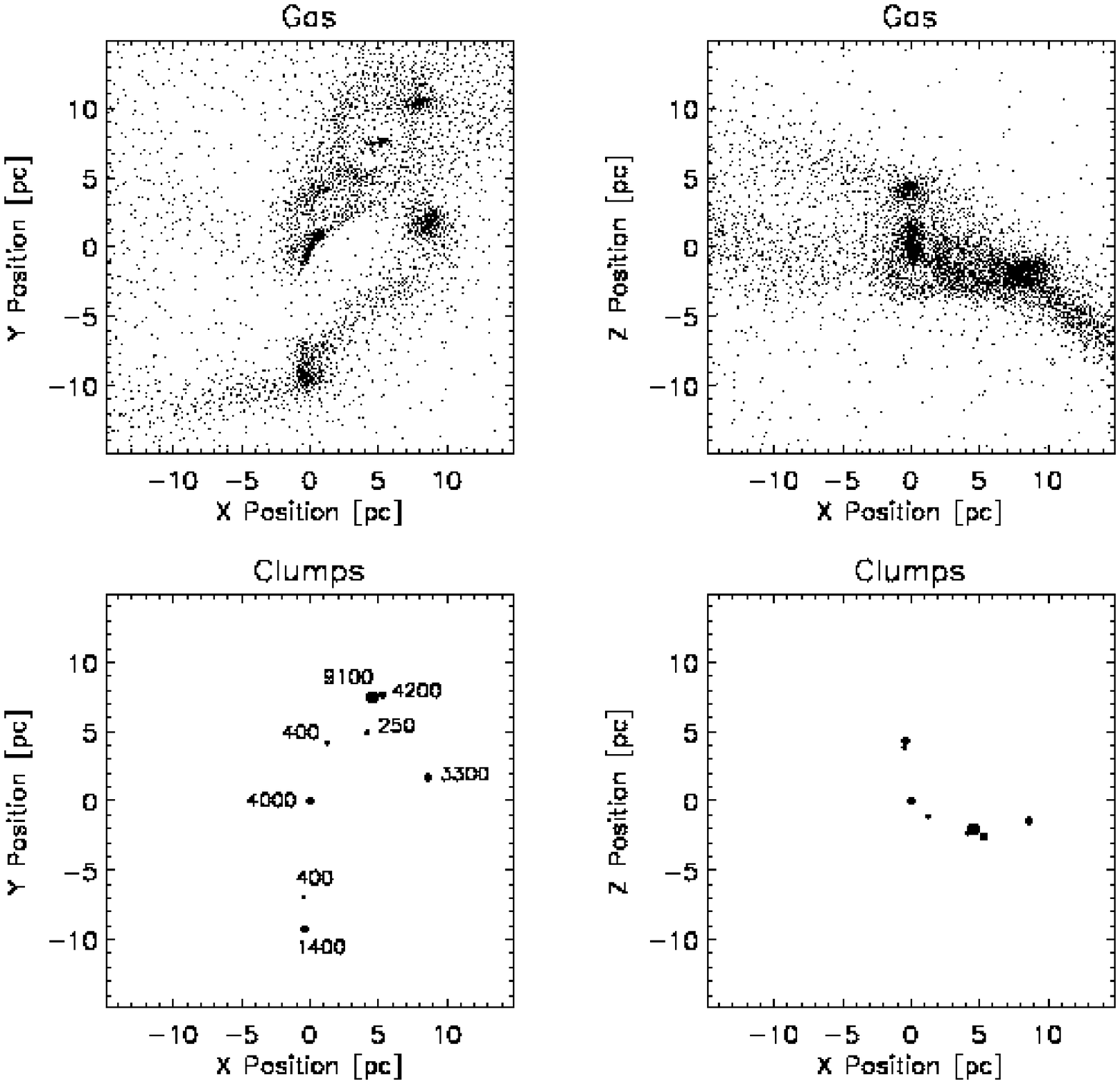,width=8.4cm,height=7.56cm}
\figcaption{
Run D: High-spin case at $z=25.5$.
The manner of presentation is the same as in Figure 22.
Notice that here the situation is shown at a much later instant, 
reflecting the delayed conversion of the diffuse gas into clumps.
The gas morphology is highly irregular and dispersed. In this case,
no very massive clumps have formed, but instead a number of clumps
with masses of a few $10^{3}M_{\odot}$.
\label{fig24}}
\end{center} 

The total angular momentum of a halo proves to be a very important parameter
in determining the mass spectrum of the clumps. To demonstrate this, we consider
the simulations with low, intermediate, and high initial spin in Figures 22,
23, and 24, respectively.
As can be clearly seen, the mass of the dominating clump decreases with
increasing spin. This clump mass - spin relation has a straightforward
explanation in the lower central density of the higher spin simulations, and
the corresponding reduction in the merger rate. Hence, high-density
clumps with masses $M\ga 50,000M_{\odot}$ can form in the center of the 
lowest-spin halos. These massive clumps might conceivably lead to the formation
of seed black holes for future quasar activity. Eisenstein \& Loeb (1995) have
investigated a scenario along these lines, emphasizing the cosmological
importance of the low-spin systems.

The emergence of such very massive (cD like) clumps is due to the
efficient outward transport of angular momentum via tidal
torques. Although we do not include the effect of any external tidal
fields, and have to insert the angular momentum explicitly at the
beginning of the simulation, the subsequent dynamics does lead to
internal tidal fields. In cases where there is such a "cD behavior", the 
effect responsible for it therefore seems to be modelled in a physically
consistent way. Interestingly, a high degree of central concentration
is also found in the simulation of ABN2000 who have 
self-consistently included cosmological tidal fields.
As we discuss below, the neglect of any proto-stellar
feedback seems to be the more crucial factor in accounting for the
emergence of very massive clumps.

Earlier on, we have mentioned that the two simulations (Runs A and K) with
a different realization of the $k^{-3}$ power spectrum, and all other parameters
being equal, lead to a rather different DM and gas morphology at the 
moment of virialization. During the later stages of the evolution, however,
these simulations converge to a similar state.
In both simulations, the morphology is dominated by two clumps
with masses close to
$\sim 20,000M_{\odot}$.
The overall features of the mass spectrum, therefore, do not
seem to depend significantly on the random initialization of the DM
fluctuations.
The late-time morphology in the simulation with 
a high baryon fraction (Run L) is characterized by the largest
fraction of gas in clumps, $\sim$ 75 \%, and has a central cD clump of mass
$M\sim 200,000 M_{\odot}$, which has gobbled up a remarkable one-half of the
total gas mass. As we have pointed out before, the case of a high baryon fraction
behaves distinctly different from the low $\Omega_{\mbox{\scriptsize B}}$
simulations in that gravitational instability is already triggered during
the initial free-fall phase. Now we see that Run L is also distinguished by
an extremely top-heavy spectrum of clump masses. This case serves as an
illustration for what happens when the relative balance of gas pressure and
self-gravity is shifted in favor of the latter.

To assess the possible role of HD cooling in the evolution of the primordial
gas, we compare Run A, which does include HD cooling, and
Run F, which does not. Otherwise, the two cases have identical initial
conditions. We find that the gas and clump morphologies, as well as
the thermodynamic behavior in the $T$ vs. $n$ plane, are overall very similar.
Slight differences, however, do exist in the way the gas fragments, and in the
resulting clump masses. Cooling due to the HD molecule, and the corresponding
chemistry of its formation and destruction, should therefore be included for
completeness, even though it does not appear to be of pre-eminent importance
in the parameter regime considered here. In our simulations, we have assumed
a deuterium abundance of $n_{\mbox{\scriptsize D}}=4\times 10^{-5}
n_{\mbox{\scriptsize H}}$, but we have carried out a test calculation with
a ten times higher abundance. With such a high D abundance, the thermal evolution
of the gas proceeds very differently. Cooling now is so efficient that the gas
quickly settles down to the temperature of the CMB, and the corresponding
Jeans mass is reduced below the resolution limit of that simulation,
$M_{res}\sim 200 M_{\odot}$. If deuterium were indeed that abundant, the
characteristic mass scale of $\sim 10^{3}M_{\odot}$ would then disappear,
since it derives from the properties of H$_{2}$ (see Section 2).
Our adopted value of $n_{\mbox{\scriptsize D}}/n_{\mbox{\scriptsize H}}=
4\times 10^{-5}$, however, seems to be a conservatively high choice,
according to recent observations of high-redshift absorption systems (e.g.,
Burles \& Tytler 1998).

Finally, we discuss the cumulative distribution of clump masses.
We consider the average of all simulations with DM fluctuations
imprinted according to a $k^{-3}$ spectrum, initial spin $\omega=0.2$, and a
total halo mass of $2\times 10^{6}M_{\odot}$. As we have discussed above, 
varying these parameters leads to a significant change in the distribution
of clump masses, and we have therefore included in the averaging only the cases
with these fiducial values.
We evaluate the number of clumps per unit logarithmic mass, and compare this to
the Salpeter case of the present-day {\it stellar} IMF, $\mbox{d}N/\mbox{d}\log m
\propto m^{-1.35}$. It turns out that the resulting mass spectrum is
remarkably flat, and that the halos in our simulations are very efficient
in building up very massive clumps, up to a few times $10^{4}M_{\odot}$.

This mass spectrum has to be taken {\it cum grano salis}, and important caveats
apply. First, recall that clumps are born with masses close to $\sim 10^{3}
M_{\odot}$, and only subsequently reach higher masses through successive mergers
and accretion. For these mechanisms to be effective, there has to be
sufficient time. Typical merging timescales are of order a few $10^{6}$ yr,
during which time the first stars in the halo might have already evolved far
enough to explode as supernovae, thereby considerably disturbing the
remaining gas in the halo. In general, the neglect of any feedback effects
constitutes the major shortcoming in our treatment of these later 
evolutionary stages. Up to the point of forming the first clump, and possibly
the first few, all the relevant physics is in principle known. It is
only towards the later stages in the evolution that the physical basis of
the simulations becomes increasingly uncertain. It is also worth
remembering that only those processes are allowed for in our numerical
treatment which tend to {\it increase} the mass of a clump. This introduces
a bias towards higher mass in the resulting mass spectrum.

Having stated all these provisions, it is nevertheless remarkable how
efficient the Population III halos are in building up clumps of very high mass.
The assessment of how relevant this finding is for an understanding
of the stellar IMF, has to await a more realistic treatment of the complex
physics of accretion from a dust-free envelope, the merging of subcondensations,
and the various feedback effects.

\section{SUMMARY AND CONCLUSIONS}

We have investigated the collapse and fragmentation of primordial, metal-free
gas. The gas is embedded in dark matter halos of mass close to $\sim 10^{6}
M_{\odot}$ which virialize at redshifts $z\simeq 20 - 30$. Due to the
low virial temperatures in these halos of a few 1000 K, cooling can only 
proceed via H$_{2}$. We find that for these systems there exists a preferred
region in parameter space. The primordial gas does attain characteristic
temperatures of a few 100 K, and densities of $\sim 10^{3} - 10^{4}$ cm$^{-3}$.
These values have their physical explanation in the microphysics of H$_{2}$
cooling, related to the minimum temperature that can be reached via H$_{2}$, and
to the critical density where the cooling rate changes from being proportional
to $n^{2}$ to an only linear dependence on density. This change in the cooling
rate reflects the transition from NLTE to LTE level populations of the
hydrogen molecule. With these values of temperature and density, the corresponding
characteristic Jeans mass is $M_{J}\sim 10^{3}M_{\odot}$. At some point
during the simulation, the gas becomes gravitationally unstable, and
forms high density ($n>10^{8}$ cm$^{-3}$) clumps with initial masses
close to the characteristic Jeans scale of $\sim 10^{3}M_{\odot}$. This is
a very robust result, and is quite independent of the initial conditions.
This suggests that Population III star formation might have favored massive
stars, possibly even very massive ones with $M\ga 100 M_{\odot}$.

In contrast to this robust nature of the thermodynamics, the
resulting morphology is a somewhat accidental feature of our
simulations, varying substantially between different random realizations
of the initial DM density field.

Although the clumps form initially with similar masses close to the 
characteristic value of $\sim 10^{3}M_{\odot}$, their further evolution is
very sensitive to the initial conditions. 
Clumps grow in mass through accretion of surrounding gas,
and merging with other clumps. Both mechanisms sensitively depend
on the central gas density.
We now discuss the dependence of the resulting clump masses on the most
important parameters. 

\noindent
{\it Total halo mass.}---
With increasing mass of the DM halo, the clump masses also
tend to increase. Initial clump masses are close to $M_{Cl}=400 M_{\odot}$
in the $10^{5} M_{\odot}$ halo, and approximately twice as massive in the
$10^{6} M_{\odot}$ cases. In addition, a few clumps form almost simultaneously
in the more massive halos, whereas only a single one forms in the
center of the `marginal' ($10^{5} M_{\odot}$) halo. 

\noindent
{\it Angular momentum.}---
The growth of a clump
by accretion and merging is very sensitive to the amount of angular
momentum initially imparted to the DM halo. Lower spin halos acquire
denser central configurations which favor the frequent merging of clumps.
Clump masses can then become very large (up to a few times $10^{4} M_{\odot}$).

\noindent
{\it Collapse redshift.}---
The efficiency of accretion and merging is also increased in runs 
with higher collapse redshifts, which is again a direct consequence
of the enhanced overall density. As a result, clump masses tend
to be larger at higher redshift.

\noindent
{\it Baryon fraction.}---
In the simulation with a larger fraction
of baryons relative to the DM, runaway collapse is already triggered during
the initial free-fall collapse, and resulting clump masses are systematically
higher.

Recently, ABN2000 have used the adaptive
mesh refinement (AMR) method to study the formation of the first stars.
These authors start their calculation with cosmological initial
conditions, resulting in the most realistic treatment of the
problem to date. Since they have presented only one case, however, it is
not easy to ascertain the robustness of their result, and this
is the specific advantage of our exploratory approach.
We agree on the existence of characteristic values
for the density and temperature of the primordial gas, deriving
from the microphysics of H$_{2}$ cooling. ABN2000 give an upper limit
for the final mass of a Population III star of $200 M_{\odot}$, whereas
the distribution of clump masses in our work suggests that somewhat
higher masses, perhaps up to $1000 M_{\odot}$, are possible. ABN2000 have
criticised our earlier work (Bromm et al. 1999) in that the rather
smooth gaseous disk found in that study is unrealistic and an artefact
of the assumed top-hat initial conditions. Most of the cases presented
here, however, do not yield such a regular disk morphology, and result
in very irregular configurations that are dominated by filaments and
knots. Furthermore, our Run H, corresponding to a less massive
halo ($M_{tot}=10^{5} M_{\odot}$), is in good overall agreement with the
simulation of ABN2000 as to the morphology and the fact
that only one clump forms in the center of the DM halo.

Although one has a reasonably strong case in arguing that all the relevant
physics is in hand to follow the collapse of the primordial gas up to the
formation of the first high-density region, this fortunate circumstance
gives way to growing uncertainty afterwards.
Two of the most tentalizing questions are the following. {\it How effective
is the energy output from the protostellar core in shutting off accretion
from a very massive, dust-free envelope?} On this question hinges the
ultimate mass scale of the first stars, and it is at present far from
being answered.
In Paper II,however, we will further address the question of the upper mass
limit of a Population III star. 
This upper mass limit is important for 
testing the idea
that Population III stars might be the precursors of
black holes of mass $\sim 10^{3}M_{\odot}$, intermediate between stellar
and supermassive ones (e.g., Madau \& Rees 2001).
{\it What is the IMF of Population III stars?}
In our simulations, we have seen how the mass spectrum of clumps is built up
through the complex interplay of merging and accretion, and that the resulting
mass spectrum is very flat compared to the Salpeter IMF. It is difficult
to tell, however, how this result relates to the stellar IMF, since again we
are hampered by our neglect of any negative feedback effects, which are
almost certain to play an important role in determining the IMF.
Both questions highlight the importance of feedback effects for a deeper
understanding of the star formation process.

\acknowledgments{We are grateful to A. Ferrara, A. Loeb, M. Norman,
and M. Rees for comments and stimulating discussions.
We would like to thank Z. Haiman and A. Loeb for
providing us with their chemical reaction rates and
L. Hernquist for making available to us a version of TREESPH.
Support from the NASA ATP grant NAG5-7074 is gratefully acknowledged.
}

\begin{appendix}

\section{A. N-body Solver}

In the following, we briefly discuss the basic ideas of solving
the gravitational N-body problem, and refer the reader to
Hernquist \& Katz (1989; HK89 henceforth),
Aarseth (1994), and Barnes (1998) for
further details. In the context of TREESPH, the treatment of self-gravity
is almost identical for the dark matter and the gas, since both components
are represented by particles which constitute a Monte-Carlo sampling
of the underlying fluids. Let us first turn to the dark matter, and
describe the minor modifications for the gas at the end of this section.

Since particle methods are Lagrangian by design, mass is conserved
automatically, thus rendering the equation of continuity
superfluous. The equation of motion for DM particle $i$ simply reads:
\begin{equation}
\frac{\mbox{d}\vec{v}_{i}}{\mbox{d}t}=-\left(\nabla \Phi\right)_{i}
\mbox{\ \ \ .}
\end{equation}
The gravitational potential is given by the standard solution to Poisson' s
equation:
\begin{equation}
\Phi_{i}=-G\int \frac{\rho (\vec{r})}{|\vec{r}-\vec{r}_{i}|}\mbox{d}^{3}r
\mbox{\ \ \ .}
\end{equation}
If one now were to assume the case of true point masses, the density could
be written as
\begin{equation}
\rho (\vec{r})=\sum_{j}m_{j}\delta(\vec{r}-\vec{r}_{j})
\mbox{\ \ \ ,}
\end{equation}
with $\delta(\vec{r})$ being the Dirac delta-function. Inserting equ. (A3)
into equ. (A2), and applying the gradient operator, yields the familiar
result for the gravitational acceleration of particle $i$:
\begin{equation}
-\left(\nabla \Phi\right)_{i}=G\sum_{j\neq i}m_{j}
\frac{\vec{r}_{j}-\vec{r}_{i}}{|\vec{r}_{j}-\vec{r}_{i}|^{3}}
\mbox{\ \ \ .}
\end{equation}
This straightforward procedure has two crucial shortcomings which we now
discuss in turn, together with the adopted remedies.

First, there is the problem of discreteness on very small scales. Close
encounters of particles lead to collisional two-body relaxation, which is
clearly undesirable in modelling a smooth fluid. Therefore, gravitational
forces have to be softened on scales close to the interparticle
distance. In TREESPH, this is accomplished by replacing the singular Dirac
delta-function with the spherical spline kernel $W$, used in the SPH
formalism (see Appendix B):
\begin{equation}
\rho (\vec{r})=\sum_{j}m_{j}W(\vec{r}-\vec{r}_{j}; \epsilon)
\mbox{\ \ \ .}
\end{equation}
The result of inserting this {\it Ansatz} into equ. (A2) is given by HK89.
Due to its compact nature ($W(r)=0$ for $r > 2\epsilon$), the use
of this kernel reproduces the Kepler-potential exactly for $r > 2\epsilon$.
Each particle has its own softening length, $\epsilon_{i}$, which
is adjusted such that there is a constant number of neighbors, $N_{s}$, in
a softening volume: $N_{s}\simeq n_{i} \left(2\epsilon_{i}\right)^{3}$, where
$n_{j}$ is the local particle density. In our simulations, we have set
$N_{s}=16$.

The second problem concerns the computational expense of determining
the gravitational forces, which scales as $O(N^{2})$ for the direct
summation in equ. (A4). An ingenious way to overcome this prohibitive
cost has been developed by Barnes \& Hut (1986). Their method organizes the
particles into a recursive tree structure, such that individual particles
correspond to the leaves of the tree, neighboring particles in space to
leaves on the same branch, and the system as a whole to the root of the tree.
Each node of this tree represents a cubical cell in real space. The leaves
correspond to the smallest cells, containing only one particle, higher-level
branching points to larger cells, containing groups of particles, and
the root to the overall system box, including all the particles. This 
procedure is completely adaptive, and can accomodate arbitrary geometries.
To finally calculate  the gravitational force on a given particle, the tree
is traversed from the root down. For each node (cell) of size $s$ and
distance $d$ to the particle, an accuracy criterion is evaluated:
\begin{equation}
\frac{s}{d} < \theta
\mbox{\ \ \ ,}
\end{equation}
where the parameter $\theta$ determines the desired precision, and is
chosen to be $\theta=0.8$ in our simulations. If this criterion is fulfilled,
all the particles in the given cell are treated as a single group whose
gravitational influence is approximately described by performing a multipole
expansion of the respective potential, in our case up to quadrupole order.
If, on the other hand, the accuracy criterion is not met, the tree is
walked down to the next level of refinement. This procedure is repeated
recursively, until either the criterion in equ. (A6) is fulfilled, or the tree descent
has reached the level of individual particles (the leaves). In consequence,
forces are calculated accurately but expensively only for nearby particles,
and approximately but fast for more remote ones. Both the construction
of the tree, and the subsequent descent of it, have a computational cost
of only $O(N\mbox{log}N)$, making simulations with larger numbers of
particles possible.

The equation of motion is solved explicitly using a standard, time-centered
leapfrog integrator. The position and velocity of particle $i$ are updated
according to:
\begin{eqnarray}
\vec{r}^{n+1/2}_{i}&=&\vec{r}_{i}^{n-1/2}+\vec{v}^{n}_{i}\Delta t_{i} 
+O\left(\Delta t_{i}^{3}\right)\nonumber \\
\vec{v}^{n+1}_{i}&=&\vec{v}_{i}^{n}+\vec{a}^{n+1/2}_{i}\Delta t_{i} 
+O\left(\Delta t_{i}^{3}\right)
\end{eqnarray}
Here, superscripts denote timesteps. Velocities are stored at times
which are offset by one-half a timestep from the positions and accelerations.
This (leapfrog) characteristic guarantees the second-order accuracy of the
algorithm.
Each particle is allowed to have its individual timestep, $\Delta t_{i}$,
which is chosen to fulfill the criterion
\begin{equation}
\Delta t_{i} \leq e_{tol}\frac{E_{i}}{a_{i}v_{i}}
\mbox{\ \ \ .}
\end{equation}
$E_{i}$ is the total energy of particle $i$, and $e_{tol}$
the tolerance parameter which we have set to be $e_{tol}=0.1$.

The gaseous component is treated in exactly the same way, with the exception
that in this case the gravitational softening length is always equal to the 
SPH smoothing length (see below). In Figure 1, we demonstrate that
the code does nicely reproduce the analytical top-hat solution, with
total energy and angular momentum being conserved to better than 5\%.

\section{B. The SPH Method}

We briefly describe the basic principles of the smoothed
particle hydrodynamics (SPH) method, and give an intuitive motivation
for the resulting equations. Further details are given in, e.g., Benz (1990),
Monaghan (1992), M\"uller (1998), and HK89.

The fluid is sampled by discrete particles, representing fluid elements.
To model a continuous medium, the mass of a particle is smoothed according
to
\begin{equation}
\rho(\vec{r})=\sum_{j}m_{j}W(\vec{r}-\vec{r}_{j};h)
\end{equation}
The smoothing kernel, $W$, is strongly peaked at $\vec{r}=\vec{r}_{j}$, and
normalized to give $\int W \mbox{d}^{3}r=1$, where integration is over all
space. TREESPH implements the spherically symmetric spline kernel
\begin{equation}
W(r,h)=\frac{1}{\pi h^{3}}\left\{
\begin{array}{ll}
1-\frac{3}{2}\left(\frac{r}{h}\right)^{2}+
\frac{3}{4}\left(\frac{r}{h}\right)^{3}, & 0\leq \frac{r}{h} \leq 1 \\
\frac{1}{4}\left[2-\left(\frac{r}{h}\right)\right]^{3},  
& 1\leq \frac{r}{h} \leq 2 \\
0, & \frac{r}{h} \geq 2 \\
\end{array}
\right. 
\end{equation}
The smoothing length, $h$, describes the spatial extent of a given SPH particle.
TREESPH assigns variable smoothing lengths to each particle, thereby 
introducing an adaptive spatial resolution, such that there is a (roughly)
constant number of particles, $N_{neigh}$, within the smoothing volume.
We have adopted $N_{neigh}=32$ in our work. Now, one can easily find 
smoothed estimates for any variable $A(\vec{r})$. Starting with the 
interpolation formula
\begin{equation}
A(\vec{r})=\int A(\vec{r}\mbox{\,}')
W(\vec{r}-\vec{r}\mbox{\,}';h)\mbox{d}^{3}r'
\mbox{\ \ \ ,}
\end{equation}
and assigning an effective volume to each particle $j$ via
$m_{j}\simeq \rho(\vec{r}_{j})\Delta^{3} r_{j}$, one can approximate the 
integral as
\begin{equation}
A(\vec{r})\simeq \sum_{j} A_{j}\frac{m_{j}}{\rho_{j}}
W(\vec{r}-\vec{r}_{j};h)
\mbox{\ \ \ ,}
\end{equation}
where $A_{j}= A(\vec{r}_{j})$, and $\rho_{j}= \rho(\vec{r}_{j})$.
The crucial advantage of the SPH method lies in the fact that it does not
need a grid to evaluate spatial derivatives. To illustrate this point, let us
apply the gradient operator to equ. (B4):
\begin{equation}
\nabla_{r}A(\vec{r})\simeq \sum_{j} A_{j}\frac{m_{j}}{\rho_{j}}
\nabla_{r}W(\vec{r}-\vec{r}_{j};h)
\mbox{\ \ \ .}
\end{equation}
Consequently, taking spatial derivatives in SPH only involves the analytical
differentiation of the kernel function which is specifically chosen to admit
this.

Next, we present the SPH equations as used in our numerical work. We give
heuristic arguments for their appearance, and refer to the cited literature
for the formal derivations. The equation of continuity is again
automatically fulfilled due to the Lagrangian nature of the SPH method.
The equation of motion for particle $i$ reads
\begin{equation}
\frac{\mbox{d}\vec{v}_{i}}{\mbox{d}t}=
-\left(\nabla \Phi\right)_{i} -\frac{\nabla P_{i}}{\rho_{i}}
 + \vec{a}_{\mbox{\scriptsize sh},i}
\mbox{\ \ \ .}
\end{equation}
To find a smoothed estimate for the pressure gradient, one could naively
write, using equ. (B5):
\begin{displaymath}
\frac{\nabla P_{i}}{\rho_{i}}=
\sum_{j} m_{j}\frac{P_{j}}{\rho_{i}\rho_{j}}
\nabla_{i}W(r_{ij};h_{i})
\mbox{\ \ \ .}
\end{displaymath}
This form, however, does not conserve linear and angular momentum. To
satisfy Newton' s third law, one wants an expression which is symmetric
with respect to any given pair of particles, $i$ and $j$. From amongst the
options given in TREESPH, we have chosen to work with
\begin{equation}
\frac{\nabla P_{i}}{\rho_{i}}=
\sum_{j}2 \frac{\sqrt{P_{i}P_{j}}}{\rho_{i}\rho_{j}}m_{j}
\frac{1}{2}\left[\nabla_{i}W(r_{ij};h_{i})+
\nabla_{i}W(r_{ij};h_{j})
\right]
\mbox{\ .}
\end{equation}
The symbol $\nabla_{i}$ denotes derivation with respect to $\vec{r}_{i}$,
and $r_{ij}=|\vec{r}_{i}-\vec{r}_{j}|$. 
The pressure is given by the ideal gas law
\begin{equation}
P_{i}=
\rho_{i}
\frac{k_{\mbox{\scriptsize B}}T}{\mu m_{\mbox{\scriptsize H}}}=
(\gamma -1) u_{i} \rho_{i}
\mbox{\ \ \ .}
\end{equation}
$\mu$ is the (dimensionles)
mean molecular weight, $\gamma$ the ratio of specific heats, and
$u_{i}$ the specific internal energy (in erg g$^{-1}$).
For an almost neutral gas consisting of helium and atomic hydrogen, one
has $\mu \simeq 1.2$, and $\gamma =5/3$.
The corresponding symmetric estimate
for the viscous acceleration $\vec{a}_{\mbox{\scriptsize sh}}=-\nabla
Q_{\mbox{\scriptsize eff}}/\rho$
can be written as:
\begin{equation}
\frac{\nabla Q_{\mbox{\scriptsize eff}, i}}{\rho_{i}}=
\sum_{j}\Pi_{ij} m_{j}
\frac{1}{2}\left[\nabla_{i}W(r_{ij};h_{i})+
\nabla_{i}W(r_{ij};h_{j})
\right]
\mbox{\ .}
\end{equation}
Finding $\Pi_{ij}$, a symmetric estimate for $Q_{\mbox{\scriptsize eff}}/
\rho^{2}$, relies on the following conceptual steps. First, let us specify
the pseudo-viscous pressure, $Q_{\mbox{\scriptsize eff}}=-\rho \nu_
{\mbox{\scriptsize eff}} \nabla \cdot \vec{v}$, arising from the presence of an
artificial viscosity. The latter has two contributions: A bulk viscosity
\begin{equation}
\nu_{\mbox{\scriptsize eff}}=\alpha l c_{s}
\mbox{\ \ \ ,}
\end{equation}
and a von Neumann-Richtmyer viscosity
\begin{equation}
\nu_{\mbox{\scriptsize eff}}=-\beta l^{2} \nabla \cdot \vec{v}
\mbox{\ \ \ .}
\end{equation}
Here, $l$ is a characteristic length, $c_{s}$ the sound speed, and
$\alpha$, $\beta$ are parameters of order unity. The resulting
pseudo-viscous pressure can then be written as
\begin{equation}
Q_{\mbox{\scriptsize eff}}=-\alpha \rho c_{s} l \nabla \cdot \vec{v}
+\beta \rho l^{2} \left(\nabla \cdot \vec{v}\right)^{2}
\mbox{\ .}
\end{equation}
A symmetric estimate for $l\nabla\cdot\vec{v}$ is given by
\begin{equation}
\mu_{ij}=\left\{
\begin{array}{ll}
\frac{h_{ij}\vec{v}_{ij}\cdot\vec{r}_{ij}}{r_{ij}^{2}+0.01 h_{ij}^{2}}
& \mbox{if \ } \vec{v}_{ij}\cdot\vec{r}_{ij}\leq 0 \\
0 &  \mbox{otherwise} \\
\end{array}
\right. 
\mbox{\ \ \ ,}
\end{equation}
where $\vec{v}_{ij}=\vec{v}_{i}-\vec{v}_{j}$, and $h_{ij}=(h_{i}+h_{j})/2$.
Assembling these ingredients, one finally has
\begin{equation}
\Pi_{ij}=\frac{-\alpha \mu_{ij}c_{ij}+\beta\mu_{ij}^{2}}{\rho_{ij}}
\mbox{\ \ \ ,}
\end{equation}
where $\rho_{ij}=(\rho_{i}+\rho_{j})/2$, and $c_{ij}=(c_{s,i}+c_{s,j})/2$.
In our work, we use $\alpha = 1$ and $\beta =2$. We have experimented
with alternative prescriptions for the artificial viscosity (as given by HK89),
and find that our results do not depend on this choice.
Certain morphological features such as pronounced rings, however,
could be due to an unphysical viscosity that can arise in simulations
with a relatively small number of particles.
The equation of motion is now fully specified, and is solved with the
leapfrog integrator. The stability of the integration is enforced
by the classic Courant condition:
\begin{equation}
\Delta t_{i}\leq\frac{h_{i}}{c_{s,i}+h_{i}|\nabla\cdot\vec{v}_{i}|}
\mbox{\ \ \ .}
\end{equation}
HK89 give a version of the criterion which also takes into account the
artificial viscosity.
The appropriate timestep for a given SPH particle is
the minimum of the Courant timestep and the one given by the energy criterion,
 equ. (A8).

Along similar lines, HK89 write the smoothed form of the thermal
energy equation, equ. (5), as
\begin{eqnarray}
\frac{\mbox{d}u_{i}}{\mbox{d}t}&=&
\sum_{j}m_{j}\left(\frac{\sqrt{P_{i}P_{j}}}{\rho_{i}\rho_{j}}+\frac{1}{2}
\Pi_{ij}\right)\vec{v}_{ij}  \\
 & &
\cdot\frac{1}{2}\left[\nabla_{i}W(r_{ij};h_{i})+
\nabla_{i}W(r_{ij};h_{j})
\right]+\left(\frac{\Gamma - \Lambda}{\rho}\right)_{i} 
\mbox{\ .} \nonumber
\end{eqnarray}
In the presence of radiative cooling and heating, this equation cannot be
solved explicitly, since the corresponding radiative timescales are typically
much shorter than the dynamical time which sets the timestep for the equation
of motion. The thermal energy equation is therefore integrated implicitly,
using the standard second-order trapezoidal rule (see HK89 for further
details).
\end{appendix}

\begin{deluxetable}{rllc} 
\footnotesize
\tablewidth{13.cm}
\tablecaption{Reaction Rates for Hydrogen Species \label{tab1}}
\tablecolumns{4}
\tablehead{
\colhead{} &
\colhead{} &
\colhead{Rate Coefficient} &  
\colhead{} \\ 
\colhead{} &
\colhead{Reaction} &  
\colhead{(cm$^{3}$s$^{-1}$)} &
\colhead{Reference} 
 } 
\startdata
 (1)& H + $e^{-}$ $\rightarrow$ H$^{+}$ + $2e^{-}$ & $5.85\times 10^{-11} 
T^{1/2}\mbox{exp}(-157,809.1/T)(1 + T^{1/2}_{5})^{-1}$ & 1 \nl
 (2)& H$^{+}$ + $e^{-}$ $\rightarrow$ H + $h\nu$ & $8.40\times 10^{-11} 
T^{-1/2}T_{3}^{-0.2}(1 + T^{0.7}_{6})^{-1}$ & 1 \nl
 (3)& H + $e^{-}$ $\rightarrow$ H$^{-}$ + $h\nu$ & See expression in
reference & 2 \nl
 (4)& H + H$^{-}$ $\rightarrow$ H$_{2}$ + $e^{-}$ & $1.30\times 10^{-9}$ 
 & 1 \nl
 (5)& H$^{-}$ + H$^{+}$ $\rightarrow$ 2H & $7.00\times 10^{-7} 
T^{-1/2}$ & 1 \nl
 (6)& H$_{2}$ + $e^{-}$ $\rightarrow$ H + H$^{-}$ & $2.70\times 10^{-8} 
T^{-3/2}\mbox{exp}(-43,000/T)$ & 1 \nl
 (7)& H$_{2}$ + H $\rightarrow$ 3H & See expression in reference & 1 \nl
 (8)& H$_{2}$ + H$^{+}$ $\rightarrow$ H$^{+}_{2}$ + H & $2.40\times 10^{-9} 
\mbox{exp}(-21,200/T)$ & 1 \nl
 (9)& H$_{2}$ + $e^{-}$ $\rightarrow$ 2H + $e^{-}$ & $4.38\times 10^{-10} 
\mbox{exp}(-102,000/T)T^{0.35}$ & 1 \nl
(10)& H$^{-}$ + $e^{-}$ $\rightarrow$ H + $2e^{-}$ & $4.00\times 10^{-12} 
T\mbox{exp}(-8750/T)$ & 1 \nl
(11)& H$^{-}$ + H $\rightarrow$ 2H + $e^{-}$ & $5.30\times 10^{-20} 
T\mbox{exp}(-8750/T)$ & 1 \nl
(12)& H$^{-}$ + H$^{+}$ $\rightarrow$ H$^{+}_{2}$ + $e^{-}$ & 
See expression in reference & 1 \nl
\enddata
\tablerefs{
(1)Haiman, Thoul, \& Loeb 1996; 
(2) Abel, Anninos, Zhang, \& Norman 1997. 
}
\end{deluxetable}

\begin{deluxetable}{rllc} 
\footnotesize
\tablewidth{13.cm}
\tablecaption{Reaction Rates for Deuterium Species \label{tab2}}
\tablecolumns{4}
\tablehead{
\colhead{} &
\colhead{} &
\colhead{Rate Coefficient} &  
\colhead{} \\ 
\colhead{} &
\colhead{Reaction} &  
\colhead{(cm$^{3}$s$^{-1}$)} &
\colhead{Reference}
 } 
\startdata
 (1)& D$^{+}$ + $e^{-}$ $\rightarrow$ D + $h\nu$ &  
  $8.40\times 10^{-11} 
T^{-1/2}T_{3}^{-0.2}(1 + T^{0.7}_{6})^{-1}$ & 1 \nl
 (2)& D + H$^{+}$ $\rightarrow$ D$^{+}$ + H  & $3.70\times 10^{-10} 
T^{0.28}\mbox{exp}(-43/T)$ & 3 \nl
 (3)& D$^{+}$ + H $\rightarrow$ D + H$^{+}$ & $3.70\times 10^{-10}T^{0.28}$
& 3 \nl
 (4)& D$^{+}$ + H$_{2}$ $\rightarrow$ H$^{+}$ + HD & $2.10\times 10^{-9}$ 
 & 3 \nl
 (5)& HD + H$^{+}$ $\rightarrow$ H$_{2}$ + D$^{+}$ & $1.00\times 10^{-9} 
\mbox{exp}(-464/T)$ & 3 
\enddata
\tablerefs{
(1)Haiman, Thoul, \& Loeb 1996; (3) Galli \& Palla 1998.
}
\end{deluxetable}

\begin{deluxetable}{lllrclcr} 
\footnotesize
\tablewidth{13.cm}
\tablecaption{Parameters for the Different Runs \label{tab3}}
\tablecolumns{8}
\tablehead{
\colhead{} &
\colhead{$M / M_{\odot}$} &
\colhead{$z_{vir}$} &  
\colhead{$n$} &
\colhead{$\omega$} &
\colhead{$\Omega_{B}$} &
\colhead{HD} & 
\colhead{N$_{SPH}$}
 } 
\startdata
Run A & $2\times 10^{6}$ & 30 & -3 & 0.2 &0.05 & Yes & 16384 \nl
Run B & $2\times 10^{6}$ & 30 & -3 & 0.2 &0.05 & Yes & 131072 \nl
Run C & $2\times 10^{6}$ & 30 & -3 & 0.1 &0.05 & Yes & 16384 \nl
Run D & $2\times 10^{6}$ & 30 & -3 & 0.4 &0.05 & Yes & 16384 \nl
Run E & $2\times 10^{6}$ & 30 & 0 & 0.2 &0.05  & Yes & 16384 \nl
Run F & $2\times 10^{6}$ & 30 & -3 & 0.2 &0.05  & No & 16384 \nl
Run G & $2\times 10^{6}$ & 20 & -3 & 0.2 &0.05  & Yes & 16384 \nl
Run H & $2\times 10^{5}$ & 30 & -3 & 0.2 &0.05 & Yes & 16384 \nl
Run K$^{a}$ & $2\times 10^{6}$ & 30 & -3 & 0.2 &0.05 & Yes & 16384 \nl
Run L & $2\times 10^{5}$ & 30 & -3 & 0.2 &0.20 & Yes & 16384 \nl
\enddata
\tablenotetext{a}{Run K has the same parameters as Run A, but with a different
realization of the random density field.}
\tablecomments{$n$ is the spectral index, $\omega$ the dimensionless
angular velocity, $\Omega_{B}$ the baryon fraction, and HD refers to the
absence or presence of HD cooling.}
\end{deluxetable}

\begin{deluxetable}{llcll} 
\footnotesize
\tablewidth{13.cm}
\tablecaption{Efficiency of Forming Clumps \label{tab4}}
\tablecolumns{5}
\tablehead{
\colhead{} &
\colhead{$t_{dyn}$} &
\colhead{$M_{cool}$} &  
\colhead{$<M_{J}>$} &
\colhead{$f_{Cl}$}\\
\colhead{} &
\colhead{[yr]} &
\colhead{[$M_{\odot}$]} &  
\colhead{[$M_{\odot}$]} &  
\colhead{}
 } 
\startdata
Run A & $6.0\times 10^{6}$ & $8.0\times 10^{4}$ & $3.0\times 10^{4}$ & 0.5 \nl
Run C & $5.4\times 10^{6}$ & $9.0\times 10^{4}$ & $2.0\times 10^{4}$ & 0.7 \nl
Run E & $3.0\times 10^{6}$ & $8.5\times 10^{4}$ & $1.5\times 10^{4}$ & 0.7 \nl
\enddata
\tablecomments{Parameters for the simulations in Figure 21.
$t_{dyn}$ is the dynamical time of the DM halo at the moment
of virialization, $M_{cool}$ the total amount of gas which has been able to
cool, $<M_{J}>$ the average Jeans mass at virialization, and $f_{Cl}$ the
baryonic mass fraction in clumps.
Run K has virtually the same parameters as Run A.}
\end{deluxetable}

\begin{deluxetable}{llcll} 
\footnotesize
\tablewidth{13.cm}
\tablecaption{Expected Merging of Clumps \label{tab5}}
\tablecolumns{5}
\tablehead{
\colhead{} &
\colhead{$n_{Cl}$} &
\colhead{$v$} &  
\colhead{$t_{coll}$} &
\colhead{$N_{merger}$}\\
\colhead{} &
\colhead{[pc$^{-3}$]} &
\colhead{[km s$^{-1}$]} &  
\colhead{[yr]} &  
\colhead{}
 } 
\startdata
Run A & $5\times 10^{-3}$ & $12.3$ & $1.5\times 10^{7}$ & $\sim 1$ \nl
Run C & $3\times 10^{-2}$ & $10.3$ & $3.0\times 10^{6}$ & $\sim 5$ \nl
Run E & $3\times 10^{-2}$ & $14.3$ & $3.0\times 10^{6}$ & $\sim 5$ \nl
\enddata
\tablecomments{Explaining the merging histories in Figure 21.
$n_{Cl}$ is the number density of clumps,
$v$ the velocity dispersion,
$t_{coll}$ the collision timescale, and $N_{merger}$ the
expected number of mergers in $\Delta t\simeq 2\times 10^{7}$ yr.
Run K has virtually the same parameters as Run A.}
\end{deluxetable}


\begin{references}

\reference{}Aarseth, S. J. 1994,
in Galactic Dynamics and N-Body Simulations, ed.
G. Contopoulos, N. K. Spyrou, \& L. Vlahos
(Berlin: Springer), 277
\reference{}Abel, T., Anninos, P., Norman, M. L., \& Zhang, Y. 1998, \apj, 508, 518
\reference{}Abel, T., Bryan, G. L., \& Norman, M. L. 2000, \apj, 540, 39
\reference{}Anninos, P., \& Norman, M. L. 1996, \apj, 460, 556
\reference{}Anninos, P., Zhang, Y., Abel, T., \& Norman, M. L. 1997, NewA, 2, 209
\reference{}Barnes, J. E., \& Hut, P. 1986, Nature, 324, 446
\reference{}Barnes, J., \& Efstathiou, G. 1987, \apj, 319, 575
\reference{}Barnes, J. E. 1998,
in Saas-Fee Advanced Course 26. Galaxies: Interactions and Induced Star
Formation, ed. 
D. Friedli, L. Martinet, \& D. Pfenniger
(Berlin: Springer)
\reference{}Bate, M. R., Bonnell, I. A., \& Price, N. M. 1995, MNRAS, 277, 362
\reference{}Bate, M. R., \& Burkert, A. 1997, MNRAS, 288, 1060
\reference{}Beers, T. C. 2000,
in ``The First Stars'', ESO Astrophysics Symposia, ed. A. Weiss, T. Abel,
\& V. Hill
(Berlin: Springer), 3
\reference{}Benz, W. 1990,
in Proceedings of the NATO Advanced Research Workshop
on The Numerical Modelling of Nonlinear Stellar Pulsations, ed.
J. R. Buchler
(Dordrecht: Kluwer), 269
\reference{}Blumenthal, G. R., Faber, S. M., Primack, J. R., \& Rees, M. J.
1984, Nature, 311, 517
\reference{}Bonnell, I. A., Bate, M. R., \& Zinnecker, H. 1998, MNRAS, 298, 93
\reference{}Bromm, V., Coppi, P. S., \& Larson, R. B. 1999, \apj, 527, L5
\reference{}Burles, S., \& Tytler, D. 1998, \apj, 507, 732
\reference{}Carlberg, R. G. 1981, MNRAS, 197, 1021
\reference{}Carr, B. J., Bond, J. R., \& Arnett, W. D. 1984, \apj, 277, 445
\reference{}Cen, R. 1992, \apjs, 78, 341
\reference{}Chen, H. W., Lanzetta, K. M., \& Pascarelle, S. 1999, Nature, 398, 586
\reference{}Ciardi, B., Ferrara, A., Governato, F., \& Jenkins, A. 2000,
MNRAS, 314, 611
\reference{}Copi, C. J., Schramm, D. N., \& Turner, M. S. 1995,
Phys. Rev. Lett.,75, 3981
\reference{}Couchman, H. M. P., \& Rees, M. J. 1986, MNRAS, 221, 53
\reference{}Cowie, L. L., \& Songaila, A. 1998, Nature, 394, 44
\reference{}Dodelson, S., \& Jubas, J. M. 1995, \apj, 439, 503
\reference{}Eisenstein, D. J., \& Loeb, A. 1995, \apj, 443, 11
\reference{}Fan, X. et al. 2000, AJ, 120, 1167
\reference{}Ferrara, A. 1998, \apj, 499, L17
\reference{}Flower, D. R., Le Bourlot, J., Pineau des For\^{e}ts, G., \&
Roueff, E. 2000, MNRAS, 314, 753
\reference{}Forrey, R. C., Balakrishnan, N., Dalgarno, A., \& Lepp, S.
1997, \apj, 489, 1000
\reference{}Fuller, T. M., \& Couchman, H.M.P. 2000, \apj, 544, 6
\reference{}Galli, D., \& Palla, F. 1998, A\&A, 335, 403
\reference{}Gunn, J. E., \& Peterson, B. A. 1965, \apj, 142, 1633
\reference{}Haiman, Z., Thoul, A. A., \& Loeb, A. 1996, \apj, 464, 523
\reference{}Haiman, Z., \& Loeb, A. 1997, \apj, 483, 21
\reference{}Hernquist, L., \& Katz, N. 1989, \apjs, 70, 419
\reference{}Hindmarsh, A. C. 1983,
in Scientific Computing, ed.
R. S. Stepleman et al.
(Amsterdam: North Holland)
\reference{}Hollenbach, D., \& McKee, C. F. 1979, \apjs, 41, 555
\reference{}Hu, E. M., Cowie, L. L., \& McMahon, R. G. 1998, \apj, 502, 99
\reference{}Hutchins, J. B. 1976, \apj, 205, 103
\reference{}Jeans, J. H. 1902,
Phil. Trans. R. Soc., 199A, 49
\reference{}Kashlinsky, A., \& Rees, M. J. 1983, MNRAS, 205, 955
\reference{}Kulsrud, R. M. 1997,
in ``Critical Dialogues in Cosmology'', ed. N. Turok
(Singapore: World Scientific), 328
\reference{}Larson, R. B. 1998, MNRAS, 301, 569
\reference{}Lawrence, C. R., Scott, D., \& White, M. 1999, PASP, 111, 525
\reference{}Lepp, S., \& Shull, J.M. 1983, \apj, 270, 578
\reference{}Loeb, A., \& Haiman, Z. 1997, \apj, 490, 571
\reference{}Loeb, A. 1998,
in ASP Conf. Ser. 133, Science with the Next Generation Space Telescope, ed.
E. Smith \& A. Koratkar
(San Francisco: ASP), 73
\reference{}Loeb, A. 1999, in ASP Conf. Ser. 193,
The Hy-Redshift Universe: Galaxy Formation and Evolution at High Redshift,
ed. A.J. Bunker \& W.J.M. van Breugel
(San Francisco: ASP), 586
\reference{}Lynden-Bell, D. 1967, MNRAS, 136, 101
\reference{}Madau, P., \& Rees, M.J. 2001, ApJ, 551, L27
\reference{}Martin, P.G., Schwarz, D.H., \& Mandy, M.E. 1996, \apj, 461, 265
\reference{}McDowell, M. R. C. 1961, Observatory, 81, 240
\reference{}Miralda-Escud\'{e}, J., \& Rees, M. J. 1997, \apj, 478, 57
\reference{}Miralda-Escud\'{e}, J., Haehnelt, M., \& Rees, M. J. 2000, \apj, 530, 1
\reference{}Monaghan, J. J. 1992, ARA\&A, 30, 543
\reference{}Moore, B., Ghigna, S., Governato, F., Lake, G., Quinn, T.,
Stadel, J., \& Tozzi, P. 1999, \apj, 524, L19
\reference{}M\"uller, E. 1998,
in Saas-Fee Advanced Course 27. Computational Methods for Astrophysical
Fluid Flow,
ed. 
O. Steiner \& A. Gautschy
(Berlin: Springer), 343
\reference{}Nakamura, F., \& Umemura, M. 1999, \apj, 515, 239
\reference{}Omukai, K., \& Nishi, R. 1998, \apj, 508, 141
\reference{}Ostriker, J. P., \& Gnedin, N. Y. 1996, \apj, 472, L63
\reference{}Padmanabhan, T. 1993, Structure Formation in the Universe
(Cambridge: Cambridge Univ. Press), 273
\reference{}Palla, F., Salpeter, E. E., \& Stahler, S. W. 1983, \apj, 271, 632
\reference{}Peacock, J. A. 1999, Cosmological Physics
(Cambridge: Cambridge Univ. Press), 381
\reference{}Press, W. H., Teukolsky, S. A., Vetterling, W. T., \&  
Flannery, B. P. 1992, Numerical
Recipes in Fortran (2nd ed.; Cambridge: Cambridge Univ. Press)
\reference{}Rees, M. J., \& Ostriker, J. P. 1977, MNRAS, 179, 541
\reference{}Rees, M. J. 1999,
in AIP Conf. Proc. 470, After the Dark Ages: When Galaxies were Young (the
Universe at $2<z<5$),
ed. S. S. Holt  \& E. Smith 
(Woodbury: AIP), 13
\reference{}Rees, M. J. 2000, New Perspectives in Astrophysical Cosmology
(Cambridge: Cambridge Univ. Press), 36
\reference{}Sellwood, J. A. 1987, ARA\&A, 25, 151
\reference{}Silk, J. 1977, \apj, 211, 638
\reference{}Silk, J. 1983, MNRAS, 205, 705
\reference{}Tegmark, M., Silk, J., Rees, M. J., Blanchard, A., Abel, T., \& Palla, F.
1997, \apj, 474, 1
\reference{}Uehara, H., Susa, H., Nishi, R., Yamada, M., \& Nakamura, T.
1996, \apj, 473, L95
\reference{}Vishniac, E. T. 1987, \apj, 322, 597
\reference{}White, M., Scott, D., \& Silk, J. 1994, ARA\&A, 32, 319
\reference{}Yoneyama, T. 1972, PASJ, 24, 87
\reference{}Zel'dovich, Y. B. 1970, A\&A, 5, 84


\end{references}
\end{document}